\documentclass[aps,prd,showpacs,notitlepage,nofootinbib,preprintnumbers,amsmath,amssymb]{revtex4-1}

	\usepackage{natbib}
	\usepackage{amsmath}
	\usepackage{makeidx}
	\usepackage{amsfonts}
	\usepackage[ansinew]{inputenc}
	\usepackage[usenames,dvipsnames]{pstricks}
	\usepackage{subfigure}
	\usepackage{epsfig}
	\usepackage{pst-grad} 
	\usepackage{pst-plot} 
	\usepackage[colorlinks,hyperindex]{hyperref}
	\usepackage{mathrsfs}
	\usepackage{bbold}
	\usepackage[margin=2cm,top=2cm, bottom=2cm]{geometry}
	\usepackage{physics}
	\usepackage{tensor}
	\usepackage{slashed}
	\usepackage{tikz}
	\usepackage{tikz-feynman}
	\usetikzlibrary{positioning}
	\tikzfeynmanset{compat=1.1.0}
	\usepackage{graphicx}
	\usepackage{xcolor}

	\usepackage[normalem]{ulem}

\def\eq#1{{Eq.~(\ref{#1})}}
\def\fig#1{{Fig.~\ref{#1}}}
\newcommand{\ben}{\begin{eqnarray*}}
\newcommand{\een}{\end{eqnarray*}}
\newcommand{\un}[1]{\underline{#1}}

\newcommand{\pd}{\partial}

\newcommand{\thalf}{\tfrac{1}{2}}

\newcommand{\stackeven}[2]{{{}_{\displaystyle{#1}}\atop\displaystyle{#2}}}

\newcommand{\gsim}{\stackeven{>}{\sim}}
\newcommand{\as}{\alpha_s}

	\hypersetup
	{
		colorlinks,%
		citecolor=black,%
		linkcolor=black,%
		urlcolor=black,%
	}

\begin{document}

\title{Lensing Mechanism Meets Small-$x$ Physics: \\ Single Transverse Spin Asymmetry in $p^{\uparrow}+p$ and $p^{\uparrow}+A$ Collisions }

\author{Yuri V. Kovchegov} 
         \email[Email: ]{kovchegov.1@osu.edu}
         \affiliation{Department of Physics, The Ohio State
           University, Columbus, OH 43210, USA}

\author{M. Gabriel Santiago}
  \email[Email: ]{santiago.98@buckeyemail.osu.edu}
	\affiliation{Department of Physics, The Ohio State
           University, Columbus, OH 43210, USA}

\begin{abstract}
We calculate the single transverse spin asymmetry (STSA) in polarized proton-proton ($p^{\uparrow}+p$) and polarized proton-nucleus ($p^{\uparrow}+A$) collisions ($A_N$) generated by a partonic lensing mechanism. The polarized proton is considered in the quark-diquark model while its interaction with the unpolarized target is calculated using the small-$x$/saturation approach, which includes multiple rescatterings and small-$x$ evolution. The phase required for the asymmetry is caused by a final-state gluon exchange between the quark and diquark, as is standard in the lensing mechanism of Brodsky, Hwang and Schmidt  \cite{Brodsky:2002cx}. Our calculation combines the lensing mechanism with small-$x$ physics in the saturation framework. The expression we obtain for the asymmetry $A_N$ of the produced quarks has the following properties: (i) The asymmetry is generated by the dominant elastic scattering contribution and $1/N_c^2$ suppressed inelastic contribution (with $N_c$ the number of quark colors); (ii) The asymmetry grows or oscillates with the produced quark's transverse momentum $p_T$ until the momentum reaches the saturation scale $Q_s$, and then only falls off as $1/p_T$ for larger momenta; (iii) The asymmetry decreases with increasing atomic number $A$ of the target for $p_T$ below or near $Q_s$, but is independent of $A$ for $p_T$ significantly above $Q_s$. We discuss how these properties may be qualitatively consistent with the data on $A_N$ published by the PHENIX collaboration \cite{Aidala:2019ctp} and with the preliminary data on $A_N$ reported by the STAR collaboration \cite{Dilks:2016ufy}. 
\end{abstract}


\maketitle

\tableofcontents



\section{Introduction}
\label{sec:Int}

The recent decade and a half saw a surge of research activity at the intersection of small-$x$ and spin physics in quantum chromodynamics (QCD) \cite{Boer:2006rj,Boer:2002ij,Boer:2008ze,Balitsky:2015qba,Balitsky:2016dgz,Altinoluk:2014oxa,Kovchegov:2013cva,Altinoluk:2015gia,Boer:2015pni,Kovchegov:2015pbl,Hatta:2016aoc,Dumitru:2015gaa,Chirilli:2018kkw,Jalilian-Marian:2018iui,Kovchegov:2019rrz,Boussarie:2019icw,Jalilian-Marian:2019kaf}. Topics receiving attention include both the longitudinal \cite{Kovchegov:2015pbl,Kovchegov:2016zex,Kovchegov:2016weo,Hatta:2016aoc,Kovchegov:2017jxc,Kovchegov:2017lsr,Kovchegov:2018znm,Cougoulic:2019aja,Kovchegov:2019rrz} and transverse \cite{Kovchegov:2012ga,Kovchegov:2013cva,Zhou:2013gsa,Kovchegov:2015zha,Hatta2016a,Hatta:2016khv,Boer:2016bfj, Kovchegov:2018zeq} spin physics of the proton. Of particular interest in the transverse spin category is the single transverse spin asymmetry (STSA) $A_N$. It is measured in polarized proton-proton ($p^{\uparrow}+p$) and polarized proton-nucleus ($p^{\uparrow}+A$) collisions, where a transversely polarized proton scatters on an unpolarized proton or nucleus. The asymmetry is defined as
\begin{equation}\label{ANdef}
A_N (p_T, y) = \frac{\frac{\dd{\sigma}^{\uparrow}}{\dd[2]{p_T} \dd{y}} -  \frac{\dd{\sigma}^{\downarrow}}{\dd[2]{p_T} \dd{y}}}{\frac{\dd{\sigma}^{\uparrow} }{\dd[2]{p_T} \dd{y}} +  \frac{\dd{\sigma}^{\downarrow}}{\dd[2]{p_T} \dd{y}}},  
\end{equation}
where $p_T$ and $y$ are the produced hadron's transverse momentum and rapidity respectively, and the arrows indicate the polarization of the (projectile) proton. As follows from its definition \eqref{ANdef}, the asymmetry measures the correlation between the transverse spin of the proton and the transverse momentum of the produced hadron. It is proportional to ${\vec p} \cdot ({\vec S} \times {\vec P})$, where $\vec P$ is the 3-momentum of the incoming polarized proton with spin $\vec S$.

The single transverse spin asymmetry in $p^{\uparrow}+p$ collisions has a rich history of experimental and theoretical study, beginning with the observations by the E581 and E704 collaborations at Fermilab \cite{Adams:1991rw,Adams:1991cs} and continuing with the more recent measurements by the PHENIX and STAR collaborations at RHIC \cite{Abelev:2008af,Adler:2005in}. At Fermilab, $A_N$ was observed to be much larger in magnitude than the original theoretical prediction in \cite{Kane:1978nd}, and was reported to grow with increasing Feynman $x$ and with increasing $p_T$. RHIC measurements have confirmed the earlier Fermilab findings. In addition, after extending the measured $p_T$ range for $A_N$, STAR collaboration found that the growth of $A_N$ flattened at higher $p_T$ \cite{Heppelmann:2013DIS,Aschenauer:2013woa}, but did not observe any significant falloff of $A_N$ with $p_T$ which one may expect theoretically. The asymmetry has other puzzling properties which have been observed experimentally. For one, $A_N$ in $p^{\uparrow}+p$ collisions was shown in \cite{Dilks:2016ufy} to be larger in processes where fewer photons were produced, thus suggesting that the asymmetry grows with increasing elasticity of the scattering. Another curious feature is that in $p^{\uparrow}+A$ collisions the asymmetry appears to either be suppressed for larger nuclear atomic numbers $A$ or remain unaffected by such increase in $A$ depending on the kinematic regime in which it is studied \cite{Dilks:2016ufy, Aidala:2019ctp}.

Several mechanisms have been proposed as theoretical explanations of STSA (for a review see \cite{DAlesio:2007bjf}). Since the transverse spin dependence enters a scattering amplitude with an imaginary factor $i$, for the corresponding contribution to the cross section to be nonzero one needs to generate a phase difference between the amplitude and the complex conjugate amplitude. Without such a phase difference the transverse spin dependence would simply cancel between the amplitude and the complex conjugate amplitude. The phase difference can be generated in several ways. In the Sivers effect the phase is a result of partonic final state interactions between the produced parton and the remnants of the projectile proton \cite{Sivers:1989cc,Sivers:1990fh}. The Sivers effect is often realized in theoretical calculations via the partonic lensing mechanism \cite{Brodsky:2002cx,Burkardt:2003uw} and leads to the well-known sign-reversal prediction between the asymmetry in semi-inclusive deep inelastic scattering (SIDIS) and in the Drell-Yan process (DY) \cite{Collins:2002kn,Brodsky:2002rv,Brodsky:2013oya}. Another mechanism, the Collins effect, generates the asymmetry through similar partonic interactions occurring during hadronization of a transversely polarized quark, with the phase-producing interaction being contained in the Collins fragmentation function \cite{Collins:1992kk}. In the framework of collinear factorization the phase difference and, hence, the asymmetry is generated using the higher-twist Efremov--Teryaev--Qiu--Sterman (ETQS) function \cite{Efremov:1981sh,Efremov:1984ip,Qiu:1991pp,Qiu:1998ia} or by employing the higher-twist fragmentation functions \cite{Kanazawa:2014dca,Metz:2012ct}.

Since the STSA is measured at RHIC in high-energy $p^{\uparrow}+p$ and $p^{\uparrow}+A$ collisions, it is natural to wonder whether the small-$x$ effects in the wave function of the unpolarized proton or nucleus (henceforth referred summarily as the target) may affect the asymmetry. While indeed $A_N$ is large mainly in the forward direction corresponding to probing large-$x$ partons in the polarized proton wave function, the forward direction also probes small-$x$ gluons (and quarks) in the unpolarized target. At small $x$ in the target one expects strong gluon fields leading to the phenomenon of gluon saturation (see \cite{Iancu:2003xm,Weigert:2005us,JalilianMarian:2005jf,Gelis:2010nm,Albacete:2014fwa,Kovchegov:2012mbw} for reviews). These strong gluon fields are likely to affect the $p_T$-distribution of the partons they knock out of the polarized proton wave function, therefore affecting $A_N$. For some of the previous efforts to incorporate small-$x$ effects in the $A_N$ calculations see \cite{Kang:2011ni,Kovchegov:2012ga,Schafer:2014zea,Zhou:2015ima,Hatta2016a,Hatta:2016khv}.

In \cite{Kovchegov:2012ga} the asymmetry was studied in the context of perturbative scattering using the small-$x$/saturation framework \cite{Iancu:2003xm,Weigert:2005us,JalilianMarian:2005jf,Gelis:2010nm,Albacete:2014fwa,Kovchegov:2012mbw} to account for the interactions with the target. Unlike any of the mechanisms outlined above, the phase needed to generate STSA came from the inclusion of an odderon exchange in the interaction with the target \cite{Kovchegov:2003dm,Hatta:2005as}. One can think of this STSA-generating mechanism as being similar to lensing, but with the phase-generating rescattering happening on the unpolarized target instead of the polarized projectile. The resulting STSA 
grows with momentum $p_T$ for low momenta, $p_T \ll Q_s$ with $Q_s$ the saturation scale, but falls off quickly, $A_N (p_T) \sim p_T^{-5}$ for $p_T \gg Q_s$. This mechanism also gave an asymmetry which was significantly suppressed for large nuclear targets, scaling as $A_N \sim A^{-\frac{7}{6}}$ with the atomic number $A$. 

In the quasi-classical power counting of the McLerran--Venugopalan (MV) model \cite{McLerran:1993ni,McLerran:1993ka,McLerran:1994vd}, the interactions with the unpolarized target resum powers of $\as^2 \, A^{1/3}$ \cite{Kovchegov:1997pc,Kovchegov:1996ty} with $\as$ the strong coupling constant. The usual saturation power counting assumes that $\as^2 \, A^{1/3} \sim 1$ such that all these exchanges are order-one. In this power counting, the STSA-generating quark production cross section calculated in \cite{Kovchegov:2012ga} is of the order $\as^2$, with one power of $\as$ needed to emit the quark to be measured, and another power of $\as$ arising due to the phase-generating odderon exchange \cite{Kovchegov:2003dm,Hatta:2005as}. Inclusion of small-$x$ evolution corrections \cite{Balitsky:1995ub,Balitsky:1998ya,Kovchegov:1999yj,Kovchegov:1999ua,Jalilian-Marian:1997dw,Jalilian-Marian:1997gr,Weigert:2000gi,Iancu:2001ad,Iancu:2000hn,Ferreiro:2001qy} in the rapidity interval between the produced quark and the target would resum powers of $\as \, \ln (1/x) \sim 1$, leaving the above parametric estimate the same. However, in a completely perturbative framework, the lensing mechanism of \cite{Brodsky:2002cx} comes into the quark production cross section also at order-$\as^2$: again one power of $\as$ is due to quark production, while another $\as$ is due to the lensing rescattering on the breakup products of the polarized proton, if it is modeled by a single gluon exchange. Hence, to complete the STSA calculation in $p^{\uparrow}+p$ and $p^{\uparrow}+A$ collisions started in \cite{Kovchegov:2012ga} at the same order in $\as$ one needs to include the lensing mechanism into the saturation picture of high energy scattering. This is the goal of this work. 

To include the lensing mechanism \cite{Brodsky:2002cx,Burkardt:2003uw} into the saturation framework, we will utilize the same quark--diquark model of the polarized proton as employed in \cite{Brodsky:2002cx}. The incoming proton splits into a quark--diquark pair, which then scatters on the eikonal gluon field of the unpolarized target. To generate the STSA these interactions are followed by a final-state rescattering between the quark and diquark, taken for simplicity to be a single gluon exchange. The STSA is generated by the interference of the process we have just described with the same process but without the final-state quark--diquark rescattering, by direct analogy to \cite{Brodsky:2002cx}. 

Below we calculate the lensing contribution to the quark production cross section in the saturation framework. The main result is given in \eq{sigma_pol2}. While proper phenomenological applications of our approach are left for future work, we try to analyze the qualitative properties of the result and compare them with the trends in the data. We find that, for a dilute unpolarized target and in the large-$N_c$ limit, the lensing mechanism gives an STSA generated solely by elastic scattering on the target. In real life this implies dominance of elastic events in generating $A_N$, in qualitative agreement with the preliminary findings by the STAR collaboration \cite{Dilks:2016ufy}. While our $A_N (p_T)$ is not flat in $p_T$ at high $p_T$,  as the preliminary STAR data appears to indicate \cite{Heppelmann:2013DIS,Aschenauer:2013woa}, our quark asymmetry grows or oscillates with $p_T$ for $p_T \ll Q_s$ and then falls off rather mildly as $A_N (p_T) \sim 1/p_T$ for $p_T \gg N_c \, Q_s$. (This high-$p_T$ fall-off is due to the $N_c^2$-suppressed inelastic contribution to $A_N$ which becomes important for $p_T \gg N_c \, Q_s$.) Indeed, the fragmentation effects not included into our calculation may further affect the $p_T$ dependence of $A_N (p_T)$. Finally, the $A$-dependence of our $A_N$ is complicated: for $p_T \lesssim Q_s$ the asymmetry decreases with increasing atomic number $A$, while for $p_T \gg N_c \, Q_s$ the asymmetry is approximately $A$-independent. The results of our calculation and the qualitative analysis appear to suggest that a more detailed phenomenology based on the predictions of the lensing mechanism combined with small-$x$ dynamics may be able to successfully describe the emerging $A_N$ data at RHIC.   

The structure of the paper is as follows: In Sec.~\ref{sec:Exact} we calculate the asymmetry-generating quark production cross section in the quark--diquark model of the polarized proton, using the saturation formalism to describe the interaction with the unpolarized target. In Sec.~\ref{sec:prop} we study the properties of the obtained STSA: we demonstrate dominance of the elastic contribution to $A_N$ in Sec.~\ref{sec:elastic}, evaluate the asymmetry coming from the large-$N_c$ (elastic) term in the cross section using the quasi-classical Glauber-Mueller approximation \cite{Mueller:1989st} for the target in Sec.~\ref{sec:est} while also comparing the qualitative trends in our results to experimental observations, and evaluate the contribution of the subleading-$N_c$ (inelastic) term to $A_N$ at high transverse momentum in Sec.~\ref{sec:estQ}, also comparing our conclusions to the trends found in the data. In Sec.~\ref{sec:Concl} we summarize our results and consider directions for future study.


\section{Single Transverse Spin Asymmetry in $p^{\uparrow} + p$ and $p^{\uparrow}+A$ Collisions from the Lensing Mechanism}
\label{sec:Exact}

\subsection{Quark Production at Leading Order}

\begin{figure}[ht]
\begin{center}
\includegraphics[width= 0.9 \textwidth]{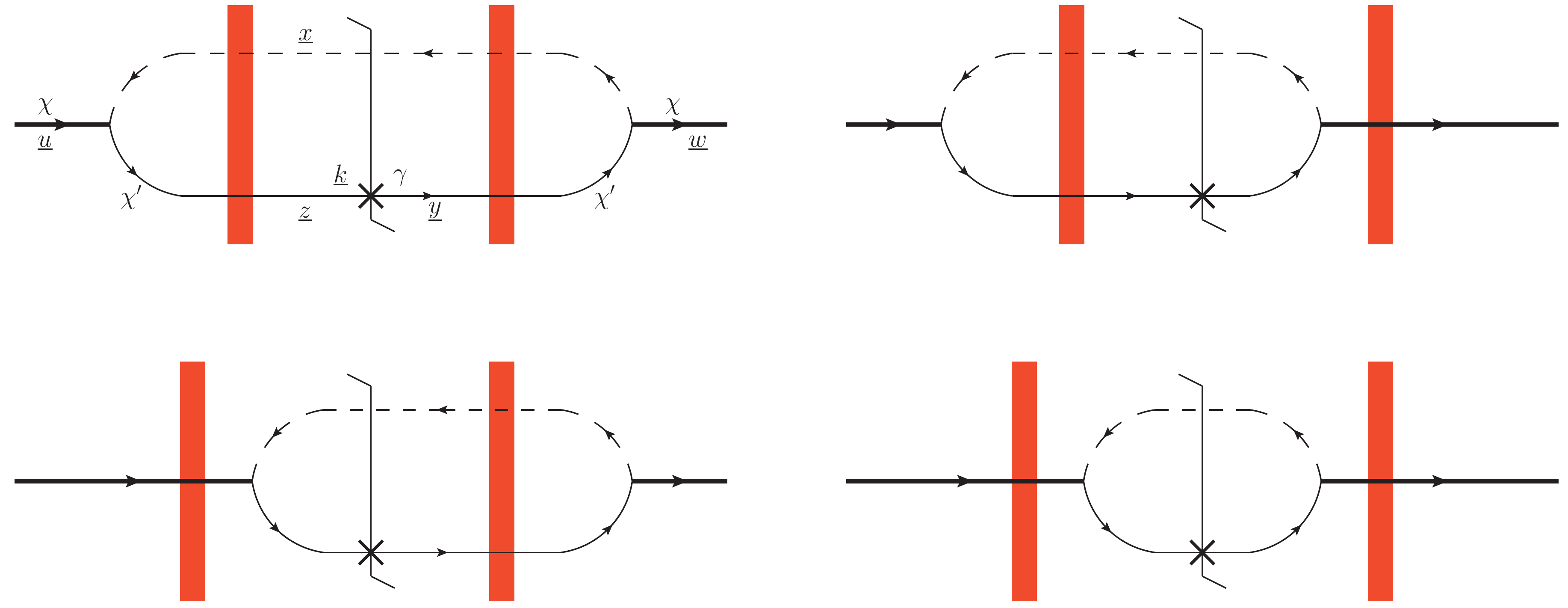} 
\caption{Quark production in $p+p$ and $p+A$ collisions in the saturation framework. Shaded rectangles represent the target shock wave.}
\label{fig:pA_unpol}
\end{center}
\end{figure}

We begin by studying quark production in $p+p$ and $p+A$ collisions using the saturation framework. The relevant diagrams are shown in \fig{fig:pA_unpol}. The projectile proton is considered in the quark--diquark model with the Yukawa-type interaction between the quark ($\psi_q$), proton ($\psi_P$) and diquark ($\varphi$) fields, $\mathscr{L}_{int} = G \, \varphi^{* \, i} \, {\bar \psi}^i_q \, \psi_P$+c.c., where $i$ is the quark and diquark fundamental color index and the asterisk denotes complex conjugation. The proton is depicted by the thick solid line in \fig{fig:pA_unpol}, the quark is shown by the thin solid line, and the scalar diquark is shown by the dashed line. The thin vertical line denotes the final-state cut, and the produced quark is labeled by the cross. For simplicity we will take the quarks to be massless, $m = 0$, and put the masses of the proton ($M_p$) and the diquark ($M$) equal to each other, $M = M_P$.

Interaction with the unpolarized proton or nuclear target is denoted by the shaded rectangles representing the shock wave in \fig{fig:pA_unpol}. The saturation framework allows us to treat the interaction with the shock wave perturbatively. We will work in light cone perturbation theory (LCPT) \cite{Lepage:1980fj,Brodsky:1997de} with the metric $\dd{s}^2 = \dd{x}^+\dd{x}^-  - \dd{x}_{\perp}^2$. In this notation the light-cone coordinates are $x^\pm = t \pm z$ and transverse vectors are denoted by ${\un v} = (v^x, v^y) = (v^1, v^2)$ with their magnitude $v_T = |{\un v}|$. We take the polarized projectile proton to be moving in the $x^+$ direction with large momentum $P^+$, having transverse spin $S$ parallel to the $x$-axis with transverse polarization $\chi$. The unpolarized target proton or nucleus (the shock wave) is moving in the $x^-$ direction with large momentum $P_{target}^-$. Throughout the paper we will be working in $A^+ =0$ light-cone gauge.

Using the standard way of calculating particle production in the saturation framework (see e.g. \cite{Kovchegov:1998bi,Kovchegov:2006qn,Kovchegov:2012ga,Kovchegov:2012mbw}), we write the expression for the quark production in the process depicted in \fig{fig:pA_unpol},
\begin{align}\label{sigma_unpol}
\frac{d \sigma}{d^2 k_T dy} = \frac{1}{2 (2 \pi)^3} \, \frac{1}{1-\gamma} \, \int d^2 x_\perp \, d^2 y_\perp \, d^2 z_\perp \, e^{- i {\un k} \cdot ({\un z} - {\un y})} d^2 u_\perp \, d^2 w_\perp \, \sum_{\chi'} \, \psi_{\chi \chi'} ({\un x}, {\un z}, {\un u}, \gamma) \, \psi^*_{\chi \chi'} ({\un x}, {\un y}, {\un w}, \gamma) \\ \times \left\langle \tr \left[ \left( V_{\un x}^\dagger \, V_{\un z} - 1 \right) \, \left( V_{\un y}^\dagger \, V_{\un x} - 1 \right) \right] \right\rangle_{y} . \notag
\end{align}
Transverse positions and polarizations employed in \eq{sigma_unpol} are shown in the upper left panel of \fig{fig:pA_unpol} along with $\gamma = k^+/P^+$. (Note that the produced quark rapidity $y$ is related to $\gamma$ via $y = \ln (\gamma P^+/k_T)$.) The light-cone wave function \cite{Lepage:1980fj,Brodsky:1997de} for the proton $\to$ quark+diquark splitting is denoted by $\psi_{\chi \chi'} ({\un x}, {\un z}, {\un u}, \gamma)$ in the transverse coordinate space. It is calculated in Appendix~\ref{sec:AppA} and is given by
\begin{align}
\label{MRWF}
\psi_{\chi \chi'} ({\un x}, {\un z}, {\un u}, \alpha) & = \frac{G \tilde{m}_{\alpha} \sqrt{\alpha} (1 - \alpha )  }{2 \pi} \ \delta^{(2)} \left( {\un x} - {\un u} + \alpha \, {\un z} - \alpha \, {\un x} \right)	\\
& \times \left[\delta_{\chi, \chi'}  K_0 (\tilde{m}_{\alpha} \abs{ {\un z} - {\un x}})  - \frac{ i \chi ( z_{\perp}^i - x_{\perp}^i ) }{ \abs{{\un z} - {\un x}}}  K_1 (\tilde{m}_{\alpha} \abs{{\un z} - {\un x}}) (i \delta_{\chi, \chi'} \delta^{i 2} - \delta_{\chi, - \chi'} \delta^{i 1} ) \right] \notag
\end{align}
with 
\begin{align}
\tilde{m}_{\alpha} \equiv  \alpha M_P .
\end{align}
Let us point out again that the proton's transverse spin $S$ is quantized along the $x$-axis. 

The interactions of the quark and diquark with the target are eikonal in \eq{sigma_unpol}, described by the fundamental-representation Wilson lines \cite{Balitsky:1995ub} and their hermitian conjugates. For a quark with transverse position ${\un x}$ the target interaction is then 
\begin{equation}\label{Wline}
V_{\un x} =  \mathcal{P} \textrm{exp} \Big[ \frac{ig}{2} \int_{- \infty}^{\infty} \dd{x^+} t^a A^{- a} (x^+, x^- = 0, {\un x}) \Big]
\end{equation}
with $t^a$ the fundamental generators of SU($N_c$), where $N_c$ is the number of quark colors. The gluon field $A^{- a}$ is generated by the target shock wave. The angle brackets $\langle \ldots \rangle_y$ denote the averaging in the target state with the rapidity interval $y$ between the particles represented by Wilson lines and the target \cite{Iancu:2003xm,Weigert:2005us,JalilianMarian:2005jf,Gelis:2010nm,Albacete:2014fwa,Kovchegov:2012mbw}. Expectation values of Wilson lines include both the multiple Glauber-Mueller scatterings in the target nucleus \cite{Mueller:1989st} along with the nonlinear small-$x$ evolution \cite{Balitsky:1995ub,Balitsky:1998ya,Kovchegov:1999yj,Kovchegov:1999ua,Jalilian-Marian:1997dw,Jalilian-Marian:1997gr,Weigert:2000gi,Iancu:2001ad,Iancu:2000hn,Ferreiro:2001qy}. 

Defining the dipole $S$-matrix expectation value for the scattering on the target 
\begin{align}
S_{{\un x} {\un y}} (Y) \equiv \left\langle \frac{1}{N_c} \tr \left[ V_{\un y}^\dagger \, V_{\un x} \right] \right\rangle_Y
\end{align}
we rewrite \eq{sigma_unpol} as
\begin{align}\label{sigma_unpol2}
\frac{d \sigma}{d^2 k_T dy} = \frac{1}{2 (2 \pi)^3} \, \frac{N_c}{1-\gamma} \, \int d^2 x_\perp \, d^2 y_\perp \, d^2 z_\perp \, e^{- i {\un k} \cdot ({\un z} - {\un y})} d^2 u_\perp \, d^2 w_\perp \, \sum_{\chi'} \, \psi_{\chi \chi'} ({\un x}, {\un z}, {\un u}, \gamma) \, \psi^*_{\chi \chi'} ({\un x}, {\un y}, {\un w}, \gamma) \\ \times \left( 1 + S_{{\un z} {\un y}} - S_{{\un x} {\un y}} - S_{{\un z} {\un x}} \right) , \notag
\end{align}
where we suppressed rapidity dependence in the arguments of the $S$-matrices for simplicity. Substituting the wave function \eqref{MRWF} into \eq{sigma_unpol2}, integrating out $\un u$ and $\un w$, and summing over $\chi'$ yields
\begin{align}\label{sigma_unpol3}
\frac{d \sigma}{d^2 k_T dy} = & \frac{G^2 \, N_c \, \gamma^3 \, (1-\gamma) \, M_P^2}{2 (2 \pi)^5} \,  \int d^2 x_\perp \, d^2 y_\perp \, d^2 z_\perp \, e^{- i {\un k} \cdot ({\un z} - {\un y})}  \left( 1 + S_{{\un z} {\un y}} - S_{{\un x} {\un y}} - S_{{\un z} {\un x}} \right)   \\ & \times \left[  K_0 (\tilde{m}_{\gamma} \abs{ {\un z} - {\un x}})  K_0 (\tilde{m}_{\gamma} \abs{ {\un y} - {\un x}}) + \frac{ ( {\un z} - {\un x} ) \cdot ( {\un y} - {\un x} ) }{ \abs{{\un z} - {\un x}} \, \abs{{\un y} - {\un x}}}   K_1 (\tilde{m}_{\gamma} \abs{ {\un z} - {\un x}})  K_1 (\tilde{m}_{\gamma} \abs{ {\un y} - {\un x}}) \right. \notag \\ & \left. + \chi \, \left(  \frac{y_{\perp}^2 - x_{\perp}^2}{ \abs{{\un y} - {\un x}}} K_0 (\tilde{m}_{\gamma} \abs{ {\un z} - {\un x}})  K_1 (\tilde{m}_{\gamma} \abs{ {\un y} - {\un x}}) + \frac{ z_{\perp}^2 - x_{\perp}^2 }{ \abs{{\un z} - {\un x}}} K_1 (\tilde{m}_{\gamma} \abs{ {\un z} - {\un x}})  K_0 (\tilde{m}_{\gamma} \abs{ {\un y} - {\un x}})\right) \right]. \notag
\end{align}
The incoming proton's polarization $\chi$ dependence only appears in the last line of \eq{sigma_unpol3}. Due to the $y_{\perp}^2 - x_{\perp}^2$ and $z_{\perp}^2 - x_{\perp}^2$ structures multiplying this $\chi$-dependent term, we expect that the resulting contribution to the cross section coming from this term would be proportional to ${\hat S} \times {\un k} = k^y$ (where ${\hat S}$ is a unit 3-vector in the direction of the proton spin, ${\hat S} = \hat x$, and the cross product is defined by ${\un u} \times {\un v} = u^x v^y - u^y v^x$). This means that the term should be odd under ${\un k} \to - {\un k}$. At the same time, if we perform the ${\un k} \to - {\un k}$ replacement in \eq{sigma_unpol3}, simultaneously swapping ${\un z} \leftrightarrow {\un y}$, the expression in the square brackets would remain invariant. Further, is we assume that $S_{{\un x} {\un y}} = S_{{\un y} {\un x}}$, the whole integrand of \eq{sigma_unpol3} would be invariant under ${\un k} \to - {\un k}$ and ${\un z} \leftrightarrow {\un y}$. Since the $\chi$-dependent term in \eq{sigma_unpol3} has to change sign under ${\un k} \to - {\un k}$, this means that it gives zero contribution to the cross section. In other words, the only way the $\chi$-dependent term in \eq{sigma_unpol3} can give a non-zero contribution to the cross section, and, therefore, generate the STSA, is if $S_{{\un x} {\un y}} \neq S_{{\un y} {\un x}}$ \cite{Kovchegov:2012ga}. The difference $S_{{\un x} {\un y}} - S_{{\un y} {\un x}}$ is non-zero due to the QCD odderon interaction with the target \cite{Kovchegov:2003dm,Hatta:2005as}: hence, the STSA in \cite{Kovchegov:2012ga} was generated via such an odderon exchange. 

Our goal here is to find the contribution to $A_N$ due to the lensing mechanism \cite{Brodsky:2002cx}. We, therefore, neglect the odderon contribution by assuming that $S_{{\un x} {\un y}} = S_{{\un y} {\un x}}$. Equation \eqref{sigma_unpol3} then simplifies to 
\begin{align}\label{sigma_unpol4}
\frac{d \sigma}{d^2 k_T dy} = & \frac{G^2 \, N_c \, \gamma^3 \, (1-\gamma) \, M_P^2}{2 (2 \pi)^5} \,  \int d^2 x_\perp \, d^2 y_\perp \, d^2 z_\perp \, e^{- i {\un k} \cdot ({\un z} - {\un y})}  \left( 1 + S_{{\un z} {\un y}} - S_{{\un x} {\un y}} - S_{{\un z} {\un x}} \right)   \\ & \times \left[  K_0 (\tilde{m}_{\gamma} \abs{ {\un z} - {\un x}})  K_0 (\tilde{m}_{\gamma} \abs{ {\un y} - {\un x}}) + \frac{ ( {\un z} - {\un x} ) \cdot ( {\un y} - {\un x} ) }{ \abs{{\un z} - {\un x}} \, \abs{{\un y} - {\un x}}}   K_1 (\tilde{m}_{\gamma} \abs{ {\un z} - {\un x}})  K_1 (\tilde{m}_{\gamma} \abs{ {\un y} - {\un x}}) \right] \notag
\end{align}
and becomes independent of the proton polarization $\chi$. This is the unpolarized quark production cross section in $p+p$ and $p+A$ collisions. It does not generate a non-zero STSA.


\subsection{Quark Production with Lensing}

\begin{figure}[ht]
\begin{center}
\includegraphics[width= 0.9 \textwidth]{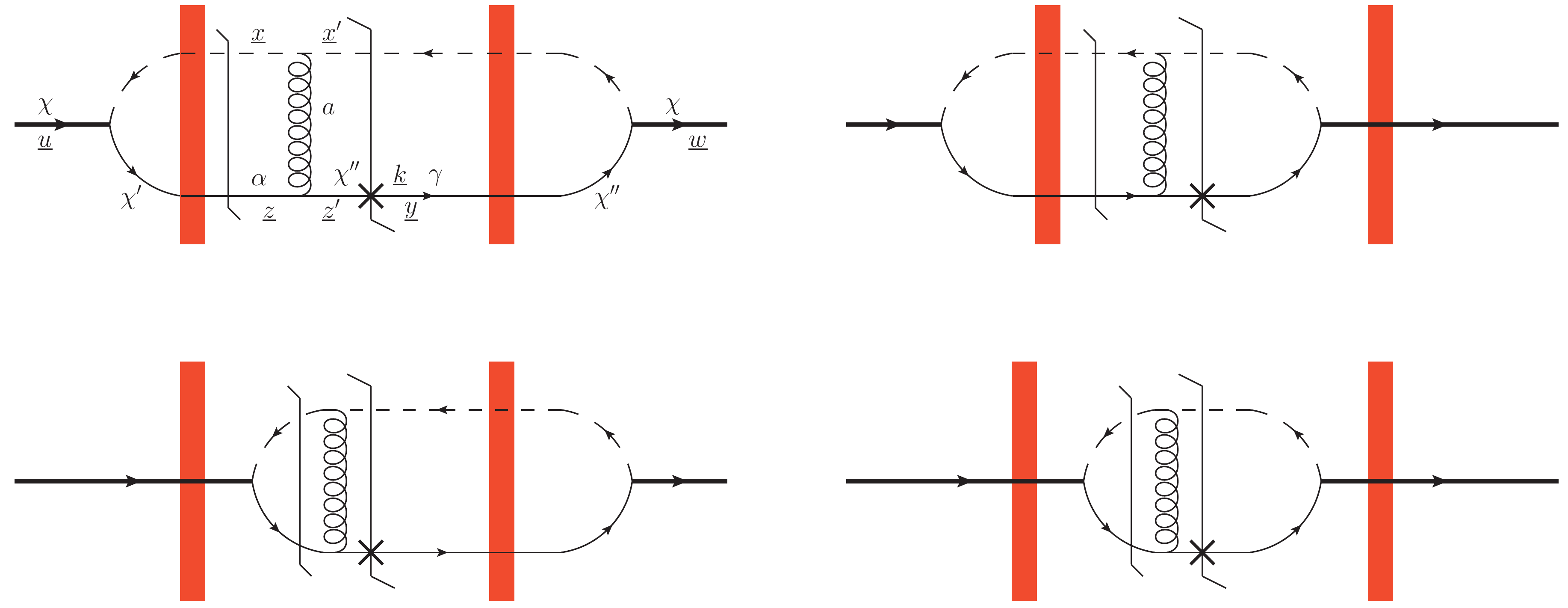} 
\caption{Quark production in $p+p$ and $p+A$ collisions in the saturation framework, now with the lensing exchange of a final-state gluon. Complex conjugates of all the diagrams need to be added in the calculation, but are not shown explicitly in this figure.}
\label{fig:pA_pol}
\end{center}
\end{figure}

It is clear from the above calculation that we need further interactions in order to generate STSA. The option we want to pursue here is the lensing mechanism \cite{Brodsky:2002cx}. In SIDIS it is realized via a final-state interaction between the outgoing quark and diquark. By analogy to that, we augment the quark production process in \fig{fig:pA_unpol} with such a quark-diquark final-state interaction, which, following \cite{Brodsky:2002cx}, we model by a gluon exchange. The resulting diagrams are depicted in \fig{fig:pA_pol}, where one also has to add the complex conjugate diagrams to the ones shown to calculate the full contribution to the cross section.  

The additional gluon-exchange interaction between the quark and diquark in \fig{fig:pA_pol}, as compared to the diagrams in \fig{fig:pA_unpol}, needs to generate a phase difference between the amplitude and the complex conjugate amplitude in order to give a non-zero STSA. The interactions of the quark--diquark system with the unpolarized target in \fig{fig:pA_pol} will give us correlators of Wilson lines, which will be real if we again neglect the odderon exchange contribution which was already included in \cite{Kovchegov:2012ga}. Therefore, the only remaining source of the phase difference is due to an additional gluon interaction in the quark-diquark system. Using the  $\as^2 \, A^{1/3} \sim 1$ power counting described above, we see that a single-gluon correction to the diagrams in  \fig{fig:pA_unpol} involving the quark and/or diquark contributes at the same order in $\as$ as the odderon exchange (order-$\as G^2$ in the diquark model at hand). If this gluon emission and/or absorption occurs inside of the shock wave, then the process would be suppressed by a factor of $1/s$ with $s$ the center-of-mass energy squared for the scattering process at hand. Physically this is due to the high scattering energy leading to the $x^+$-width of the shock wave being rather short, making gluon emission and absorption by the quark and the diquark inside the shock wave very unlikely. Hence we need to consider the gluon emission and absorption by the quark and the diquark happening before and after the shock wave, and see which ones give the phase difference between the amplitude and the complex conjugate amplitude required for STSA.

\begin{figure}[ht]
\begin{center}
\includegraphics[width= 0.75 \textwidth]{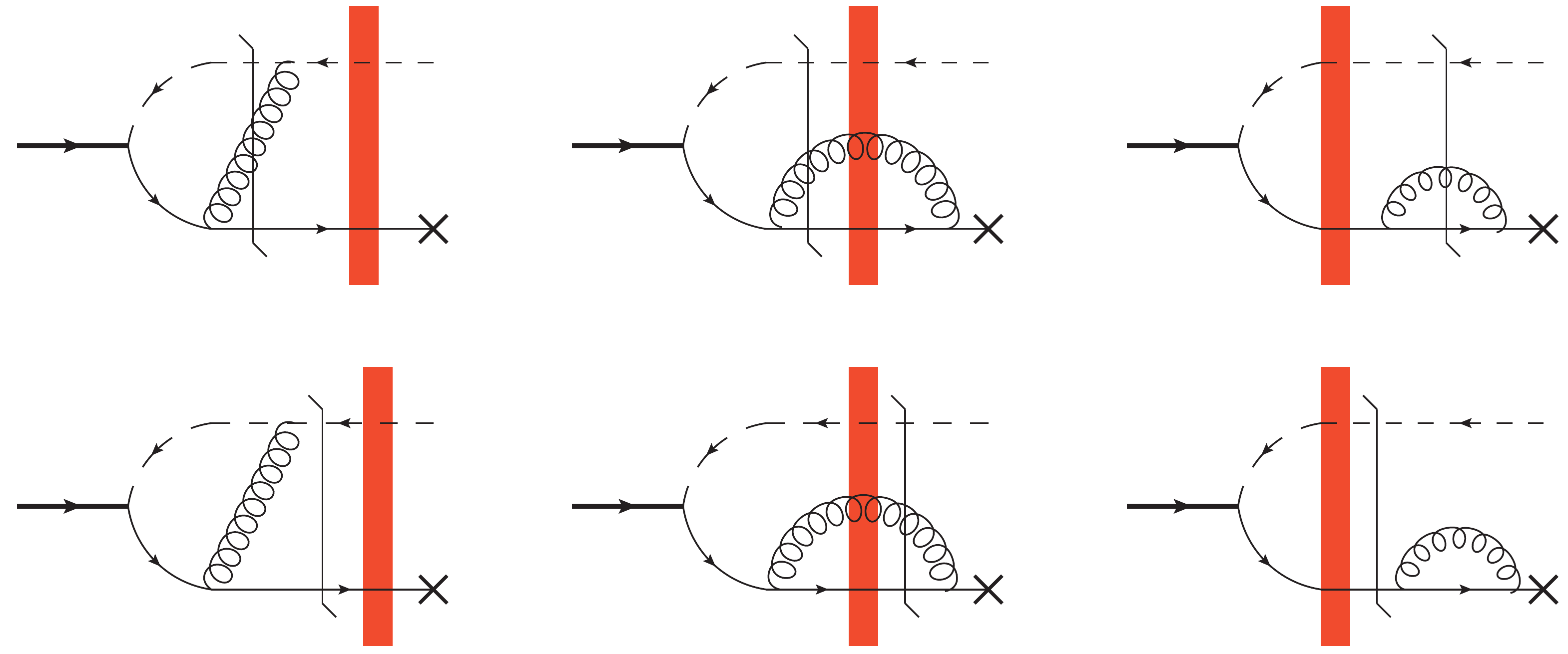} 
\caption{Examples of one-gluon corrections to the quark production in $p+p$ and $p+A$ collisions from \fig{fig:pA_unpol} which do not contribute to STSA. These diagrams do not generate a phase needed for STSA, since the contributions of the cuts (shown by solid vertical lines) are zero.}
\label{fig:pA_pol2}
\end{center}
\end{figure}

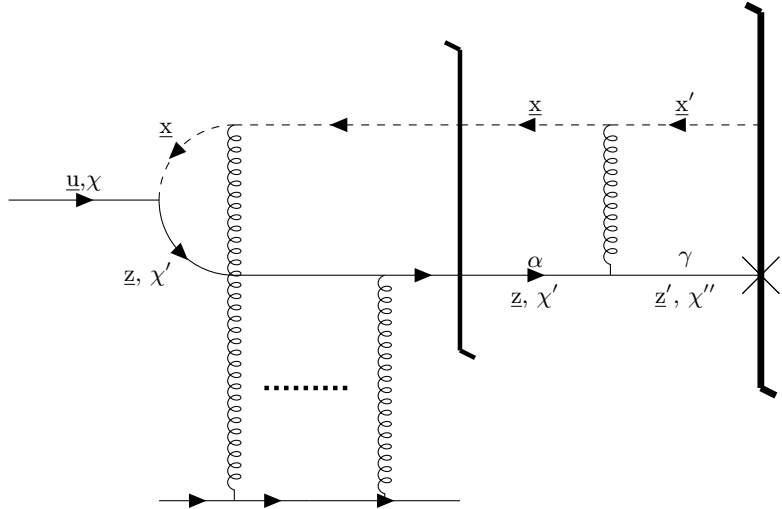
\begin{figure}[ht]
\begin{center}
\begin{tikzpicture}
\begin{feynman}
\vertex(a);
\vertex[right=2.0cm of a] (b);
\vertex[below right=4.0cm and 2.0cm of a] (c);
\vertex[above right=1.0cm and 1.0cm of b] (d);
\vertex[right=1.0cm of c] (e);
\vertex[above right=1.0cm and 4.0cm of b] (f);
\vertex[below right=1.0cm and 3.0cm of b] (g);
\vertex[right=3.0cm of c] (h);
\vertex[below right=1.0cm and 4.0cm of b] (i);
\vertex[right=2.0cm of c](j);
\vertex[right=4.0cm of c](k);
\vertex[below right=1.0cm and 1.0cm of b](w);
\vertex[above=1.0cm of f](x);
\vertex[above left= 0.1cm and 0.2cm of x](xl);
\vertex[below=1.0cm of i](y);
\vertex[below right=0.1cm and 0.2cm of y](yl);

\vertex[right=2.0cm of f](l);
\vertex[right=2.0cm of l](m);
\vertex[right=2.0cm of i](n);
\vertex[right=2.0cm of n](o);
\vertex[above=1.5cm of m](p);
\vertex[above left= 0.1cm and 0.2cm of p](pl);
\vertex[below=1.5cm of o](q);
\vertex[below right=0.1cm and 0.2cm of q](ql);
\vertex[above right=1.5cm and 0.4cm of e](r);
\vertex[above left=1.5cm and 0.45cm of h](s);

\diagram* [vertical=f to h, vertical=d to e]{
(a) -- [fermion, edge label={$\underbar{u}$,$\chi$}] (b) --[anti charged scalar,quarter left, edge label={$\underbar{x}$}](d) --[anti charged scalar](f) --[anti charged scalar, edge label={$\underbar{x}$}](l) --[anti charged scalar, edge label={$\underbar{x}'$}](m),
(b) --[fermion,quarter right, edge label'={$\underbar{z}$, $\chi'$}](w) --(g) --[fermion](i) --[fermion, edge label'={$\underbar{z}$, $\chi'$},edge label={$\alpha$}](n) --[fermion,edge label'={$\underbar{z}'$, $\chi''$},edge label={$\gamma$},insertion={[size=7pt]1.00}](o),
(c)--[fermion](e)--[fermion](j) --[fermion](k),
(d)--[gluon](e),
(g) --[gluon](h),
(xl)--[line width=0.60mm](x),
(x)--[line width=0.60mm](y),
(y)--[line width=0.60mm](yl),
(l) --[gluon](n),
(pl)--[line width=0.90mm](p),
(p) --[line width=0.90mm](q),
(q)--[line width=0.90mm](ql),
(r)--[dotted, line width=0.6mm](s),
};
\end{feynman}
\end{tikzpicture}
\caption{A more detailed depiction of the diagram contributing to the amplitude to the left of the final-state cut in the upper-left panel of \fig{fig:pA_pol}. The transverse positions and polarizations of all lines are labeled explicitly: the proton at transverse position $u_{\perp}$ splits into the quark and diquark with positions $z_{\perp}$ and $x_{\perp}$ respectively. The interaction of the quark and diquark with the target shock-wave is shown by multiple gluon exchanges: it does not alter the transverse positions of the quark and diquark. After this interaction, the final state gluon exchange happens between the quark and the diquark resulting in an outgoing quark and diquark with transverse positions $z'_{\perp}$ and $x'_{\perp}$ respectively. The final state cut and the secondary cut generating STSA are shown by vertical solid lines.}
\label{FIG:CoordDiagram}
\end{center}
\end{figure}

\begin{figure}[h]
\centering
\begin{tikzpicture}
\begin{feynman}
\vertex(a){$\underline{x}$};
\vertex[right=3.0cm of a](b);
\vertex[right=6.0cm of a](c){$\underline{x}'$};
\vertex[below=3.0cm of a](d){$\underline{z}$};
\vertex[right=3.0cm of d](e);
\vertex[right=6.0cm of d](f){$\underline{z}'$};

\diagram*{
(a)--[anti charged scalar, momentum={$p-k$}](b)--[anti charged scalar, momentum={$p-k-r$}](c),
(b)--[gluon, momentum={$r$}](e),
(d)--[fermion, momentum={$k$,$ \chi'$}](e)--[fermion, momentum={$k+r$,$\chi''$}](f),

};
\end{feynman}
\end{tikzpicture}
\caption{The diagram for $M_{FSI}$ (see text).}
\label{FIG:MFSI}
\end{figure}
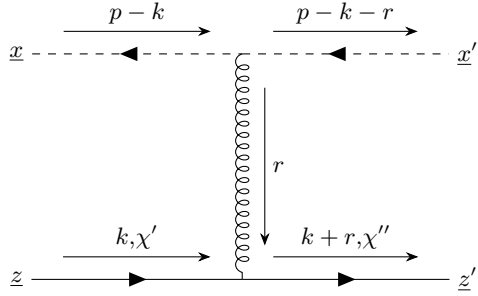

An analysis of all the possible single-gluon corrections to the diagrams in \fig{fig:pA_unpol} (outside the shock wave) shows that the only other remaining source of the phase difference is the imaginary part of the amplitude with the additional final-state gluon exchange (the diagrams left of the main final-state cut in \fig{fig:pA_pol}).  According to Cutkosky rules, such an imaginary part can be denoted by placing an additional cut through the amplitude, as shown by a somewhat shorter cut in \fig{fig:pA_pol}. This technique has been already employed in \cite{Brodsky:2013oya} where it was helpful in understanding the diagrammatic origin of STSA in SIDIS and DY processes. In \fig{fig:pA_pol2} we illustrate this technique to show some of the one-gluon correction diagrams which do not contribute to STSA. Note that the additional cut cannot be placed to the left of the shock wave, since this would lead to proton decay diagrams, which are prohibited in QCD (see the left two graphs in \fig{fig:pA_pol2} along with the middle graph in the top row). This additional cut can only be placed after (to the right of) the shock wave, as shown in \fig{fig:pA_pol}, where it generates a $2 \to 2$ on-shell scattering sub-process (the cut going through the shock wave can only generate the STSA phase due to the odderon contribution considered earlier in \cite{Kovchegov:2012ga}). Only the gluon exchange diagrams shown in \fig{fig:pA_pol} can give a non-zero contribution to the additional cut. As shown by the lower-row middle graph and the right two graphs of \fig{fig:pA_pol2}, diagrams with the gluon emitted before the shock wave and absorbed after, along with the diagrams where the extra gluon is emitted and absorbed by the quark (diquark) to the right of the shock wave, cannot give a non-trivial contribution to the second cut, and, hence, to STSA. The second cut, when applied to those diagrams, generates either $2 \to 1$ or $1 \to 1$ on-shell scattering sub-processes (as can be seen in \fig{fig:pA_pol2}), which are zero. Thus, the STSA-generating phase can arise only from the diagrams with the final-state gluon exchange between the quark and diquark via an additional cut placed after the shock wave, as depicted in \fig{fig:pA_pol}. The amplitude left of the final-state cut in the upper left panel of \fig{fig:pA_pol} is redrawn in more detail in \fig{FIG:CoordDiagram} for illustration purposes. The additional cut separates the amplitude left of the main final-state cut in the graphs of \fig{fig:pA_pol} into the same amplitude left-of-cut as we had in the diagrams of \fig{fig:pA_unpol} and the gluon-exchange $2 \to 2$ scattering amplitude between the quark and diquark pictured below in \fig{FIG:MFSI}. This latter amplitude will be denoted $M_{FSI}$ since it contains the final-state interaction. Note that $M_{FSI}$ is real, since one cannot cut the diagram in \fig{FIG:MFSI}.

In calculating the diagrams in Figs.~\ref{fig:pA_pol} or \ref{FIG:CoordDiagram} using LCPT rules \cite{Lepage:1980fj,Brodsky:1997de} we encounter an additional intermediate quark--diquark state which we cut: this means we need to keep only the imaginary part of the light-cone energy denominator corresponding to this intermediate state. (The real part of the energy denominator contributes an order-$\as$ correction to the diagrams in \fig{fig:pA_unpol}, but does not generate STSA and is, hence, discarded.) This means that when calculating the diagrams in \fig{fig:pA_pol} we need to replace the energy denominator by
\begin{align}\label{Im_repl}
\frac{1}{ p_{out}^- - p_{intermediate}^- + i\epsilon} \ \longrightarrow \ i \, \Im \left[ \frac{1}{ p_{out}^- - p_{intermediate}^- + i\epsilon} \right] = - i \pi \, \delta (  p_{out}^- - p_{intermediate}^-)
\end{align}
with $p_{out}^-$ and $p_{intermediate}^-$ denoting the light-cone energy of the outgoing and intermediate quark--diquark states. Below we will include the factor in \eq{Im_repl} into our definition of the final-state rescattering amplitude $M_{FSI}$, thus making it imaginary.

Similar to \eq{sigma_unpol} we write
\begin{align}\label{sigma_pol1}
\frac{d \sigma_\chi}{d^2 k_T dy} = \frac{1}{2 (2 \pi)^3} \, \frac{1}{1-\gamma} \! \int\limits_0^1 \frac{d \alpha}{4 \pi} & \! \int d^2 x_\perp \, d^2 x'_\perp \, d^2 y_\perp \, d^2 z_\perp \, d^2 z'_\perp  e^{- i {\un k} \cdot ({\un z}' - {\un y})} d^2 u_\perp \, d^2 w_\perp \!\! \sum_{\chi', \chi''} \! \psi_{\chi \chi'} ({\un x}, {\un z}, {\un u}, \alpha) \, \psi^*_{\chi \chi''} ({\un x}', {\un y}, {\un w}, \gamma) \notag \\ & \times \left\langle \tr \left[ t^a \left( V_{\un x}^\dagger \, V_{\un z} - 1 \right)  t^a \left( V_{\un y}^\dagger \, V_{{\un x}'} - 1 \right) \right] \right\rangle_y \, \left(- M_{FSI}^{\chi' \chi''} ({\un x}', {\un z}'; {\un x}, {\un z}; \alpha, \gamma) \right) + \mbox{c.c.}. 
\end{align}
Here $\alpha$ is the fraction of the proton's plus momentum carried by the quark before the gluon exchange with diquark, as shown in Figs.~\ref{fig:pA_pol} and \ref{FIG:CoordDiagram}. The wave function $\psi_{\chi \chi'}$ is the same as given above in \eq{MRWF} while the Wilson lines $V$ are also defined above in \eq{Wline}. The subscript $\chi$ in $\sigma_\chi$ indicates that we are only interested in the polarization-dependent part of the cross section, and thus, as we will see, only the $\chi$ dependent part of the wave function product $\psi_{\chi \chi'}  \, \psi^*_{\chi \chi''}$ contributes in \eq{sigma_pol1}. The minus sign in front of $M_{FSI}^{\chi' \chi''}$ in \eq{sigma_pol1} is due to the fact that the standard LCPT rules \cite{Lepage:1980fj,Brodsky:1997de} give a negative of the scattering amplitude.

The only ingredient in \eq{sigma_pol1} that we have not yet found is the final-state rescattering amplitude $M_{FSI}^{\chi' \chi''}$. It is depicted in \fig{FIG:MFSI}. Since all the external lines of this amplitude are on mass shell, we can calculate it using the covariant Feynman perturbation theory.  Absorbing the $(-i \pi)$ and the light-cone energy delta-function from \eq{Im_repl} into $M_{FSI}$ we get the amplitude in the mixed representation (in the longitudinal momentum space and transverse coordinate space)
\begin{align}\label{mfsi}
i M_{FSI}^{\chi' \chi''} ({\un x}', {\un z}'; {\un x}, {\un z}; \alpha, \gamma) = & \int \frac{d^2 p_{\perp}}{(2 \pi)^2} \, \frac{d^2 k_{\perp}}{(2 \pi)^2} \, \frac{d^2 r_{\perp}}{(2 \pi)^2} \, e^{i (\un{p} - \un{k} - \un{r}) \cdot {\un x}' - i (\un{p} - \un{k})\cdot {\un x} + i ({\un k} + {\un r}) \cdot {\un z}' - i {\un k} \cdot {\un z}} \frac{p^+ }{k^+ (p-k)^+} \frac{-\pi g^2 }{r^+ r^- - r_{\perp}^2 + i\epsilon} \notag \\
& \times \bar{u}_{\chi''} (k+r) [ 2(\slashed{p}-\slashed{k}) - \slashed{r} ] u_{\chi'}(k) \ \delta( (p-k-r)^- + (k+r)^- - (p-k)^- - k^-)	,
\end{align}
where the color factor has been removed from $M_{FSI}^{\chi' \chi''}$ since it was already incorporated into \eq{sigma_pol1}. Equation \eqref{mfsi} involves transverse spinors which are defined in terms of the Brodsky--Lepage helicity basis spinors as $u_{\chi} = \frac{1}{\sqrt{2}} [u_z + \chi u_{-z}]$ \cite{Kovchegov:2012ga}. The minus components of momenta in the argument of the delta-function should be understood as $k^- = k_\perp^2 /k^+$, as is standard in LCPT. Note that $p^+ = P^+$, which is the large momentum component of the incoming proton. In terms of the momentum labels in \fig{FIG:MFSI} the longitudinal momentum fractions are $\alpha = k^+/P^+$ and $\gamma= (k+r)^+ /P^+$.

In arriving at \eq{mfsi} we assumed that the diquark--gluon interactions result from the ``scalar QCD" Lagrangian $\mathscr{L}_{scalar \, QCD} = \left( D_\mu \phi \right)^\dagger \cdot D^\mu \phi - M^2 \, \phi^\dagger \cdot \phi$ with the covariant derivative $D_\mu = \pd_\mu - i (-g) A_\mu$ and the gluon field $A_\mu = \sum_a t^a A_\mu^a$. Note that the diquark has the color quantum numbers of an anti-quark, which generates an extra minus sign in the diquark--gluon coupling. Finally, the amplitude $M_{FSI}^{\chi' \chi''}$ is indeed gauge-invariant, so the gauge choice for the gluon propagator is not important. 

Evaluating the spinor products in \eq{mfsi} and Fourier-transforming the result into transverse coordinate space is rather involved. The main steps of the calculation are outlined in Appendix~\ref{sec:AppB}. In the end one obtains
\begin{align}
\label{mrMFSI}
i M_{FSI}^{\chi' \chi''} ({\un x}', {\un z}'; {\un x}, {\un z}; \alpha, \gamma) = \frac{g^2}{2 \, \pi} \,  \frac{1}{  \alpha \sqrt{\alpha\gamma} \, |{\un z}' - {\un z}|^2} \, \delta^{(2)}[ (1-\gamma) {\un x}' - (1-\alpha) {\un x} + \gamma {\un z}' - \alpha {\un z} ] \notag	\\ \times
 \delta \left[  \frac{({\un x}' - {\un z}')^2}{\alpha (1-\alpha)}  - \frac{({\un x} - {\un z})^2}{\gamma (1-\gamma)}   \right]	
\, \left[ \delta_{\chi',\chi''}   ({\un x} - {\un z} ) \cdot   ( {\un x} - {\un z}' )  - i\delta_{\chi',-\chi''} ({\un x} - {\un z}) \times  ( {\un x} - {\un z}' ) \right] .
\end{align}
In arriving at \eq{mrMFSI} we have put the quark mass to zero, $m=0$, and expanded the result to the lowest order in the diquark mass $M$, which turned out to be $M^0$: higher powers of $M$ bring no spin-dependent contributions and can be discarded if we assume that $k_T \gg M$. This is the assumption we will make from this point on. Note that discarding the quark mass terms makes $M_{FSI}$ spin independent, so we indeed only need the $\chi$-dependent part of the wave function product $\psi_{\chi \chi'}  \, \psi^*_{\chi \chi''}$ in $\sigma_{\chi}$.

Finally, substituting $M_{FSI}^{\chi' \chi''}$ from \eq{mrMFSI} into \eq{sigma_pol1} and using the wave functions \eqref{MRWF} in the latter, while keeping only the $\chi$-dependent term in $\psi_{\chi \chi'}  \, \psi^*_{\chi \chi''}$, we arrive at
\begin{align}\label{sigma_pol2}
& \frac{d \sigma_\chi}{d^2 k_T dy} = \chi \, \frac{i \as G^2 \, M_p^2 \, \gamma}{(2 \pi)^6} \! \int\limits_0^1  d \alpha \, (1 - \alpha) \!\! \int \! d^2 x_\perp \, d^2 x'_\perp \, d^2 y_\perp \, d^2 z_\perp \, d^2 z'_\perp \,  \frac{e^{- i {\un k} \cdot ({\un z}' - {\un y})}}{   |{\un z}' - {\un z}|^2} \, \delta^{(2)}[ (1-\gamma) {\un x}' - (1-\alpha) {\un x} + \gamma {\un z}' - \alpha {\un z} ] \notag	\\ & \times
 \delta \left[  \frac{({\un x}' - {\un z}')^2}{\alpha (1-\alpha)}  - \frac{({\un x} - {\un z})^2}{\gamma (1-\gamma)}   \right] \left\langle \thalf \tr \left[ V_{\un x}^\dagger \, V_{\un z} - 1 \right] \, \tr \left[ V_{\un y}^\dagger \, V_{{\un x}'} - 1 \right] - \frac{1}{2 N_c} \tr \left[ \left( V_{\un x}^\dagger \, V_{\un z} - 1 \right) \, \left( V_{\un y}^\dagger \, V_{{\un x}'} - 1 \right) \right] \, \right\rangle_y  \\ & \times  
 	\Bigg\{  ({\un x} - {\un z} ) \cdot   ( {\un x} - {\un z}' ) \left[ \frac{{\hat S} \times ({\un y} - {\un x}')}{ \abs{{\un y} - {\un x}'}} K_0 (\tilde{m}_{\alpha} \abs{ {\un z} - {\un x}})  K_1 (\tilde{m}_{\gamma} \abs{ {\un y} - {\un x}'})  + \frac{{\hat S} \times ({\un z} - {\un x}) }{ \abs{{\un z} - {\un x}}} K_1 (\tilde{m}_{\alpha} \abs{ {\un z} - {\un x}})  K_0 (\tilde{m}_{\gamma} \abs{ {\un y} - {\un x}'}) \right] \notag \\ &  - ({\un x} - {\un z}) \times  ( {\un x} - {\un z}' ) \left[ \frac{{\hat S} \cdot ({\un y} - {\un x}')}{ \abs{{\un y} - {\un x}'}} \, K_0 (\tilde{m}_{\alpha} \abs{ {\un z} - {\un x}})  K_1 (\tilde{m}_{\gamma} \abs{ {\un y} - {\un x}'})  - \frac{{\hat S} \cdot ({\un z} - {\un x}) }{ \abs{{\un z} - {\un x}}} \, K_1 (\tilde{m}_{\alpha} \abs{ {\un z} - {\un x}}) \, K_0 (\tilde{m}_{\gamma} \abs{ {\un y} - {\un x}'}) \right] \Bigg\} ,  \notag
\end{align}
where we have also used the Fierz identity to simplify the color traces and doubled the expression to account for the complex conjugate term in \eq{sigma_pol1}. 

Equation \eqref{sigma_pol2} is the main general result of our calculation for the STSA-generating quark production cross section for $p^{\uparrow}+p$ and $p^{\uparrow}+A$ collisions. It can be used to construct the numerator of $A_N$ in \eq{ANdef}, while \eq{sigma_unpol4}, along with its gluon production counterpart would contribute to the denominator of $A_N$. Below we will study the properties of $A_N$ resulting from the cross-section in \eq{sigma_pol2}.


\section{Properties of the Obtained $A_N$}
\label{sec:prop}


\subsection{Elastic Dominance}
\label{sec:elastic}

One property of our main result \eqref{sigma_pol2} can be seen without doing complicated calculations.  For $p^{\uparrow}+A$ collisions with a large nucleus, $A \gg 1$, and in the large-$N_c$ limit, the interaction with the target in \eq{sigma_pol2} simplifies to \cite{Kovchegov:1999yj}
\begin{align}\label{factorzation_LR}
& \left\langle \thalf \tr \left[ V_{\un x}^\dagger \, V_{\un z} - 1 \right] \, \tr \left[ V_{\un y}^\dagger \, V_{{\un x}'} - 1 \right] - \frac{1}{2 N_c} \tr \left[ \left( V_{\un x}^\dagger \, V_{\un z} - 1 \right) \, \left( V_{\un y}^\dagger \, V_{{\un x}'} - 1 \right) \right] \, \right\rangle_y \\ & \approx \thalf  \left\langle \tr \left[ V_{\un x}^\dagger \, V_{\un z} - 1 \right] \right\rangle_y \,  \left\langle \tr \left[ V_{\un y}^\dagger \, V_{{\un x}'} - 1 \right] \right\rangle_y \notag = \frac{N_c^2}{2} \, N_{{\un z}, {\un x}} (y) \, N_{{\un x}', {\un y}} (y),
\end{align}
where the quark dipole forward scattering amplitude is defined by \cite{Kovchegov:1999yj}
\begin{align}\label{Ndef}
N_{{\un x}, {\un y}} (Y) = 1 - S_{{\un x}, {\un y}} (Y).
\end{align}
We see that the interaction with the target factorizes into an elastic interaction to the left of the final-state cut ($N_{{\un z}, {\un x}}$) and another elastic interaction to the right of the cut ($N_{{\un x}', {\un y}}$) \cite{Kovchegov:1999kx}. We conclude that the $p^{\uparrow}+A$ spin-dependent quark production cross-section \eqref{sigma_pol2} and, therefore, $A_N$ from \eq{ANdef}, are given by elastic interaction for scattering on a large nucleus and in the large-$N_c$ limit. 

In real life $N_c =3$: hence, the accuracy of the approximation in \eq{factorzation_LR} is up to corrections of the relative order $1/N_c^2 \approx 11 \%$, though for some matrix elements of Wilson lines the precision of the large-$N_c$ approximation was shown to be much higher \cite{Kovchegov:2008mk}. Therefore, our calculation embedding the lensing mechanism into the saturation framework predicts the dominance of elastic interactions in $p^{\uparrow}+A$ collisions contributing to $A_N$ at least by a ratio of $N_c^2 : 1$. 

The applicability of the approximation \eqref{factorzation_LR} to $p^{\uparrow}+p$ collisions depends on whether the unpolarized proton target can be treated as a large nucleus, that is, it depends on the extent to which the proton can be thought of as an assembly of uncorrelated color charges. While this is a rather complicated question to address, let us simply point out that the BK equation, which was originally derived for deep inelastic scattering (DIS) on a nucleus ($e+A$) \cite{Balitsky:1995ub,Balitsky:1998ya,Kovchegov:1999yj,Kovchegov:1999ua}, has been successfully applied to the data for DIS on a proton ($e+p$), see e.g. \cite{Albacete:2009fh,Albacete:2010sy}. It is, therefore, possible that our prediction of elastic dominance in $A_N$ does, in fact, apply to $p^{\uparrow}+p$ collisions by analogy to the unpolarized DIS on the proton. We then may be able to conclude that our observation of elastic dominance is qualitatively consistent with the preliminary STAR collaboration data \cite{Dilks:2016ufy}.


\subsection{Estimates of the Asymmetry: Leading-$N_c$}
\label{sec:est}

Let us continue evaluating the cross section \eqref{sigma_pol2} in the large-$N_c$ and large-$A$ approximation, following what we have already started in Sec.~\ref{sec:elastic}. We replace the interaction with the target in \eq{sigma_pol2} by $(N_c^2/2) \, N_{{\un z}, {\un x}} \, N_{{\un x}', {\un y}}$, according to the result of \eq{factorzation_LR}. Next we make a variable change
\begin{subequations}\label{pos_shifts}
\begin{align}
& \tilde{\un z} = {\un z}' - {\un x}'	 , \\
& \tilde{\un y} = {\un y} - {\un x}' , \\
& {\un \xi} = {\un z} - {\un x} , \\
& {\un r} = {\un x}' - {\un x} , \\
& {\un b} = {\un x}.
\end{align}
\end{subequations}
Simultaneously we rewrite 
\begin{align}
N_{{\un z}, {\un x}} (y) = N \left( {\un z} - {\un x}, \frac{{\un z} + {\un x}}{2}, y \right) \approx N (\xi_T, b_T, y), 
\end{align}
where the first step is simply a change in notation, while the second step is a simplification, employing the fact that for a large nucleus target one usually has $b_T \gg \xi_T$ and that the leading high-energy behavior of $N$ is independent of the angles of the dipole separation $\un \xi$ and the impact parameter $\un b$. Similarly we approximate $N_{{\un x}', {\un y}} (y) \approx N ({\tilde y}_T,  b_T, y)$. The resulting transverse polarization-dependent cross section is
\begin{align}\label{sigma_pol3}
& \frac{d \sigma_\chi}{d^2 k_T dy} = \chi \, \frac{i \as G^2 \, N_c^2 \, M_p^2 \, \gamma}{2 (2 \pi)^6} \! \int\limits_0^1  d \alpha \, (1 - \alpha) \!\! \int \! d^2 b_\perp \, d^2 \xi_\perp \, d^2 {\tilde y}_\perp \, d^2 {\tilde z}_\perp  \,  \frac{e^{- i {\un k} \cdot (\tilde{\un z} - \tilde{\un y})}}{   |(1-\alpha) \, {\un \xi} - (1-\gamma) \tilde{\un z}|^2}  \notag	\\ & \times
 \delta \left[  \frac{\tilde{z}_T^2}{\alpha (1-\alpha)}  - \frac{\xi_T^2}{\gamma (1-\gamma)}   \right] \, N (\xi_T, b_T, y) \, N ({\tilde y}_T,  b_T, y)  \\ & \times  \Bigg\{  {\un \xi}  \cdot   ( \alpha {\un \xi} + (1-\gamma) \tilde{\un z} ) \left[ \frac{{\hat S} \times \tilde{\un y} }{ {\tilde y}_T} K_0 (\tilde{m}_{\alpha} {\xi}_T)  K_1 (\tilde{m}_{\gamma} {\tilde y}_T )  + \frac{{\hat S} \times {\un \xi} }{ \xi_T} K_1 (\tilde{m}_{\alpha} {\xi}_T)  K_0 (\tilde{m}_{\gamma} \tilde{y}_T) \right] \notag \\ &  - (1-\gamma) \, {\un \xi} \times  \tilde{\un z} \left[ \frac{{\hat S} \cdot \tilde{\un y}}{ \tilde{y}_T} \, K_0 (\tilde{m}_{\alpha} \xi_T)  K_1 (\tilde{m}_{\gamma} {\tilde y}_T)  - \frac{{\hat S} \cdot {\un \xi} }{ \xi_T} \, K_1 (\tilde{m}_{\alpha} \xi_T) \, K_0 (\tilde{m}_{\gamma} {\tilde y}_T) \right] \Bigg\} ,  \notag
\end{align}
where we have integrated out the newly-defined variable $\un r$ using the two-dimensional delta-function. 

Performing the integrals over the angles of $\un \xi$ in \eq{sigma_pol3} with the help of the angular integrals listed in Eqs.~\eqref{Iint} of Appendix~\ref{sec:AppC}, integrating out $\tilde{\un z}$, and integrating over the angles of $\tilde{\un y}$ we arrive at
\begin{align}\label{sigma_pol4}
& \frac{d \sigma_\chi}{d^2 k_T dy} = \chi \, \frac{\as G^2 \, N_c^2 \, M_p^2 \, \gamma}{4 (2 \pi)^3} {\hat S} \times \hat{k} \int d^2 b_\perp \int\limits_0^1  d \alpha \,  \frac{\alpha \, (1 - \alpha)}{|\gamma - \alpha |} \\ & \times \Bigg\{ - \min \{ \alpha, \gamma \} \, f_{11} (k_T, \tilde{m}_\gamma, b_T, y) \,  f_{00} \left( k_T \sqrt{\frac{\alpha (1-\alpha)}{\gamma (1-\gamma)}}, \tilde{m}_\alpha, b_T, y \right) \notag \\ & + \sqrt{\frac{\alpha \gamma}{(1-\alpha) (1-\gamma)}} \, (1 - \max \{ \alpha, \gamma \}) \, f_{00} (k_T, \tilde{m}_\gamma, b_T, y) \, f_{11} \left( k_T \sqrt{\frac{\alpha (1-\alpha)}{\gamma (1-\gamma)}}, \tilde{m}_\alpha, b_T, y \right) \Bigg\}  \notag
\end{align}
where $\hat{k} = {\un k}/k_T$ and we have defined
\begin{align}
f_{ij} (k_T, \tilde{m}, b_T, y) = \int\limits_0^\infty d \xi_T \, \xi_T \, J_i (k_T \xi_T) \,  K_j (\tilde{m} \xi_T) \, N(\xi_T, b_T, y)
\end{align}
for $i,j = 0,1$.

For perturbatively small distances $\xi_T \sim 1/k_T \ll 1/\tilde{m}$ we can expand the modified Bessel function obtaining
\begin{subequations}\label{fs}
\begin{align}
& f_{00} (k_T, \tilde{m}, b_T, y) \approx \int\limits_0^\infty d \xi_T \, \xi_T \, J_0 (k_T \xi_T) \,  \ln \left( \frac{1}{\tilde{m} \xi_T} \right) \, N(\xi_T, b_T, y), \\
& f_{11} (k_T, \tilde{m}, b_T, y) \approx \frac{1}{\tilde{m}} \int\limits_0^\infty d \xi_T \, J_1 (k_T \xi_T) \, N(\xi_T, b_T, y). 
\end{align}
\end{subequations}

Equation \eqref{sigma_pol4} is a fairly general simplification of our main \eq{sigma_pol2}, valid in the leading high-energy approximation (that is, for sufficiently large rapidity intervals between the produced quark and the unpolarized target). It can be used for most practical applications instead of \eq{sigma_pol2}. Next we will evaluate \eq{sigma_pol4} in the quasi-classical MV/Glauber--Mueller (GM)\cite{McLerran:1993ni,McLerran:1993ka,McLerran:1994vd,Mueller:1989st} approximation to study its properties, and, separately, explore the large-$k_T$ region. But first, an aside.


\subsubsection{An Aside}
\label{sec:aside}

As an aside let us note that in the regime where $N$ is linearized (that is, expanded to the lowest non-trivial order in the interaction with the target), $f_{11}$ and $f_{00}$ can be related to the Weizs\"{a}cker--Williams ($\phi^{WW}$) and dipole ($\phi^{dip}$) unintegrated gluon distributions (also known as the unpolarized gluon transverse momentum-dependent parton distributions, gluon TMD PDFs or simply gluon TMDs) \cite{JalilianMarian:1996xn,Kovchegov:1998bi,Braun:2000bh,Kovchegov:2001sc,Kharzeev:2003wz,Dominguez:2011wm} correspondingly. 
Indeed, recall the definitions of the Weizs\"{a}cker--Williams and dipole unintegrated gluon distributions \cite{JalilianMarian:1996xn,Kovchegov:1998bi,Braun:2000bh,Kovchegov:2001sc,Kharzeev:2003wz,Dominguez:2011wm},
\begin{subequations}\label{phi_defs}
\begin{align}
& \phi^{WW} (x, k_T^2) = \frac{C_F}{\as \, 2 \pi^3} \int d^2 b_\perp \, \frac{d^2 r_\perp}{r_T^2} \, e^{i {\un k} \cdot {\un r}} \ N_G ({\un r}, {\un b}, y = \ln (1/x)), \\
& \phi^{dip} (x, k_T^2) = - \frac{C_F}{\as \, (2 \pi)^3} k_T^2 \int d^2 b_\perp \, d^2 r_\perp \, e^{i {\un k} \cdot {\un r}} \ N_G ({\un r}, {\un b}, y = \ln (1/x)),
\end{align}
\end{subequations}
where $N_G$ is the gluon (adjoint) dipole scattering amplitude on the unpolarized target. At large-$N_c$ it is related to the quark dipole amplitude in \eq{Ndef} by $N_G = 2 N - N^2$. Outside the saturation region we can drop the quadratic term and write $N_G \approx 2 N$. Employing this approximation, and further assuming that $N ({\un r}, {\un b}, y)$ does not depend on the direction of $\un r$, we can integrate in Eqs.~\eqref{phi_defs} over the angles of $\un r$, obtaining the following approximate relations,
\begin{subequations}\label{phi_f}
\begin{align}
& f_{11} (k_T, \tilde{m}, b_T, y= \ln (1/x)) \approx - \frac{\as \pi^2}{2 {\tilde m} C_F} \frac{\pd}{\pd k_T} \frac{d \phi^{WW} (x, k_T^2, {\un b})}{d^2 b_\perp} , \\
& f_{00} (k_T, \tilde{m}, b_T, y= \ln (1/x)) \approx - \frac{\as 2 \pi^2}{k_T^2 C_F} \, \ln \left( \frac{\min \{ k_T, Q_s \} }{\tilde m} \right) \ \frac{d   \phi^{dip} (x, k_T^2, {\un b})}{d^2 b_\perp}) ,
\end{align}
\end{subequations}
where we have extended the definitions \eqref{phi_defs} to the differential fixed-impact parameter form, $d \phi /d^2 b_\perp$
 (cf. \cite{Kovchegov:2013ewa}). With the help of Eqs.~\eqref{phi_f}, we see that \eq{sigma_pol4} can be rewritten in terms of two $\sim \phi^{WW} \, \phi^{dip}$ terms. Note, however, that both $\phi^{WW}$ and $\phi^{dip}$ are distributions in the unpolarized target, one to the left and one to the right of the final-state cut. Hence, re-writing \eq{sigma_pol4} in terms of $\sim \phi^{WW} \, \phi^{dip}$ terms does not constitute factorization between the projectile and the target, and is more akin to expressing a diffractive scattering cross section as proportional to the square of the target gluon PDF.


\subsubsection{Asymmetry Estimate in the Quasi--Classical Approximation}
\label{sec:QuasiClass}

In the quasi-classical MV/GM \cite{McLerran:1993ni,McLerran:1993ka,McLerran:1994vd,Mueller:1989st} approximation the quark dipole amplitude is
\begin{align}\label{N_GGM}
N (r_T, b_T, y) = 1 - e^{-\tfrac{1}{4} \, r_T^2 \, Q_s^2 \, \ln \tfrac{1}{r_T \Lambda}},
\end{align}
where $Q_s = Q_s ({\un b})$ is the (energy-independent) quasi-classical quark saturation scale of the target while $\Lambda$ is an infrared (IR) cutoff. For brevity, we will not show the $\un b$-dependence of $Q_s ({\un b})$ explicitly below. For $k_T \sim 1/r_T$ not much larger than $Q_s$, that is, for $r_T \lesssim 1/Q_s \ \& \ r_T \gg 1/Q_s$, we can approximate \eq{N_GGM} by replacing the logarithm in the exponent by an order-one constant, that is \cite{Kovchegov:1998bi},
\begin{align}\label{N_GGM2}
N (r_T, b_T, y) \approx 1 - e^{-\tfrac{1}{4} \, r_T^2 \, Q_s^2}.
\end{align}
This is also known as the Golec-Biernat--Wusthoff (GBW) \cite{GolecBiernat:1998js,GolecBiernat:1999qd} approximation.

Substituting \eq{N_GGM2} into Eqs.~\eqref{fs} and integrating over $\xi_T$ we arrive at
\begin{subequations}\label{fs_GGM}
\begin{align}
& f_{11} (k_T, \tilde{m}, b_T, y) \approx \frac{e^{-\frac{k_T^2}{Q_s^2}}}{{\tilde m} k_T}, \\
& f_{00} (k_T, \tilde{m}, b_T, y) \approx \frac{1}{k_T^2} -  \frac{e^{-\frac{k_T^2}{Q_s^2}}}{Q_s^2} \left[ \textrm{Ei} \left( \frac{k_T^2}{Q_s^2} \right) - \textrm{ln} \left( \frac{ 4 \tilde{m}^2 k_T^2}{Q_s^4} \right) \right].
\end{align}
\end{subequations}
Here again we assume that $k_T , Q_s \gg M_P$. The function Ei$(x)$ is the exponential integral. 

Employing Eqs.~\eqref{fs_GGM} in \eq{sigma_pol4} yields
\begin{align}\label{sigma_pol5}
& \frac{d \sigma_\chi}{d^2 k_T dy} = \chi \, \frac{\as G^2 \, N_c^2 \, M_p }{4 (2 \pi)^3 k_T} {\hat S} \times \hat{k} \int d^2 b_\perp \int\limits_0^1  d \alpha \,  \frac{\alpha \, (1 - \alpha)}{|\gamma - \alpha |}  \\ & \times \left\{ - \min \{ \alpha, \gamma \} \, e^{-\frac{k_T^2}{Q_s^2}} \,  \left[ \frac{\gamma (1 - \gamma)}{k_T^2 \, \alpha ( 1 - \alpha)} -  \frac{e^{-\frac{k_T^2}{Q_s^2} \frac{\alpha (1-\alpha )}{\gamma (1- \gamma)}}}{Q_s^2} \left[ \textrm{Ei} \left( \frac{k_T^2}{Q_s^2} \frac{\alpha (1-\alpha )}{\gamma (1- \gamma)} \right) - \textrm{ln} \left( \frac{4 \tilde{m}_{\alpha}^2 k_T^2}{Q_s^4} \frac{\alpha (1-\alpha )}{\gamma (1- \gamma)} \right) \right] \right] \right. \notag \\ & \left. + \frac{\gamma^2 \, (1 - \max \{ \alpha, \gamma \})}{\alpha (1-\alpha) } \, e^{-\frac{k_T^2}{Q_s^2} \frac{\alpha (1-\alpha )}{\gamma (1-\gamma)}}  \, \left[ \frac{1}{k_T^2} -  \frac{e^{-\frac{k_T^2}{Q_s^2}}}{Q_s^2} \left[ \textrm{Ei} \left( \frac{k_T^2}{Q_s^2} \right) - \textrm{ln} \left( \frac{ 4 \tilde{m}_{\gamma}^2 k_T^2}{Q_s^4} \right) \right] \right]  \right\} . \notag
\end{align}
Once again, this result is valid in the quasi-classical approximation for $k_T , Q_s \gg M_P$ and in the $k_T \gsim Q_s \ \& \ k_T \ll Q_s$ transverse momentum ranges.

To study the STSA we need to substitute \eq{sigma_pol5} into \eq{ANdef} for $A_N$, which we rewrite as
\begin{equation}\label{ANdef2}
A_N (k_T, y) = \frac{\frac{\dd{\sigma}_{\chi = +}}{\dd[2]{k_T} \dd{y}} -  \frac{\dd{\sigma}_{\chi = -}}{\dd[2]{k_T} \dd{y}}}{2 \, \frac{\dd{\sigma}_{unp} }{\dd[2]{k_T} \dd{y}} },  
\end{equation}
where $\sigma_{unp}$ is the unpolarized hadron production cross section. Our goal here is not to do proper phenomenology, but to understand the main characteristics of our result. To that end, we will not include fragmentation functions to study the hadronic $A_N$. Instead, we will study the net partonic $A_N$ due to quark production in the numerator of \eq{ANdef2}. It is tempting to also keep only quark production in the denominator of \eq{ANdef2}: however, for central rapidities $y$ gluon production dominates over quark production in $\sigma_{unp}$, since the latter is a decreasing function of $\gamma$, while the former is not. While the proper thing to do would be to add both quark and gluon unpolarized production cross sections convoluted with their respective fragmentation functions, instead we will simply add the two partonic cross sections together in the denominator of \eq{ANdef2} and thus evaluate (cf. \cite{Kovchegov:2012ga})
\begin{equation}\label{ANdef3}
A_N (k_T, y) = \frac{\frac{\dd{\sigma}_{\chi = +}}{\dd[2]{k_T} \dd{y}} -  \frac{\dd{\sigma}_{\chi = -}}{\dd[2]{k_T} \dd{y}}}{2 \, \left[ \frac{\dd{\sigma}^q_{unp} }{\dd[2]{k_T} \dd{y}} + \frac{\dd{\sigma}^G_{unp} }{\dd[2]{k_T} \dd{y}} \right]},  
\end{equation}
where the superscripts $q$ and $G$ denote the quark and gluon production cross sections correspondingly. Again, \eq{ANdef3} should be considered as an estimate of the partonic $A_N$, and is not a real calculation of the hadronic STSA.

Having obtained the numerator for $A_N$ in \eq{sigma_pol5}, we now need to find the cross sections in the denominator of \eq{ANdef3}. The unpolarized quark production cross section in the quark--diquark model for the proton was already analyzed above, resulting in  \eq{sigma_unpol4}.   We need to further evaluate this expression in the quasi-classical approximation with $k_T , Q_s \gg M_P$ and $k_T$ not much larger than $Q_s$. Starting with the expression \eqref{sigma_unpol4}, we employ \eq{N_GGM2} while remembering that $S=1-N$ to obtain
\begin{align}\label{sigma_unpol5}
\frac{d \sigma_{unp}^q}{d^2 k_T dy} \approx \frac{G^2 \, N_c \, \gamma^3 \, (1-\gamma) \, M_P^2}{2 (2 \pi)^5}  &  \,  \int d^2 x_\perp \, d^2 {\tilde y}_\perp \, d^2 {\tilde z}_\perp \, e^{- i {\un k} \cdot (\tilde{\un z} - \tilde{\un y})}  \Big( 1 + e^{-(\tilde{\un z}-\tilde{\un y})^2  \frac{Q_s^2}{4}} - e^{-\tilde{z}_T^2 \frac{Q_s^2}{4}} - e^{-\tilde{y}_T^2  \frac{Q_s^2}{4}} \Big)   \\ & \times \left[  \ln (\tilde{m}_{\gamma} \, \tilde{z}_T)  \, \ln (\tilde{m}_{\gamma} \tilde{y}_T) + \frac{1}{\tilde{m}_{\gamma}^2} \, \frac{ \tilde{\un z}  \cdot \tilde{\un y} }{ \tilde{z}_T^2 \, \tilde{y}_T^2}   \right] , \notag
\end{align}
where $\tilde{\un z}$ and $\tilde{\un y}$ are defined in \eq{pos_shifts} as before, and we have expanded the modified Bessel functions $K_0 (\tilde{m}_{\gamma} \, \tilde{z}_T) \approx \ln 1/(\tilde{m}_{\gamma} \, \tilde{z}_T)$ and $K_1 (\tilde{m}_{\gamma} \, \tilde{z}_T) \approx 1/(\tilde{m}_{\gamma} \, \tilde{z}_T)$ due to the $k_T, Q_s \gg M_P$ assumption. Integration over $\tilde{\un z}$ and $\tilde{\un y}$ is straightforward, but a little tedious. It yields
\begin{align}\label{dsunp}
\frac{\dd{\sigma}^q_{unp} }{\dd[2]{k_T} \dd{y} } \approx  & \frac{N_c G^2 \gamma (1 - \gamma )}{2 (2 \pi)^3} \int \dd[2]{b}_{\perp}  \left[ \gamma^2 M_P^2 \Bigg( \frac{k_T^2 - Q_s^2}{Q_s^6} \, e^{-\frac{k_T^2}{Q_s^2}} \,\left[ \textrm{Ei} \left(\frac{k_T^2}{Q_s^2} \right)  - \textrm{ln} \left( \frac{ 4 \tilde{m}_{\gamma}^2 k_T^2}{Q_s^4} \right)  \right]	 \right.  \\ &
\left. +  \frac{2 e^{-\frac{k_T^2}{Q_s^2}}}{Q_s^4}  - \frac{1}{Q_s^4} + \frac{1}{k_T^2} \left[ \frac{1}{k_T^2} - 2  \frac{e^{-\frac{k_T^2}{Q_s^2}}}{Q_s^2} \left[ \textrm{Ei} \left( \frac{k_T^2}{Q_s^2} \right)  - \textrm{ln} \left( \frac{4 \tilde{m}_{\gamma}^2 k_T^2}{Q_s^4} \right) \right] \right] \right)	\notag\\ &
 \left. + \frac{e^{-\frac{k_T^2}{Q_s^2}}}{Q_s^2} \left[ \textrm{Ei} \left( \frac{k_T^2}{Q_s^2} \right) + 1 - \textrm{ln} \left( \frac{ 4 \tilde{m}_{\gamma}^2 k_T^2}{Q_s^4} \right) \right] -\frac{ 1 }{k_T^2} \left( 1 - 2 e^{-\frac{k_T^2}{Q_s^2}} \right)    \right]	, \notag
\end{align}
where ${\un b} = {\un x}$, also as before. Let us also remind the reader that $\tilde{m}_{\gamma} = \gamma M_P$ for massless quarks and for the diquark having the same mass as the proton, $M=M_P$.

The contribution in \eq{dsunp} falls off $\propto \gamma$ for small $\gamma$, as expected for ``valence" quark production at small $x$, which is suppressed at central rapidity \cite{Itakura:2003jp,Albacete:2006vv}. As mentioned above, this justifies the need to include gluon production cross section into \eq{ANdef3} to get a complete picture of $A_N$. Since, as we will see below, the numerator of $A_N$ given by \eq{sigma_pol5} falls off as $\propto \gamma$ for small $\gamma$ (at low $k_T$), including gluon production this way would ensure that the asymmetry vanishes as $\gamma \to 0$, in qualitative agreement with the experimental data.   As gluon production cannot occur in the quark--diquark model at the leading order, we take the approximate unpolarized cross section for soft gluon production from \cite{Kharzeev:2003wz} (see also \cite{Kovchegov:1998bi,JalilianMarian:2005jf}) derived for the quark projectile,
\begin{equation}\label{Gprod}
\frac{\dd{\sigma}^G_{unp} }{\dd[2]{k_T} \dd{y}}  \approx \frac{\alpha_s N_c}{2\pi^2} \int d^2 b_{\perp} \Bigg[ -\frac{1}{k_T^2} + \frac{2 e^{-\frac{k_T^2}{Q_s^2}}}{k_T^2} + \frac{ e^{-\frac{k_T^2}{Q_s^2}}}{Q_s^2} \Bigg( \textrm{Ei}\bigg[\frac{k_T^2}{Q_s^2} \bigg] -  \textrm{ln} \bigg[\frac{4 \Lambda^2 k_T^2}{Q_s^4} \bigg] \Bigg) \Bigg] ,
\end{equation}
with $\Lambda$ an IR cutoff, and add it to the quark production cross section \eqref{dsunp} to get an estimate of the transverse single-spin asymmetry in our model employing \eq{ANdef3}. 


\subsubsection{Plots of the Asymmetry}
\label{sec:plots}

\begin{figure}[ht]
	\includegraphics[width=0.7 \textwidth]{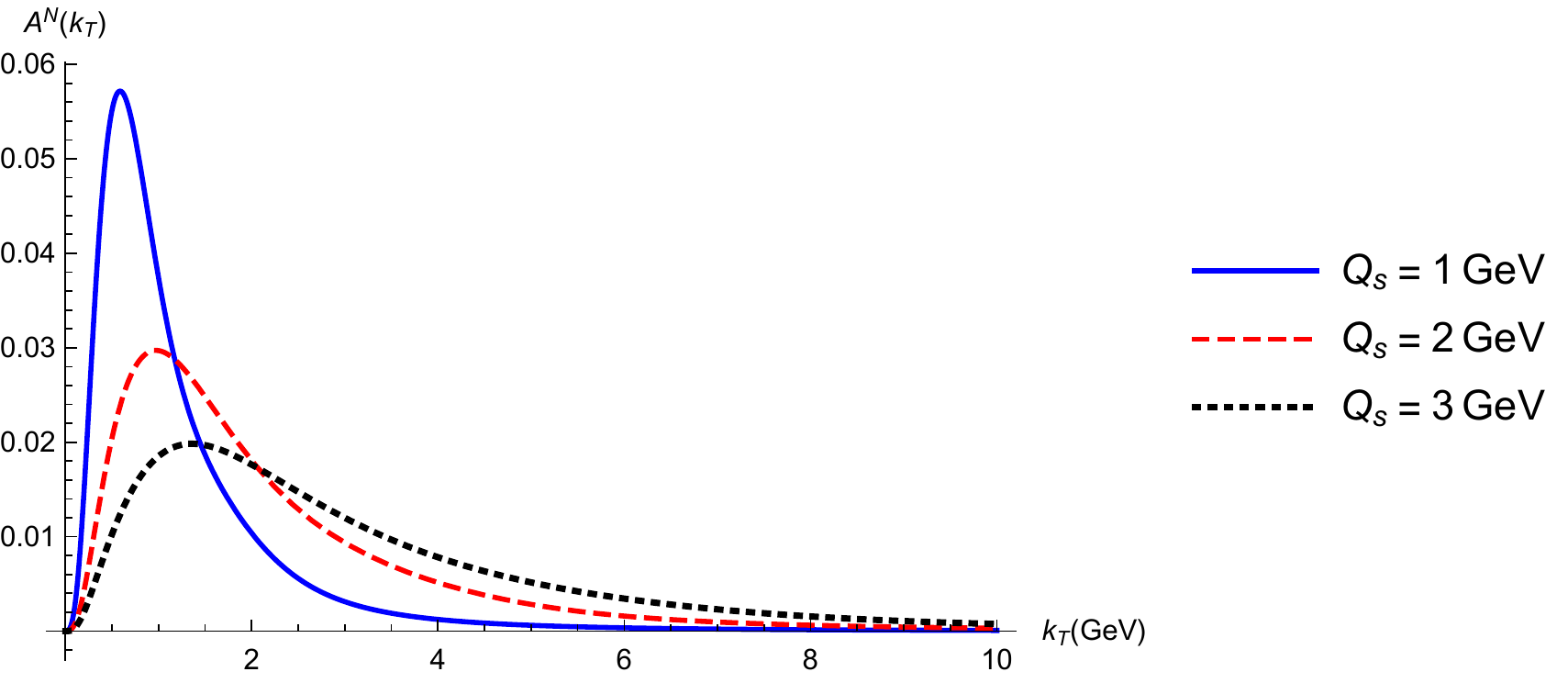}
	\caption{Plot of the leading-$N_c$ contribution to $A_N$ as a function of $k_T$ for various values of $Q_s$ and $\gamma = 0.3$. The asymmetry grows with $k_T$ at low momentum then turns over as it approaches the saturation scale, falling off quickly for $k_T \gg Q_s$.}
	\label{FIG:ANKQ2D}
\end{figure}

\begin{figure}[t]
	\includegraphics[width=0.7\textwidth]{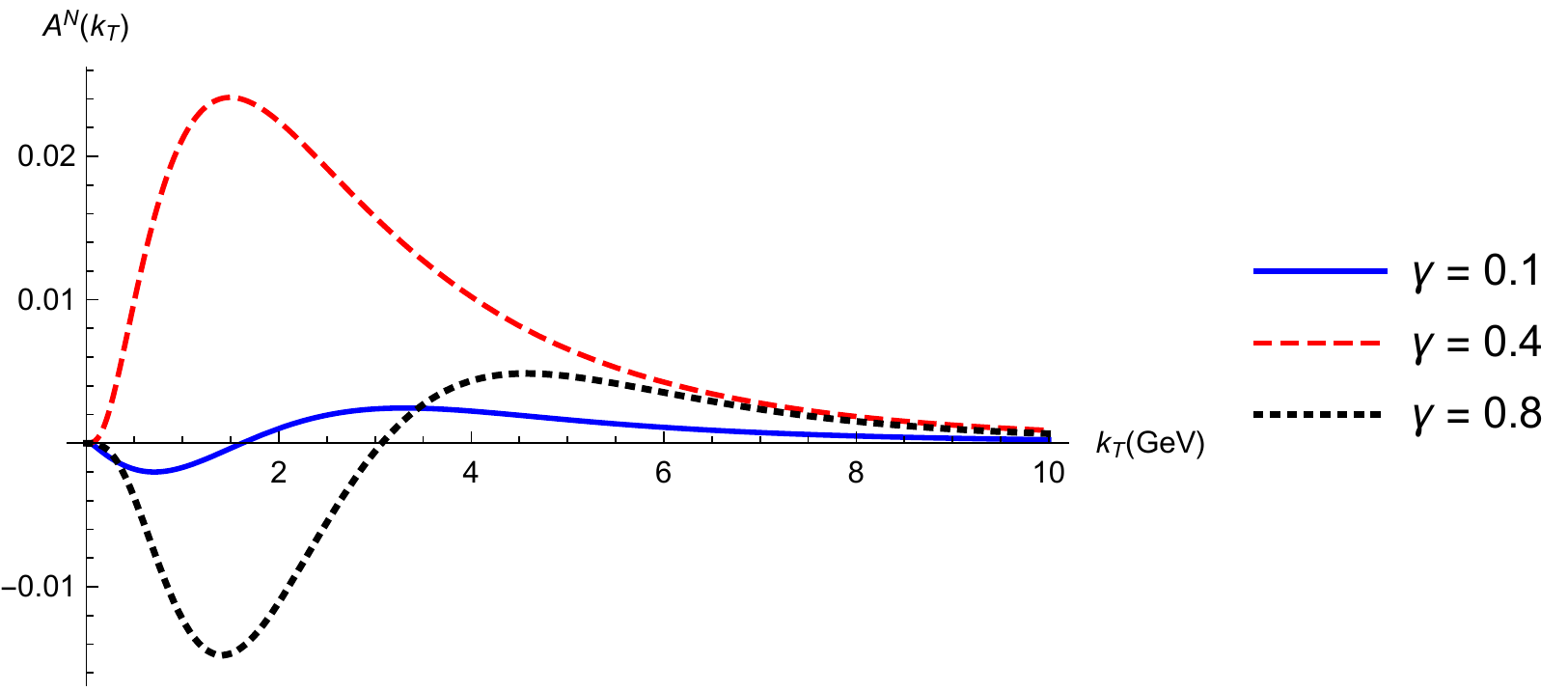}
	\caption{Plot of the leading-$N_c$ terms in $A_N$ as a function of $k_T$ for various values of $\gamma$ with $Q_s = 3$~GeV. For very large or very small $\gamma$ the asymmetry is negative at small $k_T$, and then changes sign at higher $k_T$  before falling off for $k_T \gg Q_s$.}
	\label{FIG:ANKG2D}
\end{figure}

\begin{figure}[ht]
	\includegraphics[width=0.6 \textwidth]{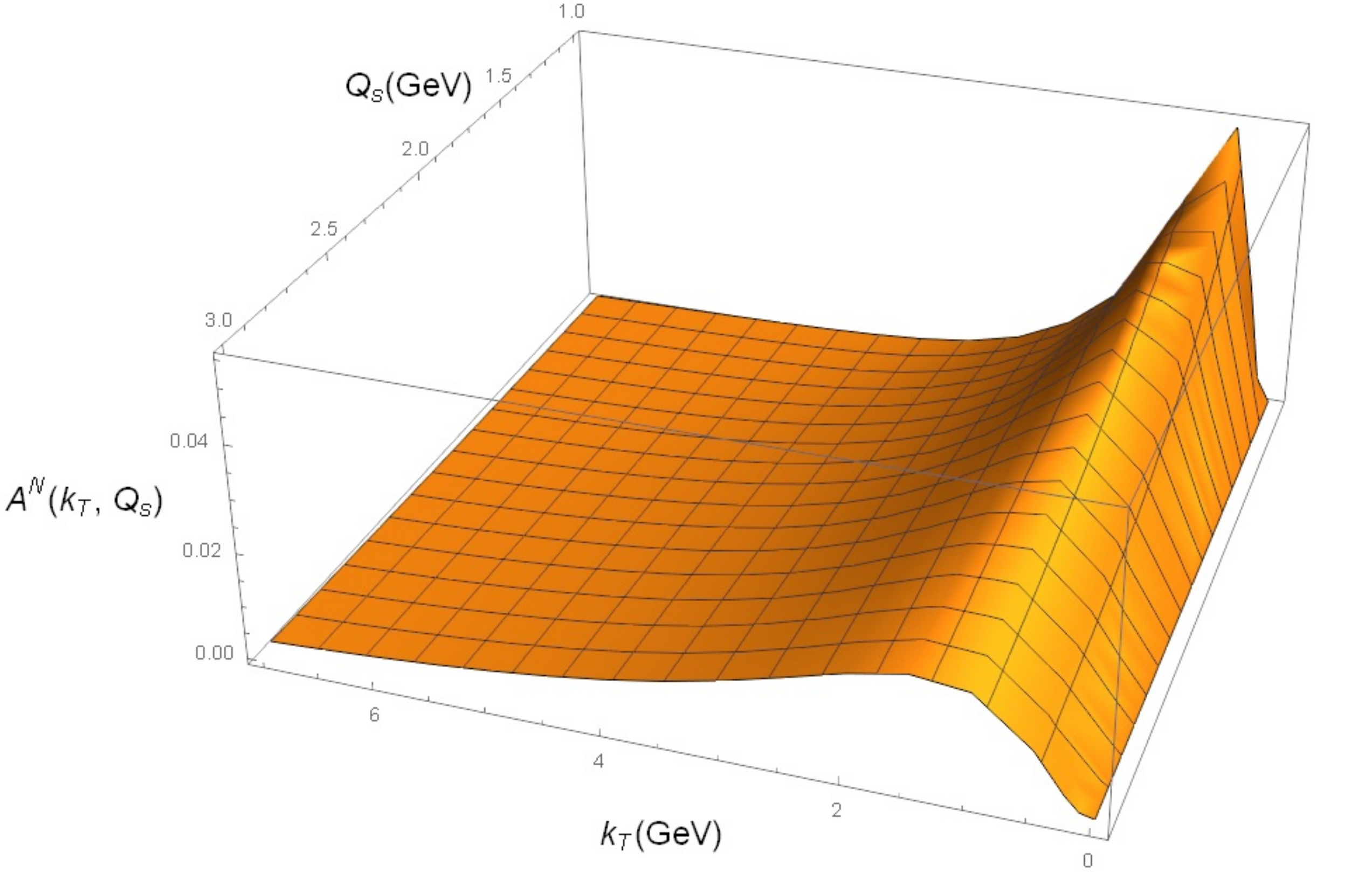}
	\caption{Plot of $A_N$ as a function of $k_T$ and $Q_s$ for $\gamma = 0.3$. The asymmetry falls off with increasing $Q_s \sim A^{1/6}$ for lower values of $k_T$, but appears to grow with $Q_s$ at higher $k_T$.}
	\label{FIG:ANKQ3D}
\end{figure}

\begin{figure}[ht]
	\includegraphics[width=0.7\textwidth]{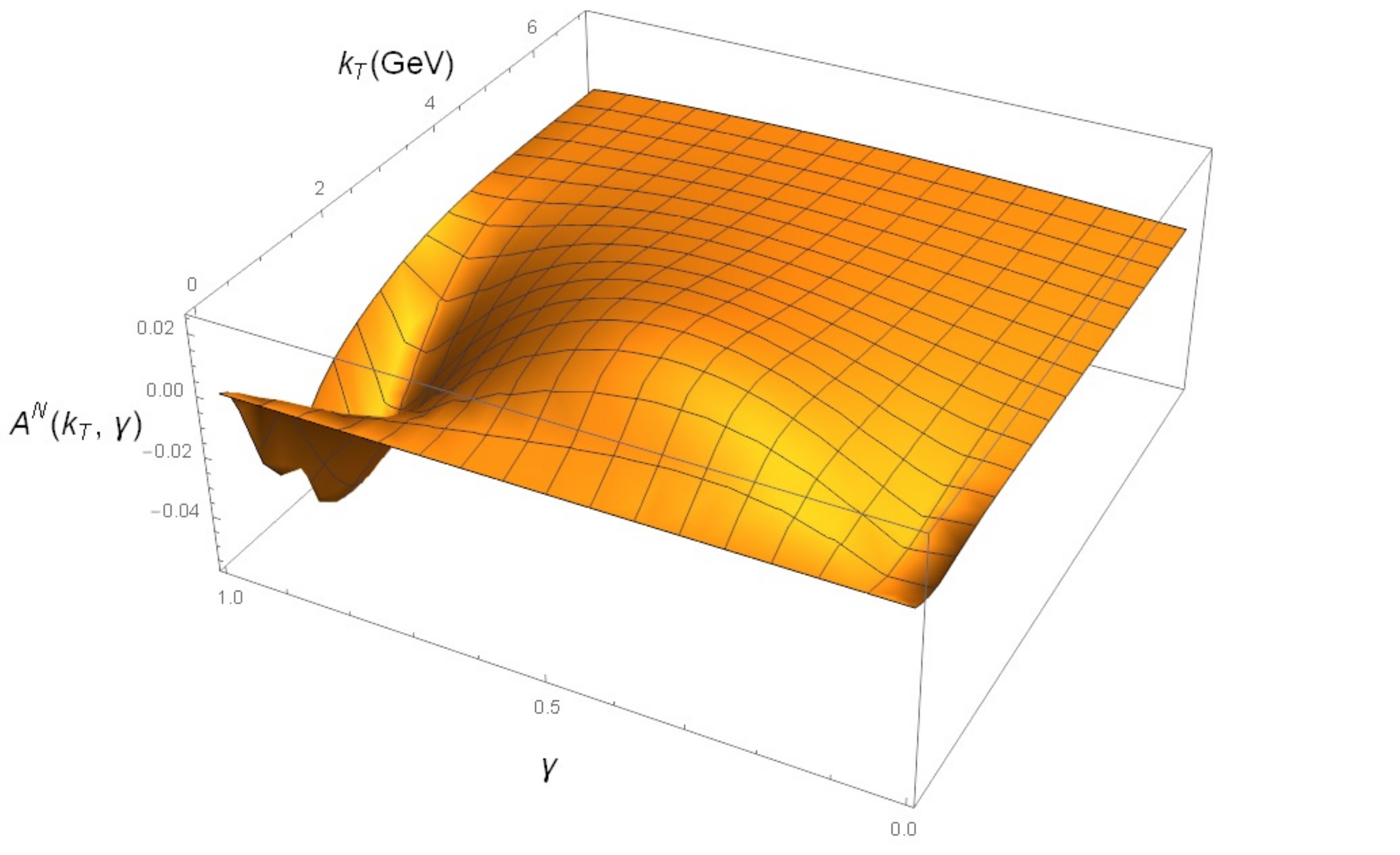}
	\caption{Plot of $A_N$ as a function of $k_T$ and $\gamma$ for $Q_s = 3$~GeV. The low-$k_T$ regime has an elaborate structure, with a distinct maximum at moderate $\gamma$ and sign-changing minima at high and low $\gamma$ which disappear as $\gamma$ nears $0$ or $1$.}
	\label{FIG:ANKG3D}
\end{figure}

We substitute Eqs.~\eqref{sigma_pol5}, \eqref{dsunp} and \eqref{Gprod} into \eq{ANdef3} and plot the resulting $A_N$ in Figures \ref{FIG:ANKQ2D}, \ref{FIG:ANKG2D}, \ref{FIG:ANKQ3D}, and \ref{FIG:ANKG3D}. In \eq{sigma_pol5} we replace
\begin{align}\label{Ssub}
{\hat S} \times \hat{k} \to -1
\end{align}
in order to adhere to the standard convention for $A_N$ where a positive asymmetry is given by the particles produced left of the polarized beam. 

We concentrate on the dependence of $A_N$ on $k_T$, $Q_s$ and $\gamma$. For simplicity we assume that $Q_s$ is a $\un b$-independent constant inside the nucleus, and is zero outside, such that the $b_\perp$-integrals in Eqs.~\eqref{sigma_pol5}, \eqref{dsunp} and \eqref{Gprod} give a factor of transverse area of the nucleus each; these factors cancel in $A_N$. Since $Q_s^2 \sim A^{1/3}$, the $Q_s$ dependence of $A_N$ probes how $A_N$ changes as the unpolarized target varies between the proton and various-size nuclei. ($A_N$ dependence on $Q_s$ may also be interpreted as centrality dependence for scattering on the same nucleus at different centrality bins.) Finally, our $\gamma$ has the meaning of the Bjorken $x$ variable in the polarized projectile proton. Since $y = \ln (\gamma P^+/k_T)$, the dependence of $A_N$ on $\gamma$ corresponds to the rapidity or Bjorken-$x$ dependence. 

We plot the asymmetry in Figures \ref{FIG:ANKQ2D}, \ref{FIG:ANKG2D}, \ref{FIG:ANKQ3D}, and \ref{FIG:ANKG3D} while taking $\alpha_s = 0.3$, $M_P=1$~GeV, $G=20$, and $\Lambda = \tilde{m}_{\gamma} $. The latter choice, $\Lambda = \tilde{m}_{\gamma} $, is done for consistency of the approach. Indeed, as follows from the wave function in \eq{MRWF}, the typical transverse size of the quark--diquark dipole is $1/{\tilde m}_\alpha \sim 1/{\tilde m}_\gamma$, making ${\tilde m}_\gamma$ the effective IR cutoff in the wave function. For consistency, we impose the same IR cutoff on other parts of the calculation by replacing $\Lambda \to \tilde{m}_{\gamma} = \gamma \, M_P$. One should worry that for small $\alpha \sim \gamma$ such IR cutoff may become small, resulting in quark--diquark dipoles becoming much larger than 1~fm. While indeed, to avoid this issue, it would be appropriate to replace ${\tilde m}_\alpha$ and ${\tilde m}_\gamma$ by something proportional to the QCD confinement scale $\Lambda_{QCD}$ for small $\alpha$ and $\gamma$ respectively, let us note that, as we will shortly see, at small $\gamma$ the asymmetry $A_N$ is also small, such that such a replacement, while justified, makes little numerical difference. Note that our Yukawa coupling in the quark--diquark model is very large, $G=20$: this coupling was chosen to get the values of $A_N$ in the same order of magnitude as the data. Our artificially high coupling $G$ presumably mimics the non-perturbative dynamics within the proton. One can also think of this large value of $G$ as simply adjusting the relative normalization between the quark \eqref{dsunp} and gluon \eqref{Gprod} contributions in the denominator of $A_N$ in \eq{ANdef3}: since the two terms were found in different models, their relative normalization is not fixed by our calculation, and quark dominance at large $\gamma$ has to be imposed by adjusting the value of $G$.

In \fig{FIG:ANKQ2D} we plot $A_N$ as a function of $k_T$ for various values of $Q_s$ with fixed $\gamma = 0.3$. We see that $A_N$ starts out growing with $k_T$, and then turns over at about $k_T \sim Q_s$ and falls off rapidly for $k_T \gg Q_s$. In addition, the magnitude of $A_N$ in the lower $k_T$ region decreases with increasing $Q_s$, corresponding to increasing atomic number $A$ of the target nucleus.  At the same time, the magnitude of $A_N$ at higher $k_T$ appears to grow with $A$. Thus, in this mechanism the low-transverse momentum asymmetry $A_N$ in $p^{\uparrow}+A$ is smaller for larger nuclei, while the higher-momentum $A_N$ is larger for higher $A$.

Similar conclusions about the $k_T$-dependence of $A_N$ can be reached from studying \fig{FIG:ANKG2D}, where we plot $A_N$ versus $k_T$ for three different values of $\gamma$ and for fixed $Q_s = 3$~GeV. While the magnitude of $A_N$ still grows with $k_T$ at $k_T \ll Q_s$, we also see that the growth is not monotonic and nodes in $A_N$ appear at certain values of $\gamma$ and $k_T$.

The conclusions we draw from Figs.~\ref{FIG:ANKQ2D} and \ref{FIG:ANKG2D} are further illustrated by the 3D plots in Figs.~\ref{FIG:ANKQ3D} and \ref{FIG:ANKG3D}.  In \fig{FIG:ANKQ3D} we plot $A_N$ as a function of $k_T$ and $Q_s \sim A^{1/6}$. Again we see growth with $k_T$ at low momenta, followed by a fall-off. The low-$k_T$ asymmetry seems to decay with increasing $Q_s$ (and, hence, $A$), while at high $k_T$ it seems to grow with $A$.

The 3D plot in \fig{FIG:ANKG3D} shows $A_N$ versus $k_T$ and $\gamma$ for $Q_s = 3$~GeV. At low $k_T$ we see the oscillations resulting in nodes in $A_N$ we have already seen in \fig{FIG:ANKG2D}. Again we observe a rapid fall-off at high $k_T$. Finally, while the behavior of $A_N$ at finite $\gamma$ is not monotonic in $k_T$ and $\gamma$, the asymmetry goes to zero as $\gamma \to 0$, as expected from \eq{sigma_pol5} and in qualitative agreement with the experiment. We note that the gluon production cross section \eqref{Gprod} we are using in the denominator of \eq{ANdef3} is $\gamma$-independent, so the $\gamma$-dependence of $A_N$ at small $\gamma$ (for $\gamma < 0.1$) where gluon production dominates in the denominator of \eqref{ANdef3} is purely determined by the polarized quark production cross section \eqref{sigma_pol5}. For larger values of $\gamma$ ($1 > \gamma >0.1$), quark production dominates in the denominator of \eqref{ANdef3}, and the gluon production cross-section is not important.

To conclude the discussion of the plots of $A_N$, let us note that while our plots here are done at the partonic level, and as such cannot be directly compared with experiment, we could still try to compare the qualitative trends in our results with those in experiment. We see that the growth with $A$ of our  $A_N$ at moderately high $k_T$ appears not to be consistent with most published experiment measurements, with the exception of perhaps \cite{Aidala:2017cnz}. (However, the measurement in \cite{Aidala:2017cnz} is performed at low $k_T$: it appears unclear at this point whether the results of  \cite{Aidala:2017cnz} can be accounted for by the growth of $A_N$ with $A$ for moderately high $k_T$ we saw in \fig{FIG:ANKQ2D} even at the qualitative level.) The above-observed suppression of the asymmetry with increasing $A$ at low $k_T$ (see Figs.~\ref{FIG:ANKQ2D} and \ref{FIG:ANKQ3D}) seems to agree with the data reported by PHENIX \cite{Aidala:2019ctp}. Furthermore, our plots do not seem to exhibit flatness at high $k_T$, as observed in \cite{Heppelmann:2013DIS,Aschenauer:2013woa}. As we will see below, at high-$k_T$ the asymmetry in our approach is dominated by the subleading-$N_c$ contribution we have not yet evaluated. So a comparison of the transverse-momentum and $A$-dependence of our $A_N$ with the data is premature at this point.


\subsubsection{High- and Low-$k_T$ STSA at Large-$N_c$}
\label{sec:high_kT1}

Let us support our conclusions obtained from the figures by analytical estimates of $A_N$ at high and low $k_T$. \\

We can find the large $k_T$ asymptotics by expanding \eq{sigma_pol5} for $k_T \gg Q_s \gg M_P$. While \eq{sigma_pol5} does not strictly-speaking apply for $k_T \gg Q_s$ due to us neglecting the logarithms in the exponent of \eq{N_GGM} when approximating it by \eq{N_GGM2}, in this case the discrepancy is logarithmic in $k_T$, and expanding \eq{sigma_pol5} for $k_T \gg Q_s$ should give us the powers of $k_T$ and of other relevant quantities correctly.  When  $k_T \gg Q_s$ we can neglect all Gaussians of $k_T$ in \eq{sigma_pol5}, unless they  are multiplied by an exponential integral of the same argument or if they contain the $\alpha(1-\alpha)$ factor, which is not suppressed in the $k_T \gg Q_s$ regime only for $\alpha (1-\alpha) \ll 1$. We get
\begin{equation}\label{eq:highkT1}
\frac{d \sigma_\chi}{d^2 k_T dy} \Bigg|_{k_T \gg Q_s} \approx  -  \chi  \frac{\as G^2 N_c^2 M_P }{4(2 \pi)^3} \, \hat{S} \cross \hat{k} \,	\int  \dd[2]{b_{\perp}} \frac{Q_s^2}{k_T^5} \int\limits_0^1 d \alpha \,\frac{ \gamma^2 (1 - \max \{ \alpha,\gamma \})}{|\alpha-\gamma|}  e^{-\frac{k_T^2}{Q_s^2}  \frac{\alpha (1-\alpha )}{\gamma (1- \gamma)}} .
\end{equation}
For $k_T \gg Q_s$ the integral is dominated by the small-$\alpha$ region where the exponential suppression is weak, since the $\alpha \to 1$ region is further suppressed by the $1 - \max \{ \alpha,\gamma \} \approx 1 - \alpha$ factor in the numerator of \eq{eq:highkT1}. We can approximate \eq{eq:highkT1} by taking $\alpha \ll 1$ and integrating over $\alpha$ from zero to infinity, obtaining  
\begin{align}\label{eq:highkT2}
\frac{d \sigma_\chi}{d^2 k_T dy} \Bigg|_{k_T \gg Q_s} & \approx  -  \chi  \frac{\as G^2 N_c^2 M_P \gamma (1-\gamma)}{4(2 \pi)^3} \, \hat{S} \cross \hat{k} \,	\int  \dd[2]{b_{\perp}} \frac{Q_s^2}{k_T^5} \int\limits_0^\infty d \alpha \,  e^{-\frac{k_T^2}{Q_s^2}  \frac{\alpha}{\gamma (1- \gamma)}} \notag \\
& = -  \chi  \frac{\as G^2 N_c^2 M_P \gamma^2 (1-\gamma)^2}{4(2 \pi)^3} \, \hat{S} \cross \hat{k} \,	\int  \dd[2]{b_{\perp}} \frac{Q_s^4}{k_T^7} \propto S_\perp \, M_P \, \frac{Q_s^4}{k_T^7},
\end{align}
where $S_\perp$ is the transverse area of the unpolarized target. 

Taking $k_T \gg Q_s \gg M_P$ we can expand Eqs.~\eqref{dsunp} and \eqref{Gprod} to derive their high-$k_T$ asymptotics
\begin{equation}\label{eq:highkT3}
\frac{\dd{\sigma}^q_{unp} }{\dd[2]{k_T} \dd{y} } \Bigg|_{k_T \gg Q_s} \approx \frac{N_c G^2 \gamma (1 - \gamma )}{ 2(2\pi)^3} S_{\perp}   \frac{Q_s^2}{k_T^4}, \ \ \ \frac{\dd{\sigma}^G_{unp} }{\dd[2]{k_T} \dd{y}}  \Bigg|_{k_T \gg Q_s} \approx \frac{\alpha_s N_c}{2\pi^2} S_{\perp} \frac{Q_s^2}{k_T^4},
\end{equation}
and see that the unpolarized production cross sections scale as $Q_s^2/k_T^{4}$. 

We conclude that at high-$k_T$ the STSA scales as
\begin{align}\label{ANhigh}
A_N (k_T, y) \Bigg|_{k_T \gg Q_s} \sim \frac{Q_s^2 \, M_P}{k_T^3}.
\end{align}
It falls off with $k_T$ and grows with the atomic number $A$ of the target nucleus since $Q_s^2 \propto A^{1/3}$, in agreement with the plot in \fig{FIG:ANKQ2D}. Unfortunately, the rapid fall-off with $k_T$ in \eq{ANhigh} appears to contradict the data \cite{Heppelmann:2013DIS,Aschenauer:2013woa}: we will return to this question in the next Subsection. Note also that the high-$k_T$ asymmetry falls off rapidly with decreasing $\gamma$, as one can see from \eq{eq:highkT2}, in agreement with the curves in \fig{FIG:ANKG2D}.

At low $k_T$ we perform a similar expansion for cross-sections in Eqs.~\eqref{sigma_pol5}, \eqref{dsunp} and \eqref{Gprod}, now assuming that $k_T \ll Q_s$ while, at the same time, $k_T \gg M_P$. For the polarization-dependent cross section we arrive at
\begin{align}\label{eq:highkT4}
\frac{d \sigma_\chi}{d^2 k_T dy} \Bigg|_{M_P \ll k_T \ll Q_s} & \approx  \chi  \frac{\as G^2 N_c^2 M_P }{4(2 \pi)^3} \, \hat{S} \cross \hat{k} \ S_{\perp} \, \frac{\gamma (3+ \gamma (14 \gamma - 15))}{3 \, k_T \, Q_s^2} \, \ln \frac{Q_s}{M_p}  ,
\end{align}
where we have also employed the $Q_s \gg M_p$ condition to drop the $\gamma$-dependent ``constant" under the logarithm. For the unpolarized cross sections we similarly obtain
\begin{equation}\label{eq:highkT5}
\frac{\dd{\sigma}^q_{unp} }{\dd[2]{k_T} \dd{y} } \Bigg|_{M_P \ll k_T \ll Q_s} \approx \frac{N_c G^2 \gamma (1 - \gamma )}{ 2(2\pi)^3} \, \frac{S_{\perp}}{k_T^2}, \ \ \ \frac{\dd{\sigma}^G_{unp} }{\dd[2]{k_T} \dd{y}}  \Bigg|_{M_P \ll k_T \ll Q_s} \approx \frac{\alpha_s N_c}{2\pi^2}  \,\frac{S_{\perp}}{k_T^2},
\end{equation}
where we have also employed the $k_T \gg M_P$ condition to drop the $\sim M_P^2/k_T^4$ term in the quark production cross section. 

Combining Eqs.~\eqref{eq:highkT4} and \eqref{eq:highkT5} we arrive at the following scaling of the STSA:
\begin{align}\label{ANlow}
A_N (k_T, y) \Bigg|_{M_P \ll k_T \ll Q_s} \sim \frac{k_T \, M_P}{Q_s^2} \, \ln \frac{Q_s}{M_p} .
\end{align}
We see that indeed $A_N \to 0$ for $k_T \to 0$. (The apparent deviations from the linear scaling $A_N \sim k_T$ at very low $k_T$ in Figs.~\ref{FIG:ANKQ2D}, \ref{FIG:ANKG2D}, and \ref{FIG:ANKQ3D} above can be attributed to the fact that our $k_T \gg M_P$ assumption used in deriving \eq{ANlow} is violated for the lowest $k_T$ values in those figures.) We also see that the low-$k_T$ $A_N$ in \eq{ANlow} is a decreasing function of the atomic number $A$, in agreement with the plots in Figs.~\ref{FIG:ANKQ2D} and \ref{FIG:ANKQ3D}.


\subsection{Estimates of the Asymmetry: Subleading-$N_c$}
\label{sec:estQ}

Let us revisit the question of high-$k_T$ asymptotics of $A_N$. As we saw in \eq{eq:highkT2}, the large-$N_c$ (double-trace) term in the polarization-dependent cross section \eqref{sigma_pol2} falls off rather fast with $k_T$, 
\begin{align}\label{eq:highkT22}
\frac{d \sigma_\chi^{\mbox{double} \ \mbox{trace}}}{d^2 k_T dy} \Bigg|_{k_T \gg Q_s} \propto S_\perp \, M_P \, N_c^2 \, \frac{Q_s^4}{k_T^7},
\end{align}
resulting in a fast fall-off of $A_N \sim 1/k_T^3$ at large $k_T$ in \eq{ANhigh}. The origin of this steep fall-off is easy to understand using our main result for polarization-dependent cross section in \eq{sigma_pol2}: there, one observes that the double-trace term is given by a 4-gluon exchange with the target at the lowest non-trivial order, which results in an additional factor of $Q_s^2/k_T^2$ suppression. At the same time, the single-trace term in \eq{sigma_pol2} starts out with a 2-gluon exchange at the lowest non-trivial order: hence one would expect that at high $k_T$ the contribution of this term to the polarization-dependent cross section in \eq{sigma_pol2} scales as $1/k_T^5$, that is,
\begin{align}\label{eq:highkT33}
\frac{d \sigma^{\mbox{single} \ \mbox{trace}}_\chi}{d^2 k_T dy} \Bigg|_{k_T \gg Q_s} \propto S_\perp \, M_P \, \frac{Q_s^2}{k_T^5}.
\end{align}
Comparing Eqs.~\eqref{eq:highkT22} and \eqref{eq:highkT33} we see that for $k_T \gg N_c \, Q_s$ the subleading-$N_c$ contribution \eqref{eq:highkT33} dominates. Hence the large-$k_T$ asymptotics of $A_N$ is given by the subleading-$N_c$ single-trace term in \eq{sigma_pol2}. To estimate this large-$k_T$ limit, study its properties, and to verify the above argument, let us evaluate the contribution of the single-trace term in \eq{sigma_pol2} at $k_T \gg N_c \, Q_s$.

The interaction with the target due to the single-trace term in \eq{sigma_pol2} is
\begin{equation}\label{Qint1}
- \frac{1}{2 N_c} \left\langle \tr \left[ \left( V_{\un x}^\dagger \, V_{\un z} - 1 \right) \, \left( V_{\un y}^\dagger \, V_{{\un x}'} - 1 \right) \right] \, \right\rangle_y  =  \frac{1}{2}  \left[ S_{{\un z}, {\un x}} (y) + S_{{\un x}', {\un y}} (y) - 1 -  Q_{{\un z}, {\un x}, {\un x}', {\un y}} (y) \right] , 
\end{equation}
where we have introduced the color-quadrupole amplitude
\begin{align}
Q_{{\un z}, {\un x}, {\un x}', {\un y}} (y) \equiv  \left\langle \frac{1}{N_c} \tr \left[V_{\un x}^\dagger \, V_{\un z} \, V_{\un y}^\dagger \, V_{{\un x}'} \right] \right\rangle_y , 
\end{align}
which was found in the MV/GM approximation to be \cite{JalilianMarian:2004da}
\begin{align}\label{Q_class}
& Q_{{\un z}, {\un x}, {\un x}', {\un y}} = e^{- \tfrac{1}{4} ({\un z} - {\un x})^2 Q_s^2 \ln \tfrac{1}{|{\un z} - {\un x}| \Lambda} - \tfrac{1}{4} ({\un y} - {\un x}')^2 Q_s^2 \ln \tfrac{1}{|{\un y} - {\un x}'| \Lambda}  }	\\
& + \frac{ ({\un z} - {\un y})^2 \ln \tfrac{1}{|{\un z} - {\un y}| \Lambda} + ({\un x} - {\un x}')^2 \ln \tfrac{1}{|{\un x} - {\un x}'| \Lambda}  - ({\un z} - {\un x}')^2 \ln \tfrac{1}{|{\un z} - {\un x}'| \Lambda}  - ({\un x} - {\un y})^2 \ln \tfrac{1}{|{\un x} - {\un y}| \Lambda} }{({\un z} - {\un x})^2 \ln \tfrac{1}{|{\un z} - {\un x}| \Lambda} + ({\un y} - {\un x}')^2 \ln \tfrac{1}{|{\un y} - {\un x}'| \Lambda}  - ({\un z} - {\un y})^2 \ln \tfrac{1}{|{\un z} - {\un y}| \Lambda} - ({\un x} - {\un x}')^2 \ln \tfrac{1}{|{\un x} - {\un x}'| \Lambda} }	\notag \\
& \times \left[ e^{- \tfrac{1}{4} ({\un z} - {\un x})^2 Q_s^2 \ln \tfrac{1}{|{\un z} - {\un x}| \Lambda} - \tfrac{1}{4} ({\un y} - {\un x}')^2 Q_s^2 \ln \tfrac{1}{|{\un y} - {\un x}'| \Lambda}  } - e^{- \tfrac{1}{4} ({\un z} - {\un y})^2 Q_s^2 \ln \tfrac{1}{|{\un z} - {\un y}| \Lambda} - \tfrac{1}{4} ({\un x} - {\un x}')^2 Q_s^2 \ln \tfrac{1}{|{\un x} - {\un x}'| \Lambda}  }  \right]. \notag
\end{align}

The lowest-order interaction with the target is obtained by expanding \eq{Qint1} to the lowest non-trivial order in $Q_s^2$. Employing the GBW approximation for the MV model again, which implies replacing all the logarithms in \eq{Q_class} by 1, we obtain
\begin{align}\label{Qint2}
\frac{1}{2}  \left[ S_{{\un z}, {\un x}} (y) + S_{{\un x}', {\un y}} (y) - 1 -  Q_{{\un z}, {\un x}, {\un x}', {\un y}} (y) \right] \approx - \frac{Q_s^2}{4} \, {\un \xi} \cdot \tilde{\un y}, 
\end{align}
where we employed the transverse vectors defined in Eqs.~\eqref{pos_shifts}. Substituting Eqs.~\eqref{Qint2} and \eqref{Qint1} into \eq{sigma_pol2}, performing the substitution \eqref{pos_shifts} and integrating out $\un r$ with the help of the delta-function yields (cf. \eq{sigma_pol3})
\begin{align}\label{sigma_Q1}
& \frac{d \sigma^{\mbox{single} \ \mbox{trace}}_\chi}{d^2 k_T dy} \Bigg|_{k_T \gg Q_s} \approx - \chi \, \frac{i \as G^2 \, M_p^2 \, \gamma}{4 (2 \pi)^6} \! \int\limits_0^1  d \alpha \, (1 - \alpha) \!\! \int \! d^2 b_\perp \, d^2 \xi_\perp \, d^2 {\tilde y}_\perp \, d^2 {\tilde z}_\perp  \,  \frac{e^{- i {\un k} \cdot (\tilde{\un z} - \tilde{\un y})}}{   |(1-\alpha) \, {\un \xi} - (1-\gamma) \tilde{\un z}|^2}  \notag	\\ & \times
 \delta \left[  \frac{\tilde{z}_T^2}{\alpha (1-\alpha)}  - \frac{\xi_T^2}{\gamma (1-\gamma)}   \right] \, Q_s^2 \, {\un \xi} \cdot \tilde{\un y}  \\ & \times  \Bigg\{  {\un \xi}  \cdot   ( \alpha {\un \xi} + (1-\gamma) \tilde{\un z} ) \left[ \frac{{\hat S} \times \tilde{\un y} }{ {\tilde y}_T} K_0 (\tilde{m}_{\alpha} {\xi}_T)  K_1 (\tilde{m}_{\gamma} {\tilde y}_T )  + \frac{{\hat S} \times {\un \xi} }{ \xi_T} K_1 (\tilde{m}_{\alpha} {\xi}_T)  K_0 (\tilde{m}_{\gamma} \tilde{y}_T) \right] \notag \\ &  - (1-\gamma) \, {\un \xi} \times  \tilde{\un z} \left[ \frac{{\hat S} \cdot \tilde{\un y}}{ \tilde{y}_T} \, K_0 (\tilde{m}_{\alpha} \xi_T)  K_1 (\tilde{m}_{\gamma} {\tilde y}_T)  - \frac{{\hat S} \cdot {\un \xi} }{ \xi_T} \, K_1 (\tilde{m}_{\alpha} \xi_T) \, K_0 (\tilde{m}_{\gamma} {\tilde y}_T) \right] \Bigg\} .  \notag
\end{align}
 
Further simplification of \eq{sigma_Q1} consists of integrating out $\tilde{\un y}$, integrating over the angles of the vector $\un \xi$ using the integrals listed in Eqs.~\eqref{I_tilde_int} of Appendix~\ref{sec:AppC}, integrating out the angles of $\tilde{\un z}$, and integrating out the magnitude $\tilde{z}_T$ with the help of the delta-function in \eq{sigma_Q1}. Finally, integrating out $\xi_T$ and $\alpha$ and again assuming that $k_T \gg Q_s \gg M_P$ we arrive at 
\begin{equation}\label{dsquad}
\frac{d \sigma^{\mbox{single} \ \mbox{trace}}_\chi}{d^2 k_T dy} \Bigg|_{k_T \gg Q_s} \approx -  \chi  \frac{\alpha_s G^2  M_P  \, \gamma^2 \ln (\gamma) }{ 4 (2 \pi)^3 \, k_T^5} \,  \hat{S} \cross \hat{k} \int d^2 b_{\perp} \, Q_s^2  =  -  \chi  \frac{\alpha_s G^2  M_P  \, \gamma^2 \ln (\gamma) }{ 4 (2 \pi)^3 \, k_T^5} \,  \hat{S} \cross \hat{k} \, S_{\perp} \, Q_s^2 . 
\end{equation}
We observe that the scaling of \eq{eq:highkT33} is indeed confirmed by our calculation. 

Substituting \eq{dsquad} into \eq{ANdef3} (while employing the substitution \eqref{Ssub} to observe that $A_N >0$), and employing Eqs.~\eqref{eq:highkT3} we see that
\begin{align}\label{AN_very_high}
A_N (k_T, y) \Bigg|_{k_T \gg N_c \, Q_s} \sim \frac{M_P}{k_T}.
\end{align}
We see that now $A_N \sim 1/k_T$, such that the fall-off with $k_T$ is very mild, in a potentially better agreement with the STAR collaboration data \cite{Heppelmann:2013DIS,Aschenauer:2013woa}. Indeed we are employing a simple quark--diquark model , so one should not expect our model to be in good quantitative agreement with the data. Parton fragmentation functions need to be included as well to do proper comparison with the data.  

Another important feature of \eq{AN_very_high} is that $A_N$ in it is independent of the target's atomic number $A$ (cf. \cite{Hatta2019a}).  This is in qualitative agreement with the preliminary results reported by STAR \cite{Dilks:2016ufy}, but seems to disagree with the PHENIX data \cite{Aidala:2019ctp} which is more in line with our low-$k_T$ result \eqref{ANlow}.


\section{Conclusions}
\label{sec:Concl}

In this paper we have calculated the STSA for quark production in $p^{\uparrow}+p$ and $p^{\uparrow}+A$ collisions resulting from the lensing mechanism embedded in the small-$x$/saturation framework, with the corresponding transverse spin-dependent cross section given by \eq{sigma_pol2}. This mechanism leads to several key features of $A_N$. First of all, the inelastic contribution is suppressed by a power of $1/N_c^2$, arising from a single-color-trace interaction as opposed to the elastic, leading-order double-color-trace interaction. This leads to an $A_N$ generated primarily in elastic collisions. Second, the asymmetry grows or oscillates with transverse momentum at $k_T \lesssim Q_s$, turning over as the momentum nears the saturation scale $Q_s$ and falling off as $1/k_T$ for very high momenta. The $1/k_T$ fall-off is driven by the inelastic $1/N_c^2$-suppressed single-trace term, which becomes dominant for $k_T \gg N_c \, Q_s$: thus, at very high $k_T$ the asymmetry is dominated by inelastic interaction, and falls off rather slowly with $k_T$. Finally, $A_N$ decreases as the target atomic number $A$ increases for $k_T$ below or near $Q_s$, while it is independent of $A$ for $k_T \gg N_c \, Q_s$. At large $N_c$ there is an intermediate region $Q_s \ll k_T \ll N_c \, Q_s$ where $A_N$ increases with increasing $A$, though phenomenological relevance of this region is not clear. 

The dominance of the elastic contributions in $A_N$ is qualitatively in agreement with the observations reported in \cite{Dilks:2016ufy}. In our calculation it arises directly from the color structure of the target interaction, where the leading-$N_c$ part of the final-state gluon exchange between the quark and diquark preferentially selects the color-singlet quark and diquark state. We believe this conclusion would remain valid even for multiple gluon exchanges between the quark and diquark in the final state, since planar large-$N_c$ diagrams would always require the quark and diquark to be in the color-singlet state. Therefore, it appears that our conclusion of the elastic dominance of the interaction is not specific for the quark--diquark model we considered here.

The $k_T$ dependence of $A_N$ far above $Q_s$ gives a plausible explanation for the slow fall-off of the asymmetry with transverse momentum which has been observed in \cite{Adams:1991cs,Abelev:2008af,Dilks:2016ufy,Heppelmann:2013DIS,Aidala:2019ctp}, though indeed a more realistic model than we have considered in this work, augmented by the proper fragmentation functions, would be needed for a detailed comparison with the data. As for the dependence on the target's atomic number $A$, at $k_T$ far above $Q_s$ our mechanism's prediction of $A$-independence of $A_N$  is in line with the experimental observations in \cite{Dilks:2016ufy} and with other theoretical results \cite{Hatta2019a}, while for lower $k_T$ we get $A_N$ suppressed for larger $A$ as observed in \cite{ Aidala:2019ctp}.

Our preliminary estimates, not shown in this work, indicate that inclusion of small-$x$ evolution effects in the interaction with the target along the lines of \cite{Kharzeev:2002pc,Kharzeev:2003wz,Albacete:2003iq,JalilianMarian:2004da,Dominguez:2011gc} is not likely to qualitatively modify our main conclusions summarized above concerning the $k_T$- and $A$-dependence of $A_N$. Mild modifications of the powers of $Q_s$ and $k_T$ in Eqs.~\eqref{ANhigh} and \eqref{ANlow} will take place due to the anomalous dimension of the Balitsky--Fadin--Kuraev--Lipatov (BFKL)
\cite{Kuraev:1977fs,Balitsky:1978ic} evolution. We expect the power of $k_T$ in \eq{AN_very_high} to be unaffected by the small-$x$ evolution.

We should note some of the limitations of our calculation coming from the simplicity of the quark--diquark model. This model has an uncertainty in the magnitude of the asymmetry, as the Yukawa coupling $G$ is not fixed to match any underlying QCD dynamics and does not drop out of the ratio \eqref{ANdef3} in the small-$x$ regime where gluons are dominant.  If the gluon production in the denominator of \eq{ANdef3} was calculated in the same quark--diquark model, as a higher-order correction, then $G$ would cancel in the ratio. However, it is not clear that this simple quark--diquark model warrants such a sophisticated calculation of a higher-order correction. Indeed, the unpolarized gluon production contribution alters the $\gamma$-dependence of $A_N$ from \eq{ANdef3} plotted in Figs.~\ref{FIG:ANKQ2D}, \ref{FIG:ANKG2D}, \ref{FIG:ANKQ3D}, and \ref{FIG:ANKG3D} only for $\gamma < 0.1$. The inclusion of gluon production in the denominator of  \eq{ANdef3} essentially serves to remove the nonphysical behavior from the unpolarized cross section, which would vanish as $\gamma \rightarrow 0$ (e.g., near mid-rapidity) if one only includes quark production. While unpolarized gluon production is important at small $\gamma$, there are many other improvements that need to be done in order to attempt to describe the data using our calculation. 

For future phenomenological applications, it will perhaps be more important to make our calculation less dependent on the specific quark-diquark model we have used here, possibly attempting to rewrite our main result \eqref{sigma_pol2} in terms of some more universal parton distributions. At the moment it is not clear how to do this. In the denominator of \eq{ANdef3} one should also find the quark and gluon production cross sections by more conventional model-independent calculations performed in the same approach, either using collinear factorization or  the small-$x$ framework , eliminating the ambiguity introduced by our use of two different models for the two cross sections.

Further limitations on this calculation can be seen from behavior of $A_N$ at the ends of the $\gamma$-range. We cannot trust our model for $\gamma \to 1$, since that is where the quark counting rules should dictate the $\gamma$-dependence. For small $\gamma$ the low-$x$ evolution between the projectile and the produced quark needs to be included. This is similar (though, perhaps, not equivalent) to determining the small-$x$ asymptotics of the Sivers TMD: first steps in that direction were made recently in \cite{Boer:2015pni,Szymanowski:2016mbq}. This small-$x$ evolution on the projectile side is likely to alter both the $k_T$ and $Q_s$ dependence of the asymmetry. Investigation of this regime, perhaps along the lines of \cite{Kovchegov:2018zeq}, are left for future work.



\section*{Acknowledgments}

YK would like to thank Matt Sievert for discussions of including the lensing mechanism into small-$x$ description of $p^{\uparrow}+p$ and $p^{\uparrow}+A$ collisions using the quark model for the projectile (polarized) proton. 

This material is based upon work supported by the U.S. Department of
Energy, Office of Science, Office of Nuclear Physics under Award
Number DE-SC0004286.



\appendix
\section{Light-cone Wave Function for the Proton $\to$ Quark+Diquark Splitting}
\label{sec:AppA}

\begin{figure}[h]
\begin{center}
\includegraphics[width= 0.4 \textwidth]{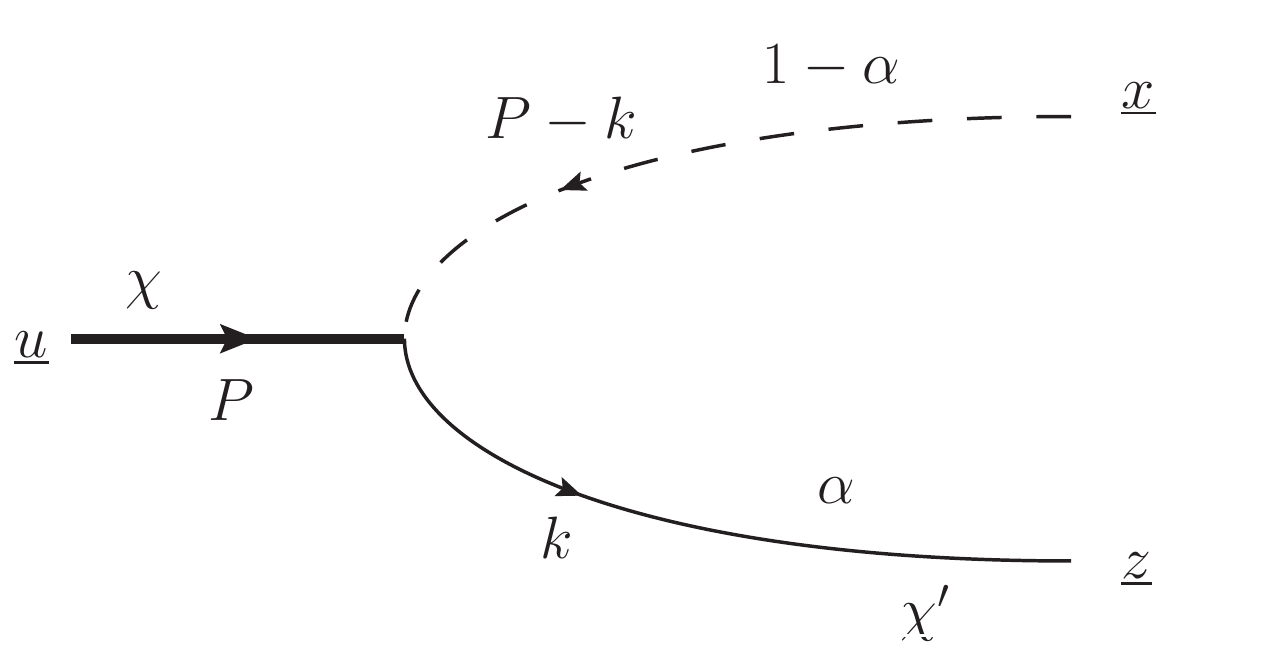} 
\caption{Light-cone wavefunction for proton splitting into a quark--diquark pair. Arrows denote the particle number flow.}
\label{FIG:InitWF}
\end{center}
\end{figure}

In this Appendix we calculate the wave function for the splitting of the proton into a quark-diquark pair given in \eq{MRWF}. The light cone wave function for the proton splitting into a quark-diquark pair is given by the diagram in \fig{FIG:InitWF}. Applying the LCPT rules \cite{Lepage:1980fj,Brodsky:1997de} we get
\begin{equation}
\label{inwf}
\psi_{\chi \chi'} (P, k; \alpha) = \frac{-G \bar u_{\chi'} (k) u_{\chi} (P)}{P^+ [P^- - k^- - (P - k)^- ]} ,
\end{equation}
with transverse spinors which are given in terms of helicity-basis Brodsky--Lepage spinors as $u_{\chi} = \frac{1}{\sqrt{2}} [u_z + \chi u_{-z}]$. (Note that our definition of the light-cone wave function is the boost-invariant definition from \cite{Kovchegov:2012mbw}.) The proton has polarization $\chi$, while the quark has polarization $\chi'$. 

Evaluating the spinor products and simplifying the energy denominator, while assuming that the quark is massless, $m=0$, yields
\begin{equation}
\psi_{\chi \chi'} (P, k; \alpha) = \frac{G \sqrt{\alpha}  (1 - \alpha )  [\delta_{\chi, \chi'} ( \alpha M_P  - i \chi (k_{\perp}^2 - \alpha P_{\perp}^2 ) )+ \delta_{\chi, - \chi'} \chi (k_{\perp}^1 - \alpha P_{\perp}^1 ) ] }{ ({\un k} - \alpha \, {\un P})^2  + \alpha M^2 - \alpha (1 - \alpha ) M_P^2 }  ,
\end{equation}
where $M_P$ is the proton mass, $M$ is the diquark mass and $\alpha = k^+/P^+$.

We want to obtain a mixed representation of the wave function with the transverse momentum components Fourier-transformed to transverse coordinate space: to do so, we perform a two dimensional Fourier transform over $k_{\perp}$ and $P_{\perp}$,  obtaining
\begin{align}\label{coord_wf}
\psi_{\chi \chi'} ({\un x}, {\un z}, {\un u}, \alpha)  & \equiv \int \frac{ \dd[2]{k_{\perp}} \dd[2]{P_{\perp}}}{(2 \pi )^4} \, e^{i {\un k} \vdot ({\un z} - {\un x}) + i {\un P} \vdot ({\un x} - {\un u})} \, \psi_{\chi \chi'} (P, k; \alpha) \\
& = \frac{G \tilde{m}_{\alpha} \sqrt{\alpha} (1 - \alpha )  }{2 \pi} \ \delta^{(2)} \left( {\un x} - {\un u} + \alpha \, {\un z} - \alpha \, {\un x} \right)	\notag \\
& \times \left[\delta_{\chi, \chi'}  K_0 (\tilde{m}_{\alpha} \abs{ {\un z} - {\un x}})  - \frac{ i \chi ( z_{\perp}^i - x_{\perp}^i ) }{ \abs{{\un z} - {\un x}}}  K_1 (\tilde{m}_{\alpha} \abs{{\un z} - {\un x}}) (i \delta_{\chi, \chi'} \delta^{i 2} - \delta_{\chi, - \chi'} \delta^{i 1} ) \right] , \notag
\end{align}
where $\tilde{m}_{\alpha}^2 =   \alpha M^2 - \alpha (1 - \alpha ) M_P^2 = \alpha^2 M_P^2$ for $M=M_P$ (cf. e.g. \cite{Meissner:2007rx}). Equation \eqref{coord_wf} is exactly \eq{MRWF} in the main text.


\section{Calculation of the Final-State Exchange Amplitude}
\label{sec:AppB}

In this Appendix we derive the final state interaction contribution to the cross section given in \eq{mrMFSI} in the main text by starting from \eq{mfsi}. First let us define the momentum-space amplitude by
\begin{align}\label{mfsi_mom}
i M_{FSI}^{\chi' \chi''} ({\un p}, {\un k}, {\un r}; \alpha, \beta) = & \ \frac{p^+ }{k^+ (p-k)^+} \frac{-\pi g^2 }{r^+ r^- - r_{\perp}^2 + i\epsilon}  \\
& \times \bar{u}_{\chi''} (k+r) [ 2(\slashed{p}-\slashed{k}) - \slashed{r} ] u_{\chi'}(k) \ \delta( (p-k-r)^- + (k+r)^- - (p-k)^- - k^-)	, \notag
\end{align}
where $\beta = r^+/P^+ = r^+/p^+$. 

Next we evaluate the spinor products in \eq{mfsi} using Brodsky--Lepage spinors \cite{Lepage:1980fj,Brodsky:1997de}. After some significant algebra we get (for massless quarks, $m=0$)
\begin{align}
& \bar{u}_{\chi''} (k+r) [ 2(\slashed{p}-\slashed{k}) - \slashed{r} ] u_{\chi'}(k) = \frac{2}{\sqrt{\alpha \, (\alpha + \beta)}} \\ & \times \left[ \frac{\delta_{\chi' \chi''}}{1-\alpha} \, ({\un k} - \alpha \, {\un p}) \cdot \left[ ({\un k} - \alpha {\un p} )  + ({\un r} - \beta {\un p} ) + (\beta {\un k} - \alpha {\un r}) \right] + i \,\delta_{\chi', - \chi''} \, ({\un r} - \beta \, {\un p}) \times ({\un k} - \alpha \, {\un p}) \right]. \notag
\end{align}
Substituting this result back into \eq{mfsi_mom}, rewriting all the minus momentum components in the argument of the delta-function in \eq{mfsi_mom} in terms of transverse and plus momentum components (e.g., $k^- = k_\perp^2 /k^+ = k_\perp^2 /(\alpha \, P^+)$), and, finally, noticing that in the gluon propagator denominator we can write $r^- = (k+r)^- - k^-$ with the momenta $(k+r)^-$ and $k^-$ also rewritten in terms of their transverse and plus components, we arrive at
\begin{align}\label{mfsi_mom2}
i M_{FSI}^{\chi' \chi''} ({\un p}, {\un k}, {\un r}; \alpha, \beta) = & \ 2 \pi \, g^2  \frac{\sqrt{\alpha \, (\alpha + \beta)}}{\alpha \, (1 - \alpha)} \, \frac{1}{(\beta {\un k} - \alpha {\un r})^2}  \, \delta \left[ \frac{ \left( ({\un k} - \alpha {\un p} )  + ({\un r} - \beta {\un p} ) \right)^2 }{(1-\alpha - \beta) \, (\alpha + \beta)} - \frac{({\un k} - \alpha {\un p} )^2}{\alpha \, (1 - \alpha)} \right] \\
& \times \left[ \frac{\delta_{\chi' \chi''}}{1-\alpha} \, ({\un k} - \alpha \, {\un p}) \cdot \left[ ({\un k} - \alpha {\un p} )  + ({\un r} - \beta {\un p} ) + (\beta {\un k} - \alpha {\un r}) \right] + i \,\delta_{\chi', - \chi''} \, ({\un r} - \beta \, {\un p}) \times ({\un k} - \alpha \, {\un p}) \right]. \notag
\end{align}

To perform the transverse Fourier transform 
\begin{align}\label{mfsi_Fourier}
i M_{FSI}^{\chi' \chi''} ({\un x}', {\un z}'; {\un x}, {\un z}; \alpha, \gamma) = & \int \frac{d^2 p_{\perp}}{(2 \pi)^2} \, \frac{d^2 k_{\perp}}{(2 \pi)^2} \, \frac{d^2 r_{\perp}}{(2 \pi)^2} \, e^{i (\un{p} - \un{k} - \un{r}) \cdot {\un x}' - i (\un{p} - \un{k})\cdot {\un x} + i ({\un k} + {\un r}) \cdot {\un z}' - i {\un k} \cdot {\un z}} \, i M_{FSI}^{\chi' \chi''} ({\un p}, {\un k}, {\un r}; \alpha, \beta) 
\end{align}
where $\gamma = \alpha + \beta = (k^+ + r^+)/P^+$, it is convenient to change the variables
\begin{equation}
\begin{cases}
\tilde{{\un k}} = {\un k} -\alpha \, {\un p} ,	\\
\tilde{{\un r}} = {\un r} - \beta \, {\un p} ,
\end{cases}
\end{equation}
which makes the amplitude independent of $\un p$, 
\begin{align}\label{mfsi_mom3}
i M_{FSI}^{\chi' \chi''} ({\un p}, \tilde{\un k}, \tilde{\un r}; \alpha, \gamma) & = \frac{2\pi g^2 \, \sqrt{\alpha\gamma}}{\alpha (1-\alpha) \left(\alpha \tilde{\un r} - (\gamma - \alpha) \tilde{\un k}\right)^2} \, \delta \left[ \frac{ ( \tilde{\un k} + \tilde{\un r} )^2 }{\gamma \, (1-\gamma)} - \frac{\tilde{\un k}^2}{\alpha \, (1 - \alpha)} \right] \\
& \times \left[ \frac{\delta_{\chi' \chi''}}{1-\alpha} \, \tilde{\un k} \cdot \left( \tilde{\un k} (1+\gamma - \alpha)  + \tilde{\un r} (1-\alpha) \right) + i \,\delta_{\chi', - \chi''} \, \tilde{\un r} \times \tilde{\un k}  \right]. \notag
\end{align}
Substituting \eq{mfsi_mom3} into \eq{mfsi_Fourier} and integrating over $\un p$, $\un k$ and $\un r$, we arrive at \eq{mrMFSI} in the main text. The following relation may be useful in performing the Fourier transforms:
\begin{align}
\int \frac{d^2 k_{\perp}}{(2 \pi)^2} \, \frac{d^2 q_{\perp}}{(2 \pi)^2} \, e^{i {\un k} \cdot {\un x} + i {\un q} \cdot {\un z} } \ \delta \left[ \frac{k_\perp^2}{\alpha \, (1-\alpha)} - \frac{q_\perp^2}{\gamma \, (1-\gamma)} \right] = \frac{1}{(2 \pi)^2} \ \delta \left[ \frac{z_\perp^2}{\alpha \, (1-\alpha)} - \frac{x_\perp^2}{\gamma \, (1-\gamma)} \right] .
\end{align}


\section{Some Useful Angular Integrals}
\label{sec:AppC}

Here is a list of useful angular integrals used in the main text. This set of integrals is done under the constraint
\begin{align}
\frac{\tilde{z}_T^2}{\alpha (1-\alpha)} = \frac{\xi_T^2}{\gamma (1-\gamma)}
\end{align} 
resulting from the delta-function in \eq{sigma_pol3}. Below $\theta_\xi$ is the angle of the vector $\un \xi$ with respect to, say, $\tilde{\un z}$. In addition, we introduced unit vectors $\hat{\xi} = {\un \xi}/\xi_T$ and $\hat{\tilde{z}} = \tilde{\un z}/\tilde{z}_T$.
\begin{subequations}\label{Iint}
\begin{align}
& I_1 \equiv \int\limits_0^{2 \pi} d \theta_\xi \, \frac{{\un \xi}  \cdot   ( \alpha {\un \xi} + (1-\gamma) \tilde{\un z} )}{   |(1-\alpha) \, {\un \xi} - (1-\gamma) \tilde{\un z}|^2} = \frac{2 \pi \min \{ \alpha, \gamma \} }{(1-\alpha) \, | \alpha - \gamma |}, \\
& I_2 \equiv \int\limits_0^{2 \pi} d \theta_\xi \, \frac{{\un \xi}  \cdot   ( \alpha {\un \xi} + (1-\gamma) \tilde{\un z} )}{   |(1-\alpha) \, {\un \xi} - (1-\gamma) \tilde{\un z}|^2} \, \hat{\xi} = \pi \frac{(\alpha + \gamma) \, (\min \{ \alpha, \gamma \} - \alpha \, \gamma)}{(1-\alpha)^{3/2} \, \sqrt{\alpha \, \gamma \, (1-\gamma)} \, | \alpha - \gamma |} \, \hat{\tilde{z}} ,  \\
& I_3 \equiv \int\limits_0^{2 \pi} d \theta_\xi \, \frac{{\un \xi}  \times \tilde{\un z}}{   |(1-\alpha) \, {\un \xi} - (1-\gamma) \tilde{\un z}|^2} = 0 , \\
& I_4 \equiv \int\limits_0^{2 \pi} d \theta_\xi \, \frac{{\un \xi}  \times \tilde{\un z}}{   |(1-\alpha) \, {\un \xi} - (1-\gamma) \tilde{\un z}|^2}  \, \hat{S} \cdot \hat{\xi} = \pi \sqrt{\frac{\alpha \, \gamma}{(1-\alpha) \, (1-\gamma)}} \, \frac{\min \{ \alpha, \gamma \} - \alpha \, \gamma}{\alpha \, \gamma \, (1-\alpha) \, (1-\gamma)} \, \hat{S} \times \hat{\tilde{z}}.
\end{align}
\end{subequations}

In the next set of integrals $i,j = 1,2$ and $\epsilon^{ij}$ is the two-dimensional Levi-Civita symbol with $\epsilon^{12} = +1$. 
\begin{subequations}\label{I_tilde_int}
\begin{align}
& \tilde{I}_1 \equiv \int\limits_0^{2 \pi} d \theta_\xi \, \frac{{\un \xi}  \cdot   ( \alpha {\un \xi} + (1-\gamma) \tilde{\un z} )}{   |(1-\alpha) \, {\un \xi} - (1-\gamma) \tilde{\un z}|^2} \, \hat{\xi} = \frac{\pi (\alpha+\gamma) \, \left[ \min \{ \alpha, \gamma \} - \alpha \, \gamma \right] }{(1-\alpha)^{3/2} \, | \alpha - \gamma | \, \sqrt{\alpha \, \gamma \, (1-\gamma)}} \, \hat{\tilde{z}}, \\
& \tilde{I}_2 \equiv \int\limits_0^{2 \pi} d \theta_\xi \, \frac{{\un \xi}  \cdot   ( \alpha {\un \xi} + (1-\gamma) \tilde{\un z} )}{   |(1-\alpha) \, {\un \xi} - (1-\gamma) \tilde{\un z}|^2} \, \hat{\xi}^i \, \hat{\xi}^j = \frac{2 \pi \min \{ \alpha, \gamma \} }{(1-\alpha) \, | \alpha - \gamma |} \, \hat{\tilde{z}}^{i} \, \hat{\tilde{z}}^{j} \\ & \hspace*{6.3cm} + \left[ \epsilon^{ik} \, \hat{\tilde{z}}^{k} \, \epsilon^{jm} \, \hat{\tilde{z}}^{m} -  \hat{\tilde{z}}^{i} \, \hat{\tilde{z}}^{j} \right] \frac{\pi}{4} \, \frac{\gamma^2 + \alpha^2 - 2 \alpha^2 \, \gamma^2 - |\alpha^2 - \gamma^2|}{\alpha \, \gamma \, (1-\alpha)^2 \, (1-\gamma)} , \notag \\
& \tilde{I}_3 \equiv \int\limits_0^{2 \pi} d \theta_\xi \, \frac{{\un \xi}  \times \tilde{\un z}}{   |(1-\alpha) \, {\un \xi} - (1-\gamma) \tilde{\un z}|^2}  \, \hat{\xi}^i  =  \frac{\pi \, \left[ \min \{ \alpha, \gamma \} - \alpha \, \gamma \right] }{(1-\alpha)^{3/2} \, (1-\gamma)^{3/2} \, \sqrt{\alpha \, \gamma }} \, \epsilon^{ij} \, \hat{\tilde{z}}^j , \\
& \tilde{I}_4 \equiv \int\limits_0^{2 \pi} d \theta_\xi \, \frac{{\un \xi}  \times \tilde{\un z}}{   |(1-\alpha) \, {\un \xi} - (1-\gamma) \tilde{\un z}|^2}  \, \hat{\xi}^i \, \hat{\xi}^j = - \left[ \hat{\tilde{z}}^{i} \, \hat{\tilde{z}}^{j} + \epsilon^{ik} \, \hat{\tilde{z}}^{k} \, \epsilon^{jm} \, \hat{\tilde{z}}^{m}  \right] \frac{\pi}{2} \frac{\alpha \, \gamma}{\left[ \max \{ \alpha, \gamma \} - \alpha \, \gamma \right]^2} .
\end{align}
\end{subequations}



\begin{thebibliography}{109}%
\makeatletter
\providecommand \@ifxundefined [1]{%
 \@ifx{#1\undefined}
}%
\providecommand \@ifnum [1]{%
 \ifnum #1\expandafter \@firstoftwo
 \else \expandafter \@secondoftwo
 \fi
}%
\providecommand \@ifx [1]{%
 \ifx #1\expandafter \@firstoftwo
 \else \expandafter \@secondoftwo
 \fi
}%
\providecommand \natexlab [1]{#1}%
\providecommand \enquote  [1]{``#1''}%
\providecommand \bibnamefont  [1]{#1}%
\providecommand \bibfnamefont [1]{#1}%
\providecommand \citenamefont [1]{#1}%
\providecommand \href@noop [0]{\@secondoftwo}%
\providecommand \href [0]{\begingroup \@sanitize@url \@href}%
\providecommand \@href[1]{\@@startlink{#1}\@@href}%
\providecommand \@@href[1]{\endgroup#1\@@endlink}%
\providecommand \@sanitize@url [0]{\catcode `\\12\catcode `\$12\catcode
  `\&12\catcode `\#12\catcode `\^12\catcode `\_12\catcode `\%12\relax}%
\providecommand \@@startlink[1]{}%
\providecommand \@@endlink[0]{}%
\providecommand \url  [0]{\begingroup\@sanitize@url \@url }%
\providecommand \@url [1]{\endgroup\@href {#1}{\urlprefix }}%
\providecommand \urlprefix  [0]{URL }%
\providecommand \Eprint [0]{\href }%
\providecommand \doibase [0]{http://dx.doi.org/}%
\providecommand \selectlanguage [0]{\@gobble}%
\providecommand \bibinfo  [0]{\@secondoftwo}%
\providecommand \bibfield  [0]{\@secondoftwo}%
\providecommand \translation [1]{[#1]}%
\providecommand \BibitemOpen [0]{}%
\providecommand \bibitemStop [0]{}%
\providecommand \bibitemNoStop [0]{.\EOS\space}%
\providecommand \EOS [0]{\spacefactor3000\relax}%
\providecommand \BibitemShut  [1]{\csname bibitem#1\endcsname}%
\let\auto@bib@innerbib\@empty
\bibitem [{\citenamefont {Brodsky}\ \emph
  {et~al.}(2002{\natexlab{a}})\citenamefont {Brodsky}, \citenamefont {Hwang},\
  and\ \citenamefont {Schmidt}}]{Brodsky:2002cx}%
  \BibitemOpen
  \bibfield  {author} {\bibinfo {author} {\bibfnamefont {S.~J.}\ \bibnamefont
  {Brodsky}}, \bibinfo {author} {\bibfnamefont {D.~S.}\ \bibnamefont {Hwang}},
  \ and\ \bibinfo {author} {\bibfnamefont {I.}~\bibnamefont {Schmidt}},\ }\href
  {\doibase 10.1016/S0370-2693(02)01320-5} {\bibfield  {journal} {\bibinfo
  {journal} {Phys.Lett.}\ }\textbf {\bibinfo {volume} {B530}},\ \bibinfo
  {pages} {99} (\bibinfo {year} {2002}{\natexlab{a}})},\ \Eprint
  {http://arxiv.org/abs/hep-ph/0201296} {arXiv:hep-ph/0201296 [hep-ph]}
  \BibitemShut {NoStop}%
\bibitem [{\citenamefont {Aidala}\ \emph {et~al.}(2019)\citenamefont {Aidala}
  \emph {et~al.}}]{Aidala:2019ctp}%
  \BibitemOpen
  \bibfield  {author} {\bibinfo {author} {\bibfnamefont {C.}~\bibnamefont
  {Aidala}} \emph {et~al.} (\bibinfo {collaboration} {PHENIX}),\ }\href
  {\doibase 10.1103/PhysRevLett.123.122001} {\bibfield  {journal} {\bibinfo
  {journal} {Phys. Rev. Lett.}\ }\textbf {\bibinfo {volume} {123}},\ \bibinfo
  {pages} {122001} (\bibinfo {year} {2019})},\ \Eprint
  {http://arxiv.org/abs/1903.07422} {arXiv:1903.07422 [hep-ex]} \BibitemShut
  {NoStop}%
\bibitem [{\citenamefont {Dilks}(2016)}]{Dilks:2016ufy}%
  \BibitemOpen
  \bibfield  {author} {\bibinfo {author} {\bibfnamefont {C.}~\bibnamefont
  {Dilks}} (\bibinfo {collaboration} {STAR}),\ }\bibfield  {booktitle} {\emph
  {\bibinfo {booktitle} {{Proceedings, 24th International Workshop on
  Deep-Inelastic Scattering and Related Subjects (DIS 2016): Hamburg, Germany,
  April 11-15, 2016}}},\ }\href {\doibase 10.22323/1.265.0212} {\bibfield
  {journal} {\bibinfo  {journal} {PoS}\ }\textbf {\bibinfo {volume}
  {DIS2016}},\ \bibinfo {pages} {212} (\bibinfo {year} {2016})},\ \Eprint
  {http://arxiv.org/abs/1805.08875} {arXiv:1805.08875 [hep-ex]} \BibitemShut
  {NoStop}%
\bibitem [{\citenamefont {Boer}\ \emph {et~al.}(2006)\citenamefont {Boer},
  \citenamefont {Dumitru},\ and\ \citenamefont {Hayashigaki}}]{Boer:2006rj}%
  \BibitemOpen
  \bibfield  {author} {\bibinfo {author} {\bibfnamefont {D.}~\bibnamefont
  {Boer}}, \bibinfo {author} {\bibfnamefont {A.}~\bibnamefont {Dumitru}}, \
  and\ \bibinfo {author} {\bibfnamefont {A.}~\bibnamefont {Hayashigaki}},\
  }\href {\doibase 10.1103/PhysRevD.74.074018} {\bibfield  {journal} {\bibinfo
  {journal} {Phys.Rev.}\ }\textbf {\bibinfo {volume} {D74}},\ \bibinfo {pages}
  {074018} (\bibinfo {year} {2006})},\ \Eprint
  {http://arxiv.org/abs/hep-ph/0609083} {arXiv:hep-ph/0609083 [hep-ph]}
  \BibitemShut {NoStop}%
\bibitem [{\citenamefont {Boer}\ and\ \citenamefont
  {Dumitru}(2003)}]{Boer:2002ij}%
  \BibitemOpen
  \bibfield  {author} {\bibinfo {author} {\bibfnamefont {D.}~\bibnamefont
  {Boer}}\ and\ \bibinfo {author} {\bibfnamefont {A.}~\bibnamefont {Dumitru}},\
  }\href {\doibase 10.1016/S0370-2693(03)00081-9} {\bibfield  {journal}
  {\bibinfo  {journal} {Phys.Lett.}\ }\textbf {\bibinfo {volume} {B556}},\
  \bibinfo {pages} {33} (\bibinfo {year} {2003})},\ \Eprint
  {http://arxiv.org/abs/hep-ph/0212260} {arXiv:hep-ph/0212260 [hep-ph]}
  \BibitemShut {NoStop}%
\bibitem [{\citenamefont {Boer}\ \emph {et~al.}(2009)\citenamefont {Boer},
  \citenamefont {Utermann},\ and\ \citenamefont {Wessels}}]{Boer:2008ze}%
  \BibitemOpen
  \bibfield  {author} {\bibinfo {author} {\bibfnamefont {D.}~\bibnamefont
  {Boer}}, \bibinfo {author} {\bibfnamefont {A.}~\bibnamefont {Utermann}}, \
  and\ \bibinfo {author} {\bibfnamefont {E.}~\bibnamefont {Wessels}},\ }\href
  {\doibase 10.1016/j.physletb.2008.12.004} {\bibfield  {journal} {\bibinfo
  {journal} {Phys.Lett.}\ }\textbf {\bibinfo {volume} {B671}},\ \bibinfo
  {pages} {91} (\bibinfo {year} {2009})},\ \Eprint
  {http://arxiv.org/abs/0811.0998} {arXiv:0811.0998 [hep-ph]} \BibitemShut
  {NoStop}%
\bibitem [{\citenamefont {Balitsky}\ and\ \citenamefont
  {Tarasov}(2015)}]{Balitsky:2015qba}%
  \BibitemOpen
  \bibfield  {author} {\bibinfo {author} {\bibfnamefont {I.}~\bibnamefont
  {Balitsky}}\ and\ \bibinfo {author} {\bibfnamefont {A.}~\bibnamefont
  {Tarasov}},\ }\href {\doibase 10.1007/JHEP10(2015)017} {\bibfield  {journal}
  {\bibinfo  {journal} {JHEP}\ }\textbf {\bibinfo {volume} {10}},\ \bibinfo
  {pages} {017} (\bibinfo {year} {2015})},\ \Eprint
  {http://arxiv.org/abs/1505.02151} {arXiv:1505.02151 [hep-ph]} \BibitemShut
  {NoStop}%
\bibitem [{\citenamefont {Balitsky}\ and\ \citenamefont
  {Tarasov}(2016)}]{Balitsky:2016dgz}%
  \BibitemOpen
  \bibfield  {author} {\bibinfo {author} {\bibfnamefont {I.}~\bibnamefont
  {Balitsky}}\ and\ \bibinfo {author} {\bibfnamefont {A.}~\bibnamefont
  {Tarasov}},\ }\href {\doibase 10.1007/JHEP06(2016)164} {\bibfield  {journal}
  {\bibinfo  {journal} {JHEP}\ }\textbf {\bibinfo {volume} {06}},\ \bibinfo
  {pages} {164} (\bibinfo {year} {2016})},\ \Eprint
  {http://arxiv.org/abs/1603.06548} {arXiv:1603.06548 [hep-ph]} \BibitemShut
  {NoStop}%
\bibitem [{\citenamefont {Altinoluk}\ \emph {et~al.}(2014)\citenamefont
  {Altinoluk}, \citenamefont {Armesto}, \citenamefont {Beuf}, \citenamefont
  {Martinez},\ and\ \citenamefont {Salgado}}]{Altinoluk:2014oxa}%
  \BibitemOpen
  \bibfield  {author} {\bibinfo {author} {\bibfnamefont {T.}~\bibnamefont
  {Altinoluk}}, \bibinfo {author} {\bibfnamefont {N.}~\bibnamefont {Armesto}},
  \bibinfo {author} {\bibfnamefont {G.}~\bibnamefont {Beuf}}, \bibinfo {author}
  {\bibfnamefont {M.}~\bibnamefont {Martinez}}, \ and\ \bibinfo {author}
  {\bibfnamefont {C.~A.}\ \bibnamefont {Salgado}},\ }\href {\doibase
  10.1007/JHEP07(2014)068} {\bibfield  {journal} {\bibinfo  {journal} {JHEP}\
  }\textbf {\bibinfo {volume} {07}},\ \bibinfo {pages} {068} (\bibinfo {year}
  {2014})},\ \Eprint {http://arxiv.org/abs/1404.2219} {arXiv:1404.2219
  [hep-ph]} \BibitemShut {NoStop}%
\bibitem [{\citenamefont {Kovchegov}\ and\ \citenamefont
  {Sievert}(2014)}]{Kovchegov:2013cva}%
  \BibitemOpen
  \bibfield  {author} {\bibinfo {author} {\bibfnamefont {Y.~V.}\ \bibnamefont
  {Kovchegov}}\ and\ \bibinfo {author} {\bibfnamefont {M.~D.}\ \bibnamefont
  {Sievert}},\ }\href {\doibase 10.1103/PhysRevD.89.054035} {\bibfield
  {journal} {\bibinfo  {journal} {Phys. Rev.}\ }\textbf {\bibinfo {volume}
  {D89}},\ \bibinfo {pages} {054035} (\bibinfo {year} {2014})},\ \Eprint
  {http://arxiv.org/abs/1310.5028} {arXiv:1310.5028 [hep-ph]} \BibitemShut
  {NoStop}%
\bibitem [{\citenamefont {Altinoluk}\ \emph {et~al.}(2016)\citenamefont
  {Altinoluk}, \citenamefont {Armesto}, \citenamefont {Beuf},\ and\
  \citenamefont {Moscoso}}]{Altinoluk:2015gia}%
  \BibitemOpen
  \bibfield  {author} {\bibinfo {author} {\bibfnamefont {T.}~\bibnamefont
  {Altinoluk}}, \bibinfo {author} {\bibfnamefont {N.}~\bibnamefont {Armesto}},
  \bibinfo {author} {\bibfnamefont {G.}~\bibnamefont {Beuf}}, \ and\ \bibinfo
  {author} {\bibfnamefont {A.}~\bibnamefont {Moscoso}},\ }\href {\doibase
  10.1007/JHEP01(2016)114} {\bibfield  {journal} {\bibinfo  {journal} {JHEP}\
  }\textbf {\bibinfo {volume} {01}},\ \bibinfo {pages} {114} (\bibinfo {year}
  {2016})},\ \Eprint {http://arxiv.org/abs/1505.01400} {arXiv:1505.01400
  [hep-ph]} \BibitemShut {NoStop}%
\bibitem [{\citenamefont {Boer}\ \emph {et~al.}(2016)\citenamefont {Boer},
  \citenamefont {Echevarria}, \citenamefont {Mulders},\ and\ \citenamefont
  {Zhou}}]{Boer:2015pni}%
  \BibitemOpen
  \bibfield  {author} {\bibinfo {author} {\bibfnamefont {D.}~\bibnamefont
  {Boer}}, \bibinfo {author} {\bibfnamefont {M.~G.}\ \bibnamefont
  {Echevarria}}, \bibinfo {author} {\bibfnamefont {P.}~\bibnamefont {Mulders}},
  \ and\ \bibinfo {author} {\bibfnamefont {J.}~\bibnamefont {Zhou}},\ }\href
  {\doibase 10.1103/PhysRevLett.116.122001} {\bibfield  {journal} {\bibinfo
  {journal} {Phys. Rev. Lett.}\ }\textbf {\bibinfo {volume} {116}},\ \bibinfo
  {pages} {122001} (\bibinfo {year} {2016})},\ \Eprint
  {http://arxiv.org/abs/1511.03485} {arXiv:1511.03485 [hep-ph]} \BibitemShut
  {NoStop}%
\bibitem [{\citenamefont {Kovchegov}\ \emph {et~al.}(2016)\citenamefont
  {Kovchegov}, \citenamefont {Pitonyak},\ and\ \citenamefont
  {Sievert}}]{Kovchegov:2015pbl}%
  \BibitemOpen
  \bibfield  {author} {\bibinfo {author} {\bibfnamefont {Y.~V.}\ \bibnamefont
  {Kovchegov}}, \bibinfo {author} {\bibfnamefont {D.}~\bibnamefont {Pitonyak}},
  \ and\ \bibinfo {author} {\bibfnamefont {M.~D.}\ \bibnamefont {Sievert}},\
  }\href {\doibase 10.1007/JHEP01(2016)072} {\bibfield  {journal} {\bibinfo
  {journal} {JHEP}\ }\textbf {\bibinfo {volume} {01}},\ \bibinfo {pages} {072}
  (\bibinfo {year} {2016})},\ \Eprint {http://arxiv.org/abs/1511.06737}
  {arXiv:1511.06737 [hep-ph]} \BibitemShut {NoStop}%
\bibitem [{\citenamefont {Hatta}\ \emph
  {et~al.}(2017{\natexlab{a}})\citenamefont {Hatta}, \citenamefont {Nakagawa},
  \citenamefont {Yuan}, \citenamefont {Zhao},\ and\ \citenamefont
  {Xiao}}]{Hatta:2016aoc}%
  \BibitemOpen
  \bibfield  {author} {\bibinfo {author} {\bibfnamefont {Y.}~\bibnamefont
  {Hatta}}, \bibinfo {author} {\bibfnamefont {Y.}~\bibnamefont {Nakagawa}},
  \bibinfo {author} {\bibfnamefont {F.}~\bibnamefont {Yuan}}, \bibinfo {author}
  {\bibfnamefont {Y.}~\bibnamefont {Zhao}}, \ and\ \bibinfo {author}
  {\bibfnamefont {B.}~\bibnamefont {Xiao}},\ }\href {\doibase
  10.1103/PhysRevD.95.114032} {\bibfield  {journal} {\bibinfo  {journal} {Phys.
  Rev.}\ }\textbf {\bibinfo {volume} {D95}},\ \bibinfo {pages} {114032}
  (\bibinfo {year} {2017}{\natexlab{a}})},\ \Eprint
  {http://arxiv.org/abs/1612.02445} {arXiv:1612.02445 [hep-ph]} \BibitemShut
  {NoStop}%
\bibitem [{\citenamefont {Dumitru}\ \emph {et~al.}(2015)\citenamefont
  {Dumitru}, \citenamefont {Lappi},\ and\ \citenamefont
  {Skokov}}]{Dumitru:2015gaa}%
  \BibitemOpen
  \bibfield  {author} {\bibinfo {author} {\bibfnamefont {A.}~\bibnamefont
  {Dumitru}}, \bibinfo {author} {\bibfnamefont {T.}~\bibnamefont {Lappi}}, \
  and\ \bibinfo {author} {\bibfnamefont {V.}~\bibnamefont {Skokov}},\ }\href
  {\doibase 10.1103/PhysRevLett.115.252301} {\bibfield  {journal} {\bibinfo
  {journal} {Phys. Rev. Lett.}\ }\textbf {\bibinfo {volume} {115}},\ \bibinfo
  {pages} {252301} (\bibinfo {year} {2015})},\ \Eprint
  {http://arxiv.org/abs/1508.04438} {arXiv:1508.04438 [hep-ph]} \BibitemShut
  {NoStop}%
\bibitem [{\citenamefont {Chirilli}(2019)}]{Chirilli:2018kkw}%
  \BibitemOpen
  \bibfield  {author} {\bibinfo {author} {\bibfnamefont {G.~A.}\ \bibnamefont
  {Chirilli}},\ }\href {\doibase 10.1007/JHEP01(2019)118} {\bibfield  {journal}
  {\bibinfo  {journal} {JHEP}\ }\textbf {\bibinfo {volume} {01}},\ \bibinfo
  {pages} {118} (\bibinfo {year} {2019})},\ \Eprint
  {http://arxiv.org/abs/1807.11435} {arXiv:1807.11435 [hep-ph]} \BibitemShut
  {NoStop}%
\bibitem [{\citenamefont
  {Jalilian-Marian}(2019{\natexlab{a}})}]{Jalilian-Marian:2018iui}%
  \BibitemOpen
  \bibfield  {author} {\bibinfo {author} {\bibfnamefont {J.}~\bibnamefont
  {Jalilian-Marian}},\ }\href {\doibase 10.1103/PhysRevD.99.014043} {\bibfield
  {journal} {\bibinfo  {journal} {Phys. Rev.}\ }\textbf {\bibinfo {volume}
  {D99}},\ \bibinfo {pages} {014043} (\bibinfo {year} {2019}{\natexlab{a}})},\
  \Eprint {http://arxiv.org/abs/1809.04625} {arXiv:1809.04625 [hep-ph]}
  \BibitemShut {NoStop}%
\bibitem [{\citenamefont {Kovchegov}(2019)}]{Kovchegov:2019rrz}%
  \BibitemOpen
  \bibfield  {author} {\bibinfo {author} {\bibfnamefont {Y.~V.}\ \bibnamefont
  {Kovchegov}},\ }\href {\doibase 10.1007/JHEP03(2019)174} {\bibfield
  {journal} {\bibinfo  {journal} {JHEP}\ }\textbf {\bibinfo {volume} {03}},\
  \bibinfo {pages} {174} (\bibinfo {year} {2019})},\ \Eprint
  {http://arxiv.org/abs/1901.07453} {arXiv:1901.07453 [hep-ph]} \BibitemShut
  {NoStop}%
\bibitem [{\citenamefont {Boussarie}\ \emph {et~al.}(2019)\citenamefont
  {Boussarie}, \citenamefont {Hatta},\ and\ \citenamefont
  {Yuan}}]{Boussarie:2019icw}%
  \BibitemOpen
  \bibfield  {author} {\bibinfo {author} {\bibfnamefont {R.}~\bibnamefont
  {Boussarie}}, \bibinfo {author} {\bibfnamefont {Y.}~\bibnamefont {Hatta}}, \
  and\ \bibinfo {author} {\bibfnamefont {F.}~\bibnamefont {Yuan}},\ }\href
  {\doibase 10.1016/j.physletb.2019.134817} {\bibfield  {journal} {\bibinfo
  {journal} {Phys. Lett.}\ }\textbf {\bibinfo {volume} {B797}},\ \bibinfo
  {pages} {134817} (\bibinfo {year} {2019})},\ \Eprint
  {http://arxiv.org/abs/1904.02693} {arXiv:1904.02693 [hep-ph]} \BibitemShut
  {NoStop}%
\bibitem [{\citenamefont
  {Jalilian-Marian}(2019{\natexlab{b}})}]{Jalilian-Marian:2019kaf}%
  \BibitemOpen
  \bibfield  {author} {\bibinfo {author} {\bibfnamefont {J.}~\bibnamefont
  {Jalilian-Marian}},\ }\href@noop {} {\  (\bibinfo {year}
  {2019}{\natexlab{b}})},\ \Eprint {http://arxiv.org/abs/1912.08878}
  {arXiv:1912.08878 [hep-ph]} \BibitemShut {NoStop}%
\bibitem [{\citenamefont {Kovchegov}\ \emph
  {et~al.}(2017{\natexlab{a}})\citenamefont {Kovchegov}, \citenamefont
  {Pitonyak},\ and\ \citenamefont {Sievert}}]{Kovchegov:2016zex}%
  \BibitemOpen
  \bibfield  {author} {\bibinfo {author} {\bibfnamefont {Y.~V.}\ \bibnamefont
  {Kovchegov}}, \bibinfo {author} {\bibfnamefont {D.}~\bibnamefont {Pitonyak}},
  \ and\ \bibinfo {author} {\bibfnamefont {M.~D.}\ \bibnamefont {Sievert}},\
  }\href {\doibase 10.1103/PhysRevD.95.014033} {\bibfield  {journal} {\bibinfo
  {journal} {Phys. Rev.}\ }\textbf {\bibinfo {volume} {D95}},\ \bibinfo {pages}
  {014033} (\bibinfo {year} {2017}{\natexlab{a}})},\ \Eprint
  {http://arxiv.org/abs/1610.06197} {arXiv:1610.06197 [hep-ph]} \BibitemShut
  {NoStop}%
\bibitem [{\citenamefont {Kovchegov}\ \emph
  {et~al.}(2017{\natexlab{b}})\citenamefont {Kovchegov}, \citenamefont
  {Pitonyak},\ and\ \citenamefont {Sievert}}]{Kovchegov:2016weo}%
  \BibitemOpen
  \bibfield  {author} {\bibinfo {author} {\bibfnamefont {Y.~V.}\ \bibnamefont
  {Kovchegov}}, \bibinfo {author} {\bibfnamefont {D.}~\bibnamefont {Pitonyak}},
  \ and\ \bibinfo {author} {\bibfnamefont {M.~D.}\ \bibnamefont {Sievert}},\
  }\href {\doibase 10.1103/PhysRevLett.118.052001} {\bibfield  {journal}
  {\bibinfo  {journal} {Phys. Rev. Lett.}\ }\textbf {\bibinfo {volume} {118}},\
  \bibinfo {pages} {052001} (\bibinfo {year} {2017}{\natexlab{b}})},\ \Eprint
  {http://arxiv.org/abs/1610.06188} {arXiv:1610.06188 [hep-ph]} \BibitemShut
  {NoStop}%
\bibitem [{\citenamefont {Kovchegov}\ \emph
  {et~al.}(2017{\natexlab{c}})\citenamefont {Kovchegov}, \citenamefont
  {Pitonyak},\ and\ \citenamefont {Sievert}}]{Kovchegov:2017jxc}%
  \BibitemOpen
  \bibfield  {author} {\bibinfo {author} {\bibfnamefont {Y.~V.}\ \bibnamefont
  {Kovchegov}}, \bibinfo {author} {\bibfnamefont {D.}~\bibnamefont {Pitonyak}},
  \ and\ \bibinfo {author} {\bibfnamefont {M.~D.}\ \bibnamefont {Sievert}},\
  }\href {\doibase 10.1016/j.physletb.2017.06.032} {\bibfield  {journal}
  {\bibinfo  {journal} {Phys. Lett.}\ }\textbf {\bibinfo {volume} {B772}},\
  \bibinfo {pages} {136} (\bibinfo {year} {2017}{\natexlab{c}})},\ \Eprint
  {http://arxiv.org/abs/1703.05809} {arXiv:1703.05809 [hep-ph]} \BibitemShut
  {NoStop}%
\bibitem [{\citenamefont {Kovchegov}\ \emph
  {et~al.}(2017{\natexlab{d}})\citenamefont {Kovchegov}, \citenamefont
  {Pitonyak},\ and\ \citenamefont {Sievert}}]{Kovchegov:2017lsr}%
  \BibitemOpen
  \bibfield  {author} {\bibinfo {author} {\bibfnamefont {Y.~V.}\ \bibnamefont
  {Kovchegov}}, \bibinfo {author} {\bibfnamefont {D.}~\bibnamefont {Pitonyak}},
  \ and\ \bibinfo {author} {\bibfnamefont {M.~D.}\ \bibnamefont {Sievert}},\
  }\href {\doibase 10.1007/JHEP10(2017)198} {\bibfield  {journal} {\bibinfo
  {journal} {JHEP}\ }\textbf {\bibinfo {volume} {10}},\ \bibinfo {pages} {198}
  (\bibinfo {year} {2017}{\natexlab{d}})},\ \Eprint
  {http://arxiv.org/abs/1706.04236} {arXiv:1706.04236 [nucl-th]} \BibitemShut
  {NoStop}%
\bibitem [{\citenamefont {Kovchegov}\ and\ \citenamefont
  {Sievert}(2019{\natexlab{a}})}]{Kovchegov:2018znm}%
  \BibitemOpen
  \bibfield  {author} {\bibinfo {author} {\bibfnamefont {Y.~V.}\ \bibnamefont
  {Kovchegov}}\ and\ \bibinfo {author} {\bibfnamefont {M.~D.}\ \bibnamefont
  {Sievert}},\ }\href {\doibase 10.1103/PhysRevD.99.054032} {\bibfield
  {journal} {\bibinfo  {journal} {Phys. Rev.}\ }\textbf {\bibinfo {volume}
  {D99}},\ \bibinfo {pages} {054032} (\bibinfo {year} {2019}{\natexlab{a}})},\
  \Eprint {http://arxiv.org/abs/1808.09010} {arXiv:1808.09010 [hep-ph]}
  \BibitemShut {NoStop}%
\bibitem [{\citenamefont {Cougoulic}\ and\ \citenamefont
  {Kovchegov}(2019)}]{Cougoulic:2019aja}%
  \BibitemOpen
  \bibfield  {author} {\bibinfo {author} {\bibfnamefont {F.}~\bibnamefont
  {Cougoulic}}\ and\ \bibinfo {author} {\bibfnamefont {Y.~V.}\ \bibnamefont
  {Kovchegov}},\ }\href {\doibase 10.1103/PhysRevD.100.114020} {\bibfield
  {journal} {\bibinfo  {journal} {Phys. Rev.}\ }\textbf {\bibinfo {volume}
  {D100}},\ \bibinfo {pages} {114020} (\bibinfo {year} {2019})},\ \Eprint
  {http://arxiv.org/abs/1910.04268} {arXiv:1910.04268 [hep-ph]} \BibitemShut
  {NoStop}%
\bibitem [{\citenamefont {Kovchegov}\ and\ \citenamefont
  {Sievert}(2012)}]{Kovchegov:2012ga}%
  \BibitemOpen
  \bibfield  {author} {\bibinfo {author} {\bibfnamefont {Y.~V.}\ \bibnamefont
  {Kovchegov}}\ and\ \bibinfo {author} {\bibfnamefont {M.~D.}\ \bibnamefont
  {Sievert}},\ }\href {\doibase 10.1103/PhysRevD.86.034028} {\bibfield
  {journal} {\bibinfo  {journal} {Phys.Rev.}\ }\textbf {\bibinfo {volume}
  {D86}},\ \bibinfo {pages} {034028} (\bibinfo {year} {2012})},\ \Eprint
  {http://arxiv.org/abs/1201.5890} {arXiv:1201.5890 [hep-ph]} \BibitemShut
  {NoStop}%
\bibitem [{\citenamefont {Zhou}(2014)}]{Zhou:2013gsa}%
  \BibitemOpen
  \bibfield  {author} {\bibinfo {author} {\bibfnamefont {J.}~\bibnamefont
  {Zhou}},\ }\href {\doibase 10.1103/PhysRevD.89.074050} {\bibfield  {journal}
  {\bibinfo  {journal} {Phys.Rev.}\ }\textbf {\bibinfo {volume} {D89}},\
  \bibinfo {pages} {074050} (\bibinfo {year} {2014})},\ \Eprint
  {http://arxiv.org/abs/1308.5912} {arXiv:1308.5912 [hep-ph]} \BibitemShut
  {NoStop}%
\bibitem [{\citenamefont {Kovchegov}\ and\ \citenamefont
  {Sievert}(2016)}]{Kovchegov:2015zha}%
  \BibitemOpen
  \bibfield  {author} {\bibinfo {author} {\bibfnamefont {Y.~V.}\ \bibnamefont
  {Kovchegov}}\ and\ \bibinfo {author} {\bibfnamefont {M.~D.}\ \bibnamefont
  {Sievert}},\ }\href {\doibase 10.1016/j.nuclphysb.2015.12.008} {\bibfield
  {journal} {\bibinfo  {journal} {Nucl. Phys.}\ }\textbf {\bibinfo {volume}
  {B903}},\ \bibinfo {pages} {164} (\bibinfo {year} {2016})},\ \Eprint
  {http://arxiv.org/abs/1505.01176} {arXiv:1505.01176 [hep-ph]} \BibitemShut
  {NoStop}%
\bibitem [{\citenamefont {Hatta}\ \emph {et~al.}(2016)\citenamefont {Hatta},
  \citenamefont {Xiao}, \citenamefont {Yoshida},\ and\ \citenamefont
  {Yuan}}]{Hatta2016a}%
  \BibitemOpen
  \bibfield  {author} {\bibinfo {author} {\bibfnamefont {Y.}~\bibnamefont
  {Hatta}}, \bibinfo {author} {\bibfnamefont {B.-W.}\ \bibnamefont {Xiao}},
  \bibinfo {author} {\bibfnamefont {S.}~\bibnamefont {Yoshida}}, \ and\
  \bibinfo {author} {\bibfnamefont {F.}~\bibnamefont {Yuan}},\ }\href {\doibase
  10.1103/PhysRevD.94.054013} {\bibfield  {journal} {\bibinfo  {journal} {Phys.
  Rev.}\ }\textbf {\bibinfo {volume} {D94}},\ \bibinfo {pages} {054013}
  (\bibinfo {year} {2016})},\ \Eprint {http://arxiv.org/abs/1606.08640}
  {arXiv:1606.08640 [hep-ph]} \BibitemShut {NoStop}%
\bibitem [{\citenamefont {Hatta}\ \emph
  {et~al.}(2017{\natexlab{b}})\citenamefont {Hatta}, \citenamefont {Xiao},
  \citenamefont {Yoshida},\ and\ \citenamefont {Yuan}}]{Hatta:2016khv}%
  \BibitemOpen
  \bibfield  {author} {\bibinfo {author} {\bibfnamefont {Y.}~\bibnamefont
  {Hatta}}, \bibinfo {author} {\bibfnamefont {B.-W.}\ \bibnamefont {Xiao}},
  \bibinfo {author} {\bibfnamefont {S.}~\bibnamefont {Yoshida}}, \ and\
  \bibinfo {author} {\bibfnamefont {F.}~\bibnamefont {Yuan}},\ }\href {\doibase
  10.1103/PhysRevD.95.014008} {\bibfield  {journal} {\bibinfo  {journal} {Phys.
  Rev.}\ }\textbf {\bibinfo {volume} {D95}},\ \bibinfo {pages} {014008}
  (\bibinfo {year} {2017}{\natexlab{b}})},\ \Eprint
  {http://arxiv.org/abs/1611.04746} {arXiv:1611.04746 [hep-ph]} \BibitemShut
  {NoStop}%
\bibitem [{\citenamefont {Boer}(2017)}]{Boer:2016bfj}%
  \BibitemOpen
  \bibfield  {author} {\bibinfo {author} {\bibfnamefont {D.}~\bibnamefont
  {Boer}},\ }\bibfield  {booktitle} {\emph {\bibinfo {booktitle} {{Proceedings,
  New Observables in Quarkonium Production: Trento, Italy, February 28-March 4,
  2016}}},\ }\href {\doibase 10.1007/s00601-016-1198-6} {\bibfield  {journal}
  {\bibinfo  {journal} {Few Body Syst.}\ }\textbf {\bibinfo {volume} {58}},\
  \bibinfo {pages} {32} (\bibinfo {year} {2017})},\ \Eprint
  {http://arxiv.org/abs/1611.06089} {arXiv:1611.06089 [hep-ph]} \BibitemShut
  {NoStop}%
\bibitem [{\citenamefont {Kovchegov}\ and\ \citenamefont
  {Sievert}(2019{\natexlab{b}})}]{Kovchegov:2018zeq}%
  \BibitemOpen
  \bibfield  {author} {\bibinfo {author} {\bibfnamefont {Y.~V.}\ \bibnamefont
  {Kovchegov}}\ and\ \bibinfo {author} {\bibfnamefont {M.~D.}\ \bibnamefont
  {Sievert}},\ }\href {\doibase 10.1103/PhysRevD.99.054033} {\bibfield
  {journal} {\bibinfo  {journal} {Phys. Rev.}\ }\textbf {\bibinfo {volume}
  {D99}},\ \bibinfo {pages} {054033} (\bibinfo {year} {2019}{\natexlab{b}})},\
  \Eprint {http://arxiv.org/abs/1808.10354} {arXiv:1808.10354 [hep-ph]}
  \BibitemShut {NoStop}%
\bibitem [{\citenamefont {Adams}\ \emph
  {et~al.}(1991{\natexlab{a}})\citenamefont {Adams} \emph
  {et~al.}}]{Adams:1991rw}%
  \BibitemOpen
  \bibfield  {author} {\bibinfo {author} {\bibfnamefont {D.}~\bibnamefont
  {Adams}} \emph {et~al.} (\bibinfo {collaboration} {E581, E704}),\ }\href
  {\doibase 10.1016/0370-2693(91)91351-U} {\bibfield  {journal} {\bibinfo
  {journal} {Phys.Lett.}\ }\textbf {\bibinfo {volume} {B261}},\ \bibinfo
  {pages} {201} (\bibinfo {year} {1991}{\natexlab{a}})}\BibitemShut {NoStop}%
\bibitem [{\citenamefont {Adams}\ \emph
  {et~al.}(1991{\natexlab{b}})\citenamefont {Adams} \emph
  {et~al.}}]{Adams:1991cs}%
  \BibitemOpen
  \bibfield  {author} {\bibinfo {author} {\bibfnamefont {D.}~\bibnamefont
  {Adams}} \emph {et~al.} (\bibinfo {collaboration} {FNAL-E704}),\ }\href
  {\doibase 10.1016/0370-2693(91)90378-4} {\bibfield  {journal} {\bibinfo
  {journal} {Phys.Lett.}\ }\textbf {\bibinfo {volume} {B264}},\ \bibinfo
  {pages} {462} (\bibinfo {year} {1991}{\natexlab{b}})}\BibitemShut {NoStop}%
\bibitem [{\citenamefont {Abelev}\ \emph {et~al.}(2008)\citenamefont {Abelev}
  \emph {et~al.}}]{Abelev:2008af}%
  \BibitemOpen
  \bibfield  {author} {\bibinfo {author} {\bibfnamefont {B.~I.}\ \bibnamefont
  {Abelev}} \emph {et~al.} (\bibinfo {collaboration} {STAR}),\ }\href {\doibase
  10.1103/PhysRevLett.101.222001} {\bibfield  {journal} {\bibinfo  {journal}
  {Phys. Rev. Lett.}\ }\textbf {\bibinfo {volume} {101}},\ \bibinfo {pages}
  {222001} (\bibinfo {year} {2008})},\ \Eprint {http://arxiv.org/abs/0801.2990}
  {arXiv:0801.2990 [hep-ex]} \BibitemShut {NoStop}%
\bibitem [{\citenamefont {Adler}\ \emph {et~al.}(2005)\citenamefont {Adler}
  \emph {et~al.}}]{Adler:2005in}%
  \BibitemOpen
  \bibfield  {author} {\bibinfo {author} {\bibfnamefont {S.}~\bibnamefont
  {Adler}} \emph {et~al.} (\bibinfo {collaboration} {PHENIX}),\ }\href
  {\doibase 10.1103/PhysRevLett.95.202001} {\bibfield  {journal} {\bibinfo
  {journal} {Phys.Rev.Lett.}\ }\textbf {\bibinfo {volume} {95}},\ \bibinfo
  {pages} {202001} (\bibinfo {year} {2005})},\ \Eprint
  {http://arxiv.org/abs/hep-ex/0507073} {arXiv:hep-ex/0507073 [hep-ex]}
  \BibitemShut {NoStop}%
\bibitem [{\citenamefont {Kane}\ \emph {et~al.}(1978)\citenamefont {Kane},
  \citenamefont {Pumplin},\ and\ \citenamefont {Repko}}]{Kane:1978nd}%
  \BibitemOpen
  \bibfield  {author} {\bibinfo {author} {\bibfnamefont {G.~L.}\ \bibnamefont
  {Kane}}, \bibinfo {author} {\bibfnamefont {J.}~\bibnamefont {Pumplin}}, \
  and\ \bibinfo {author} {\bibfnamefont {W.}~\bibnamefont {Repko}},\ }\href
  {\doibase 10.1103/PhysRevLett.41.1689} {\bibfield  {journal} {\bibinfo
  {journal} {Phys.Rev.Lett.}\ }\textbf {\bibinfo {volume} {41}},\ \bibinfo
  {pages} {1689} (\bibinfo {year} {1978})}\BibitemShut {NoStop}%
\bibitem [{\citenamefont {Heppelmann}(2013)}]{Heppelmann:2013DIS}%
  \BibitemOpen
  \bibfield  {author} {\bibinfo {author} {\bibfnamefont {S.}~\bibnamefont
  {Heppelmann}} (\bibinfo {collaboration} {STAR}),\ }\bibfield  {booktitle}
  {\emph {\bibinfo {booktitle} {{Proceedings, 21st International Workshop on
  Deep-Inelastic Scattering and Related Subjects (DIS 2013): Marseilles,
  France, April 22-26, 2013}}},\ }\href {\doibase 10.22323/1.191.0240}
  {\bibfield  {journal} {\bibinfo  {journal} {PoS}\ }\textbf {\bibinfo {volume}
  {DIS2013}} (\bibinfo {year} {2013}),\ 10.22323/1.191.0240}\BibitemShut
  {NoStop}%
\bibitem [{\citenamefont {Aschenauer}\ \emph {et~al.}(2013)\citenamefont
  {Aschenauer} \emph {et~al.}}]{Aschenauer:2013woa}%
  \BibitemOpen
  \bibfield  {author} {\bibinfo {author} {\bibfnamefont {E.~C.}\ \bibnamefont
  {Aschenauer}} \emph {et~al.},\ }\href@noop {} {\  (\bibinfo {year} {2013})},\
  \Eprint {http://arxiv.org/abs/1304.0079} {arXiv:1304.0079 [nucl-ex]}
  \BibitemShut {NoStop}%
\bibitem [{\citenamefont {D'Alesio}\ and\ \citenamefont
  {Murgia}(2008)}]{DAlesio:2007bjf}%
  \BibitemOpen
  \bibfield  {author} {\bibinfo {author} {\bibfnamefont {U.}~\bibnamefont
  {D'Alesio}}\ and\ \bibinfo {author} {\bibfnamefont {F.}~\bibnamefont
  {Murgia}},\ }\href {\doibase 10.1016/j.ppnp.2008.01.001} {\bibfield
  {journal} {\bibinfo  {journal} {Prog. Part. Nucl. Phys.}\ }\textbf {\bibinfo
  {volume} {61}},\ \bibinfo {pages} {394} (\bibinfo {year} {2008})},\ \Eprint
  {http://arxiv.org/abs/0712.4328} {arXiv:0712.4328 [hep-ph]} \BibitemShut
  {NoStop}%
\bibitem [{\citenamefont {Sivers}(1990)}]{Sivers:1989cc}%
  \BibitemOpen
  \bibfield  {author} {\bibinfo {author} {\bibfnamefont {D.~W.}\ \bibnamefont
  {Sivers}},\ }\href {\doibase 10.1103/PhysRevD.41.83} {\bibfield  {journal}
  {\bibinfo  {journal} {Phys.Rev.}\ }\textbf {\bibinfo {volume} {D41}},\
  \bibinfo {pages} {83} (\bibinfo {year} {1990})}\BibitemShut {NoStop}%
\bibitem [{\citenamefont {Sivers}(1991)}]{Sivers:1990fh}%
  \BibitemOpen
  \bibfield  {author} {\bibinfo {author} {\bibfnamefont {D.~W.}\ \bibnamefont
  {Sivers}},\ }\href {\doibase 10.1103/PhysRevD.43.261} {\bibfield  {journal}
  {\bibinfo  {journal} {Phys.Rev.}\ }\textbf {\bibinfo {volume} {D43}},\
  \bibinfo {pages} {261} (\bibinfo {year} {1991})}\BibitemShut {NoStop}%
\bibitem [{\citenamefont {Burkardt}(2004)}]{Burkardt:2003uw}%
  \BibitemOpen
  \bibfield  {author} {\bibinfo {author} {\bibfnamefont {M.}~\bibnamefont
  {Burkardt}},\ }\href {\doibase 10.1016/j.nuclphysa.2004.02.008} {\bibfield
  {journal} {\bibinfo  {journal} {Nucl. Phys.}\ }\textbf {\bibinfo {volume}
  {A735}},\ \bibinfo {pages} {185} (\bibinfo {year} {2004})},\ \Eprint
  {http://arxiv.org/abs/hep-ph/0302144} {arXiv:hep-ph/0302144 [hep-ph]}
  \BibitemShut {NoStop}%
\bibitem [{\citenamefont {Collins}(2002)}]{Collins:2002kn}%
  \BibitemOpen
  \bibfield  {author} {\bibinfo {author} {\bibfnamefont {J.~C.}\ \bibnamefont
  {Collins}},\ }\href {\doibase 10.1016/S0370-2693(02)01819-1} {\bibfield
  {journal} {\bibinfo  {journal} {Phys.Lett.}\ }\textbf {\bibinfo {volume}
  {B536}},\ \bibinfo {pages} {43} (\bibinfo {year} {2002})},\ \Eprint
  {http://arxiv.org/abs/hep-ph/0204004} {arXiv:hep-ph/0204004 [hep-ph]}
  \BibitemShut {NoStop}%
\bibitem [{\citenamefont {Brodsky}\ \emph
  {et~al.}(2002{\natexlab{b}})\citenamefont {Brodsky}, \citenamefont {Hwang},\
  and\ \citenamefont {Schmidt}}]{Brodsky:2002rv}%
  \BibitemOpen
  \bibfield  {author} {\bibinfo {author} {\bibfnamefont {S.~J.}\ \bibnamefont
  {Brodsky}}, \bibinfo {author} {\bibfnamefont {D.~S.}\ \bibnamefont {Hwang}},
  \ and\ \bibinfo {author} {\bibfnamefont {I.}~\bibnamefont {Schmidt}},\ }\href
  {\doibase 10.1016/S0550-3213(02)00617-X} {\bibfield  {journal} {\bibinfo
  {journal} {Nucl.Phys.}\ }\textbf {\bibinfo {volume} {B642}},\ \bibinfo
  {pages} {344} (\bibinfo {year} {2002}{\natexlab{b}})},\ \Eprint
  {http://arxiv.org/abs/hep-ph/0206259} {arXiv:hep-ph/0206259 [hep-ph]}
  \BibitemShut {NoStop}%
\bibitem [{\citenamefont {Brodsky}\ \emph {et~al.}(2013)\citenamefont
  {Brodsky}, \citenamefont {Hwang}, \citenamefont {Kovchegov}, \citenamefont
  {Schmidt},\ and\ \citenamefont {Sievert}}]{Brodsky:2013oya}%
  \BibitemOpen
  \bibfield  {author} {\bibinfo {author} {\bibfnamefont {S.~J.}\ \bibnamefont
  {Brodsky}}, \bibinfo {author} {\bibfnamefont {D.~S.}\ \bibnamefont {Hwang}},
  \bibinfo {author} {\bibfnamefont {Y.~V.}\ \bibnamefont {Kovchegov}}, \bibinfo
  {author} {\bibfnamefont {I.}~\bibnamefont {Schmidt}}, \ and\ \bibinfo
  {author} {\bibfnamefont {M.~D.}\ \bibnamefont {Sievert}},\ }\href {\doibase
  10.1103/PhysRevD.88.014032} {\bibfield  {journal} {\bibinfo  {journal}
  {Phys.Rev.}\ }\textbf {\bibinfo {volume} {D88}},\ \bibinfo {pages} {014032}
  (\bibinfo {year} {2013})},\ \Eprint {http://arxiv.org/abs/1304.5237}
  {arXiv:1304.5237 [hep-ph]} \BibitemShut {NoStop}%
\bibitem [{\citenamefont {Collins}(1993)}]{Collins:1992kk}%
  \BibitemOpen
  \bibfield  {author} {\bibinfo {author} {\bibfnamefont {J.~C.}\ \bibnamefont
  {Collins}},\ }\href {\doibase 10.1016/0550-3213(93)90262-N} {\bibfield
  {journal} {\bibinfo  {journal} {Nucl.Phys.}\ }\textbf {\bibinfo {volume}
  {B396}},\ \bibinfo {pages} {161} (\bibinfo {year} {1993})},\ \Eprint
  {http://arxiv.org/abs/hep-ph/9208213} {arXiv:hep-ph/9208213 [hep-ph]}
  \BibitemShut {NoStop}%
\bibitem [{\citenamefont {Efremov}\ and\ \citenamefont
  {Teryaev}(1982)}]{Efremov:1981sh}%
  \BibitemOpen
  \bibfield  {author} {\bibinfo {author} {\bibfnamefont {A.}~\bibnamefont
  {Efremov}}\ and\ \bibinfo {author} {\bibfnamefont {O.}~\bibnamefont
  {Teryaev}},\ }\href@noop {} {\bibfield  {journal} {\bibinfo  {journal}
  {Sov.J.Nucl.Phys.}\ }\textbf {\bibinfo {volume} {36}},\ \bibinfo {pages}
  {140} (\bibinfo {year} {1982})}\BibitemShut {NoStop}%
\bibitem [{\citenamefont {Efremov}\ and\ \citenamefont
  {Teryaev}(1985)}]{Efremov:1984ip}%
  \BibitemOpen
  \bibfield  {author} {\bibinfo {author} {\bibfnamefont {A.}~\bibnamefont
  {Efremov}}\ and\ \bibinfo {author} {\bibfnamefont {O.}~\bibnamefont
  {Teryaev}},\ }\href {\doibase 10.1016/0370-2693(85)90999-2} {\bibfield
  {journal} {\bibinfo  {journal} {Phys.Lett.}\ }\textbf {\bibinfo {volume}
  {B150}},\ \bibinfo {pages} {383} (\bibinfo {year} {1985})}\BibitemShut
  {NoStop}%
\bibitem [{\citenamefont {Qiu}\ and\ \citenamefont
  {Sterman}(1991)}]{Qiu:1991pp}%
  \BibitemOpen
  \bibfield  {author} {\bibinfo {author} {\bibfnamefont {J.-w.}\ \bibnamefont
  {Qiu}}\ and\ \bibinfo {author} {\bibfnamefont {G.~F.}\ \bibnamefont
  {Sterman}},\ }\href {\doibase 10.1103/PhysRevLett.67.2264} {\bibfield
  {journal} {\bibinfo  {journal} {Phys.Rev.Lett.}\ }\textbf {\bibinfo {volume}
  {67}},\ \bibinfo {pages} {2264} (\bibinfo {year} {1991})}\BibitemShut
  {NoStop}%
\bibitem [{\citenamefont {Qiu}\ and\ \citenamefont
  {Sterman}(1998)}]{Qiu:1998ia}%
  \BibitemOpen
  \bibfield  {author} {\bibinfo {author} {\bibfnamefont {J.-w.}\ \bibnamefont
  {Qiu}}\ and\ \bibinfo {author} {\bibfnamefont {G.~F.}\ \bibnamefont
  {Sterman}},\ }\href {\doibase 10.1103/PhysRevD.59.014004} {\bibfield
  {journal} {\bibinfo  {journal} {Phys.Rev.}\ }\textbf {\bibinfo {volume}
  {D59}},\ \bibinfo {pages} {014004} (\bibinfo {year} {1998})},\ \Eprint
  {http://arxiv.org/abs/hep-ph/9806356} {arXiv:hep-ph/9806356 [hep-ph]}
  \BibitemShut {NoStop}%
\bibitem [{\citenamefont {Kanazawa}\ \emph {et~al.}(2014)\citenamefont
  {Kanazawa}, \citenamefont {Koike}, \citenamefont {Metz},\ and\ \citenamefont
  {Pitonyak}}]{Kanazawa:2014dca}%
  \BibitemOpen
  \bibfield  {author} {\bibinfo {author} {\bibfnamefont {K.}~\bibnamefont
  {Kanazawa}}, \bibinfo {author} {\bibfnamefont {Y.}~\bibnamefont {Koike}},
  \bibinfo {author} {\bibfnamefont {A.}~\bibnamefont {Metz}}, \ and\ \bibinfo
  {author} {\bibfnamefont {D.}~\bibnamefont {Pitonyak}},\ }\href {\doibase
  10.1103/PhysRevD.89.111501} {\bibfield  {journal} {\bibinfo  {journal} {Phys.
  Rev.}\ }\textbf {\bibinfo {volume} {D89}},\ \bibinfo {pages} {111501}
  (\bibinfo {year} {2014})},\ \Eprint {http://arxiv.org/abs/1404.1033}
  {arXiv:1404.1033 [hep-ph]} \BibitemShut {NoStop}%
\bibitem [{\citenamefont {Metz}\ and\ \citenamefont
  {Pitonyak}(2013)}]{Metz:2012ct}%
  \BibitemOpen
  \bibfield  {author} {\bibinfo {author} {\bibfnamefont {A.}~\bibnamefont
  {Metz}}\ and\ \bibinfo {author} {\bibfnamefont {D.}~\bibnamefont
  {Pitonyak}},\ }\href {\doibase 10.1016/j.physletb.2013.05.043,
  10.1016/j.physletb.2016.10.011} {\bibfield  {journal} {\bibinfo  {journal}
  {Phys. Lett.}\ }\textbf {\bibinfo {volume} {B723}},\ \bibinfo {pages} {365}
  (\bibinfo {year} {2013})},\ \bibinfo {note} {[Erratum: Phys.
  Lett.B762,549(2016)]},\ \Eprint {http://arxiv.org/abs/1212.5037}
  {arXiv:1212.5037 [hep-ph]} \BibitemShut {NoStop}%
\bibitem [{\citenamefont {Iancu}\ and\ \citenamefont
  {Venugopalan}(2003)}]{Iancu:2003xm}%
  \BibitemOpen
  \bibfield  {author} {\bibinfo {author} {\bibfnamefont {E.}~\bibnamefont
  {Iancu}}\ and\ \bibinfo {author} {\bibfnamefont {R.}~\bibnamefont
  {Venugopalan}},\ }\href@noop {} {\  (\bibinfo {year} {2003})},\ \Eprint
  {http://arxiv.org/abs/hep-ph/0303204} {hep-ph/0303204} \BibitemShut {NoStop}%
\bibitem [{\citenamefont {Weigert}(2005)}]{Weigert:2005us}%
  \BibitemOpen
  \bibfield  {author} {\bibinfo {author} {\bibfnamefont {H.}~\bibnamefont
  {Weigert}},\ }\href@noop {} {\bibfield  {journal} {\bibinfo  {journal} {Prog.
  Part. Nucl. Phys.}\ }\textbf {\bibinfo {volume} {55}},\ \bibinfo {pages}
  {461} (\bibinfo {year} {2005})},\ \Eprint
  {http://arxiv.org/abs/hep-ph/0501087} {hep-ph/0501087} \BibitemShut {NoStop}%
\bibitem [{\citenamefont {Jalilian-Marian}\ and\ \citenamefont
  {Kovchegov}(2006)}]{JalilianMarian:2005jf}%
  \BibitemOpen
  \bibfield  {author} {\bibinfo {author} {\bibfnamefont {J.}~\bibnamefont
  {Jalilian-Marian}}\ and\ \bibinfo {author} {\bibfnamefont {Y.~V.}\
  \bibnamefont {Kovchegov}},\ }\href {\doibase 10.1016/j.ppnp.2005.07.002}
  {\bibfield  {journal} {\bibinfo  {journal} {Prog. Part. Nucl. Phys.}\
  }\textbf {\bibinfo {volume} {56}},\ \bibinfo {pages} {104} (\bibinfo {year}
  {2006})},\ \Eprint {http://arxiv.org/abs/hep-ph/0505052}
  {arXiv:hep-ph/0505052 [hep-ph]} \BibitemShut {NoStop}%
\bibitem [{\citenamefont {Gelis}\ \emph {et~al.}(2010)\citenamefont {Gelis},
  \citenamefont {Iancu}, \citenamefont {Jalilian-Marian},\ and\ \citenamefont
  {Venugopalan}}]{Gelis:2010nm}%
  \BibitemOpen
  \bibfield  {author} {\bibinfo {author} {\bibfnamefont {F.}~\bibnamefont
  {Gelis}}, \bibinfo {author} {\bibfnamefont {E.}~\bibnamefont {Iancu}},
  \bibinfo {author} {\bibfnamefont {J.}~\bibnamefont {Jalilian-Marian}}, \ and\
  \bibinfo {author} {\bibfnamefont {R.}~\bibnamefont {Venugopalan}},\ }\href
  {\doibase 10.1146/annurev.nucl.010909.083629} {\bibfield  {journal} {\bibinfo
   {journal} {Ann.Rev.Nucl.Part.Sci.}\ }\textbf {\bibinfo {volume} {60}},\
  \bibinfo {pages} {463} (\bibinfo {year} {2010})},\ \Eprint
  {http://arxiv.org/abs/1002.0333} {arXiv:1002.0333 [hep-ph]} \BibitemShut
  {NoStop}%
\bibitem [{\citenamefont {Albacete}\ and\ \citenamefont
  {Marquet}(2014)}]{Albacete:2014fwa}%
  \BibitemOpen
  \bibfield  {author} {\bibinfo {author} {\bibfnamefont {J.~L.}\ \bibnamefont
  {Albacete}}\ and\ \bibinfo {author} {\bibfnamefont {C.}~\bibnamefont
  {Marquet}},\ }\href {\doibase 10.1016/j.ppnp.2014.01.004} {\bibfield
  {journal} {\bibinfo  {journal} {Prog.Part.Nucl.Phys.}\ }\textbf {\bibinfo
  {volume} {76}},\ \bibinfo {pages} {1} (\bibinfo {year} {2014})},\ \Eprint
  {http://arxiv.org/abs/1401.4866} {arXiv:1401.4866 [hep-ph]} \BibitemShut
  {NoStop}%
\bibitem [{\citenamefont {Kovchegov}\ and\ \citenamefont
  {Levin}(2012)}]{Kovchegov:2012mbw}%
  \BibitemOpen
  \bibfield  {author} {\bibinfo {author} {\bibfnamefont {Y.~V.}\ \bibnamefont
  {Kovchegov}}\ and\ \bibinfo {author} {\bibfnamefont {E.}~\bibnamefont
  {Levin}},\ }\href {http://www.cambridge.org/de/knowledge/isbn/item6803159}
  {\emph {\bibinfo {title} {{Quantum chromodynamics at high energy}}}},\
  Vol.~\bibinfo {volume} {33}\ (\bibinfo  {publisher} {Cambridge University
  Press},\ \bibinfo {year} {2012})\BibitemShut {NoStop}%
\bibitem [{\citenamefont {Kang}\ and\ \citenamefont
  {Yuan}(2011)}]{Kang:2011ni}%
  \BibitemOpen
  \bibfield  {author} {\bibinfo {author} {\bibfnamefont {Z.-B.}\ \bibnamefont
  {Kang}}\ and\ \bibinfo {author} {\bibfnamefont {F.}~\bibnamefont {Yuan}},\
  }\href {\doibase 10.1103/PhysRevD.84.034019} {\bibfield  {journal} {\bibinfo
  {journal} {Phys.Rev.}\ }\textbf {\bibinfo {volume} {D84}},\ \bibinfo {pages}
  {034019} (\bibinfo {year} {2011})},\ \Eprint {http://arxiv.org/abs/1106.1375}
  {arXiv:1106.1375 [hep-ph]} \BibitemShut {NoStop}%
\bibitem [{\citenamefont {Sch{\"a}fer}\ and\ \citenamefont
  {Zhou}(2014)}]{Schafer:2014zea}%
  \BibitemOpen
  \bibfield  {author} {\bibinfo {author} {\bibfnamefont {A.}~\bibnamefont
  {Sch{\"a}fer}}\ and\ \bibinfo {author} {\bibfnamefont {J.}~\bibnamefont
  {Zhou}},\ }\href {\doibase 10.1103/PhysRevD.90.034016} {\bibfield  {journal}
  {\bibinfo  {journal} {Phys. Rev. D}\ }\textbf {\bibinfo {volume} {90}},\
  \bibinfo {pages} {034016} (\bibinfo {year} {2014})},\ \Eprint
  {http://arxiv.org/abs/1404.5809} {arXiv:1404.5809 [hep-ph]} \BibitemShut
  {NoStop}%
\bibitem [{\citenamefont {Zhou}(2015)}]{Zhou:2015ima}%
  \BibitemOpen
  \bibfield  {author} {\bibinfo {author} {\bibfnamefont {J.}~\bibnamefont
  {Zhou}},\ }\href {\doibase 10.1103/PhysRevD.92.014034} {\bibfield  {journal}
  {\bibinfo  {journal} {Phys. Rev. D}\ }\textbf {\bibinfo {volume} {92}},\
  \bibinfo {pages} {014034} (\bibinfo {year} {2015})},\ \Eprint
  {http://arxiv.org/abs/1502.02457} {arXiv:1502.02457 [hep-ph]} \BibitemShut
  {NoStop}%
\bibitem [{\citenamefont {Kovchegov}\ \emph {et~al.}(2004)\citenamefont
  {Kovchegov}, \citenamefont {Szymanowski},\ and\ \citenamefont
  {Wallon}}]{Kovchegov:2003dm}%
  \BibitemOpen
  \bibfield  {author} {\bibinfo {author} {\bibfnamefont {Y.~V.}\ \bibnamefont
  {Kovchegov}}, \bibinfo {author} {\bibfnamefont {L.}~\bibnamefont
  {Szymanowski}}, \ and\ \bibinfo {author} {\bibfnamefont {S.}~\bibnamefont
  {Wallon}},\ }\href {\doibase 10.1016/j.physletb.2004.02.036} {\bibfield
  {journal} {\bibinfo  {journal} {Phys.Lett.}\ }\textbf {\bibinfo {volume}
  {B586}},\ \bibinfo {pages} {267} (\bibinfo {year} {2004})},\ \bibinfo {note}
  {dedicated to the memory of Jan Kwiecinski},\ \Eprint
  {http://arxiv.org/abs/hep-ph/0309281} {arXiv:hep-ph/0309281 [hep-ph]}
  \BibitemShut {NoStop}%
\bibitem [{\citenamefont {Hatta}\ \emph {et~al.}(2005)\citenamefont {Hatta},
  \citenamefont {Iancu}, \citenamefont {Itakura},\ and\ \citenamefont
  {McLerran}}]{Hatta:2005as}%
  \BibitemOpen
  \bibfield  {author} {\bibinfo {author} {\bibfnamefont {Y.}~\bibnamefont
  {Hatta}}, \bibinfo {author} {\bibfnamefont {E.}~\bibnamefont {Iancu}},
  \bibinfo {author} {\bibfnamefont {K.}~\bibnamefont {Itakura}}, \ and\
  \bibinfo {author} {\bibfnamefont {L.}~\bibnamefont {McLerran}},\ }\href
  {\doibase 10.1016/j.nuclphysa.2005.05.163} {\bibfield  {journal} {\bibinfo
  {journal} {Nucl.Phys.}\ }\textbf {\bibinfo {volume} {A760}},\ \bibinfo
  {pages} {172} (\bibinfo {year} {2005})},\ \Eprint
  {http://arxiv.org/abs/hep-ph/0501171} {arXiv:hep-ph/0501171 [hep-ph]}
  \BibitemShut {NoStop}%
\bibitem [{\citenamefont {McLerran}\ and\ \citenamefont
  {Venugopalan}(1994{\natexlab{a}})}]{McLerran:1993ni}%
  \BibitemOpen
  \bibfield  {author} {\bibinfo {author} {\bibfnamefont {L.~D.}\ \bibnamefont
  {McLerran}}\ and\ \bibinfo {author} {\bibfnamefont {R.}~\bibnamefont
  {Venugopalan}},\ }\href@noop {} {\bibfield  {journal} {\bibinfo  {journal}
  {Phys. Rev.}\ }\textbf {\bibinfo {volume} {D49}},\ \bibinfo {pages} {2233}
  (\bibinfo {year} {1994}{\natexlab{a}})},\ \Eprint
  {http://arxiv.org/abs/hep-ph/9309289} {hep-ph/9309289} \BibitemShut {NoStop}%
\bibitem [{\citenamefont {McLerran}\ and\ \citenamefont
  {Venugopalan}(1994{\natexlab{b}})}]{McLerran:1993ka}%
  \BibitemOpen
  \bibfield  {author} {\bibinfo {author} {\bibfnamefont {L.~D.}\ \bibnamefont
  {McLerran}}\ and\ \bibinfo {author} {\bibfnamefont {R.}~\bibnamefont
  {Venugopalan}},\ }\href@noop {} {\bibfield  {journal} {\bibinfo  {journal}
  {Phys. Rev.}\ }\textbf {\bibinfo {volume} {D49}},\ \bibinfo {pages} {3352}
  (\bibinfo {year} {1994}{\natexlab{b}})},\ \Eprint
  {http://arxiv.org/abs/hep-ph/9311205} {hep-ph/9311205} \BibitemShut {NoStop}%
\bibitem [{\citenamefont {McLerran}\ and\ \citenamefont
  {Venugopalan}(1994{\natexlab{c}})}]{McLerran:1994vd}%
  \BibitemOpen
  \bibfield  {author} {\bibinfo {author} {\bibfnamefont {L.~D.}\ \bibnamefont
  {McLerran}}\ and\ \bibinfo {author} {\bibfnamefont {R.}~\bibnamefont
  {Venugopalan}},\ }\href@noop {} {\bibfield  {journal} {\bibinfo  {journal}
  {Phys. Rev.}\ }\textbf {\bibinfo {volume} {D50}},\ \bibinfo {pages} {2225}
  (\bibinfo {year} {1994}{\natexlab{c}})},\ \Eprint
  {http://arxiv.org/abs/hep-ph/9402335} {hep-ph/9402335} \BibitemShut {NoStop}%
\bibitem [{\citenamefont {Kovchegov}(1997)}]{Kovchegov:1997pc}%
  \BibitemOpen
  \bibfield  {author} {\bibinfo {author} {\bibfnamefont {Y.~V.}\ \bibnamefont
  {Kovchegov}},\ }\href@noop {} {\bibfield  {journal} {\bibinfo  {journal}
  {Phys. Rev.}\ }\textbf {\bibinfo {volume} {D55}},\ \bibinfo {pages} {5445}
  (\bibinfo {year} {1997})},\ \Eprint {http://arxiv.org/abs/hep-ph/9701229}
  {hep-ph/9701229} \BibitemShut {NoStop}%
\bibitem [{\citenamefont {Kovchegov}(1996)}]{Kovchegov:1996ty}%
  \BibitemOpen
  \bibfield  {author} {\bibinfo {author} {\bibfnamefont {Y.~V.}\ \bibnamefont
  {Kovchegov}},\ }\href@noop {} {\bibfield  {journal} {\bibinfo  {journal}
  {Phys. Rev.}\ }\textbf {\bibinfo {volume} {D54}},\ \bibinfo {pages} {5463}
  (\bibinfo {year} {1996})},\ \Eprint {http://arxiv.org/abs/hep-ph/9605446}
  {hep-ph/9605446} \BibitemShut {NoStop}%
\bibitem [{\citenamefont {Balitsky}(1996)}]{Balitsky:1995ub}%
  \BibitemOpen
  \bibfield  {author} {\bibinfo {author} {\bibfnamefont {I.}~\bibnamefont
  {Balitsky}},\ }\href {\doibase 10.1016/0550-3213(95)00638-9} {\bibfield
  {journal} {\bibinfo  {journal} {Nucl. Phys.}\ }\textbf {\bibinfo {volume}
  {B463}},\ \bibinfo {pages} {99} (\bibinfo {year} {1996})},\ \Eprint
  {http://arxiv.org/abs/hep-ph/9509348} {arXiv:hep-ph/9509348 [hep-ph]}
  \BibitemShut {NoStop}%
\bibitem [{\citenamefont {Balitsky}(1999)}]{Balitsky:1998ya}%
  \BibitemOpen
  \bibfield  {author} {\bibinfo {author} {\bibfnamefont {I.}~\bibnamefont
  {Balitsky}},\ }\href@noop {} {\bibfield  {journal} {\bibinfo  {journal}
  {Phys. Rev.}\ }\textbf {\bibinfo {volume} {D60}},\ \bibinfo {pages} {014020}
  (\bibinfo {year} {1999})},\ \Eprint {http://arxiv.org/abs/hep-ph/9812311}
  {hep-ph/9812311} \BibitemShut {NoStop}%
\bibitem [{\citenamefont {Kovchegov}(1999)}]{Kovchegov:1999yj}%
  \BibitemOpen
  \bibfield  {author} {\bibinfo {author} {\bibfnamefont {Y.~V.}\ \bibnamefont
  {Kovchegov}},\ }\href@noop {} {\bibfield  {journal} {\bibinfo  {journal}
  {Phys. Rev.}\ }\textbf {\bibinfo {volume} {D60}},\ \bibinfo {pages} {034008}
  (\bibinfo {year} {1999})},\ \Eprint {http://arxiv.org/abs/hep-ph/9901281}
  {hep-ph/9901281} \BibitemShut {NoStop}%
\bibitem [{\citenamefont {Kovchegov}(2000)}]{Kovchegov:1999ua}%
  \BibitemOpen
  \bibfield  {author} {\bibinfo {author} {\bibfnamefont {Y.~V.}\ \bibnamefont
  {Kovchegov}},\ }\href@noop {} {\bibfield  {journal} {\bibinfo  {journal}
  {Phys. Rev.}\ }\textbf {\bibinfo {volume} {D61}},\ \bibinfo {pages} {074018}
  (\bibinfo {year} {2000})},\ \Eprint {http://arxiv.org/abs/hep-ph/9905214}
  {hep-ph/9905214} \BibitemShut {NoStop}%
\bibitem [{\citenamefont {Jalilian-Marian}\ \emph
  {et~al.}(1998{\natexlab{a}})\citenamefont {Jalilian-Marian}, \citenamefont
  {Kovner},\ and\ \citenamefont {Weigert}}]{Jalilian-Marian:1997dw}%
  \BibitemOpen
  \bibfield  {author} {\bibinfo {author} {\bibfnamefont {J.}~\bibnamefont
  {Jalilian-Marian}}, \bibinfo {author} {\bibfnamefont {A.}~\bibnamefont
  {Kovner}}, \ and\ \bibinfo {author} {\bibfnamefont {H.}~\bibnamefont
  {Weigert}},\ }\href@noop {} {\bibfield  {journal} {\bibinfo  {journal} {Phys.
  Rev.}\ }\textbf {\bibinfo {volume} {D59}},\ \bibinfo {pages} {014015}
  (\bibinfo {year} {1998}{\natexlab{a}})},\ \Eprint
  {http://arxiv.org/abs/hep-ph/9709432} {hep-ph/9709432} \BibitemShut {NoStop}%
\bibitem [{\citenamefont {Jalilian-Marian}\ \emph
  {et~al.}(1998{\natexlab{b}})\citenamefont {Jalilian-Marian}, \citenamefont
  {Kovner}, \citenamefont {Leonidov},\ and\ \citenamefont
  {Weigert}}]{Jalilian-Marian:1997gr}%
  \BibitemOpen
  \bibfield  {author} {\bibinfo {author} {\bibfnamefont {J.}~\bibnamefont
  {Jalilian-Marian}}, \bibinfo {author} {\bibfnamefont {A.}~\bibnamefont
  {Kovner}}, \bibinfo {author} {\bibfnamefont {A.}~\bibnamefont {Leonidov}}, \
  and\ \bibinfo {author} {\bibfnamefont {H.}~\bibnamefont {Weigert}},\
  }\href@noop {} {\bibfield  {journal} {\bibinfo  {journal} {Phys. Rev.}\
  }\textbf {\bibinfo {volume} {D59}},\ \bibinfo {pages} {014014} (\bibinfo
  {year} {1998}{\natexlab{b}})},\ \Eprint {http://arxiv.org/abs/hep-ph/9706377}
  {hep-ph/9706377} \BibitemShut {NoStop}%
\bibitem [{\citenamefont {Weigert}(2002)}]{Weigert:2000gi}%
  \BibitemOpen
  \bibfield  {author} {\bibinfo {author} {\bibfnamefont {H.}~\bibnamefont
  {Weigert}},\ }\href@noop {} {\bibfield  {journal} {\bibinfo  {journal} {Nucl.
  Phys.}\ }\textbf {\bibinfo {volume} {A703}},\ \bibinfo {pages} {823}
  (\bibinfo {year} {2002})},\ \Eprint {http://arxiv.org/abs/hep-ph/0004044}
  {hep-ph/0004044} \BibitemShut {NoStop}%
\bibitem [{\citenamefont {Iancu}\ \emph
  {et~al.}(2001{\natexlab{a}})\citenamefont {Iancu}, \citenamefont {Leonidov},\
  and\ \citenamefont {McLerran}}]{Iancu:2001ad}%
  \BibitemOpen
  \bibfield  {author} {\bibinfo {author} {\bibfnamefont {E.}~\bibnamefont
  {Iancu}}, \bibinfo {author} {\bibfnamefont {A.}~\bibnamefont {Leonidov}}, \
  and\ \bibinfo {author} {\bibfnamefont {L.~D.}\ \bibnamefont {McLerran}},\
  }\href {\doibase 10.1016/S0370-2693(01)00524-X} {\bibfield  {journal}
  {\bibinfo  {journal} {Phys. Lett.}\ }\textbf {\bibinfo {volume} {B510}},\
  \bibinfo {pages} {133} (\bibinfo {year} {2001}{\natexlab{a}})}\BibitemShut
  {NoStop}%
\bibitem [{\citenamefont {Iancu}\ \emph
  {et~al.}(2001{\natexlab{b}})\citenamefont {Iancu}, \citenamefont {Leonidov},\
  and\ \citenamefont {McLerran}}]{Iancu:2000hn}%
  \BibitemOpen
  \bibfield  {author} {\bibinfo {author} {\bibfnamefont {E.}~\bibnamefont
  {Iancu}}, \bibinfo {author} {\bibfnamefont {A.}~\bibnamefont {Leonidov}}, \
  and\ \bibinfo {author} {\bibfnamefont {L.~D.}\ \bibnamefont {McLerran}},\
  }\href@noop {} {\bibfield  {journal} {\bibinfo  {journal} {Nucl. Phys.}\
  }\textbf {\bibinfo {volume} {A692}},\ \bibinfo {pages} {583} (\bibinfo {year}
  {2001}{\natexlab{b}})},\ \Eprint {http://arxiv.org/abs/hep-ph/0011241}
  {hep-ph/0011241} \BibitemShut {NoStop}%
\bibitem [{\citenamefont {Ferreiro}\ \emph {et~al.}(2002)\citenamefont
  {Ferreiro}, \citenamefont {Iancu}, \citenamefont {Leonidov},\ and\
  \citenamefont {McLerran}}]{Ferreiro:2001qy}%
  \BibitemOpen
  \bibfield  {author} {\bibinfo {author} {\bibfnamefont {E.}~\bibnamefont
  {Ferreiro}}, \bibinfo {author} {\bibfnamefont {E.}~\bibnamefont {Iancu}},
  \bibinfo {author} {\bibfnamefont {A.}~\bibnamefont {Leonidov}}, \ and\
  \bibinfo {author} {\bibfnamefont {L.}~\bibnamefont {McLerran}},\ }\href@noop
  {} {\bibfield  {journal} {\bibinfo  {journal} {Nucl. Phys.}\ }\textbf
  {\bibinfo {volume} {A703}},\ \bibinfo {pages} {489} (\bibinfo {year}
  {2002})},\ \Eprint {http://arxiv.org/abs/hep-ph/0109115} {hep-ph/0109115}
  \BibitemShut {NoStop}%
\bibitem [{\citenamefont {Mueller}(1990)}]{Mueller:1989st}%
  \BibitemOpen
  \bibfield  {author} {\bibinfo {author} {\bibfnamefont {A.~H.}\ \bibnamefont
  {Mueller}},\ }\href@noop {} {\bibfield  {journal} {\bibinfo  {journal} {Nucl.
  Phys.}\ }\textbf {\bibinfo {volume} {B335}},\ \bibinfo {pages} {115}
  (\bibinfo {year} {1990})}\BibitemShut {NoStop}%
\bibitem [{\citenamefont {Lepage}\ and\ \citenamefont
  {Brodsky}(1980)}]{Lepage:1980fj}%
  \BibitemOpen
  \bibfield  {author} {\bibinfo {author} {\bibfnamefont {G.~P.}\ \bibnamefont
  {Lepage}}\ and\ \bibinfo {author} {\bibfnamefont {S.~J.}\ \bibnamefont
  {Brodsky}},\ }\href@noop {} {\bibfield  {journal} {\bibinfo  {journal} {Phys.
  Rev.}\ }\textbf {\bibinfo {volume} {D22}},\ \bibinfo {pages} {2157} (\bibinfo
  {year} {1980})}\BibitemShut {NoStop}%
\bibitem [{\citenamefont {Brodsky}\ \emph {et~al.}(1998)\citenamefont
  {Brodsky}, \citenamefont {Pauli},\ and\ \citenamefont
  {Pinsky}}]{Brodsky:1997de}%
  \BibitemOpen
  \bibfield  {author} {\bibinfo {author} {\bibfnamefont {S.~J.}\ \bibnamefont
  {Brodsky}}, \bibinfo {author} {\bibfnamefont {H.-C.}\ \bibnamefont {Pauli}},
  \ and\ \bibinfo {author} {\bibfnamefont {S.~S.}\ \bibnamefont {Pinsky}},\
  }\href {\doibase 10.1016/S0370-1573(97)00089-6} {\bibfield  {journal}
  {\bibinfo  {journal} {Phys.Rept.}\ }\textbf {\bibinfo {volume} {301}},\
  \bibinfo {pages} {299} (\bibinfo {year} {1998})},\ \Eprint
  {http://arxiv.org/abs/hep-ph/9705477} {arXiv:hep-ph/9705477 [hep-ph]}
  \BibitemShut {NoStop}%
\bibitem [{\citenamefont {Kovchegov}\ and\ \citenamefont
  {Mueller}(1998)}]{Kovchegov:1998bi}%
  \BibitemOpen
  \bibfield  {author} {\bibinfo {author} {\bibfnamefont {Y.~V.}\ \bibnamefont
  {Kovchegov}}\ and\ \bibinfo {author} {\bibfnamefont {A.~H.}\ \bibnamefont
  {Mueller}},\ }\href@noop {} {\bibfield  {journal} {\bibinfo  {journal} {Nucl.
  Phys.}\ }\textbf {\bibinfo {volume} {B529}},\ \bibinfo {pages} {451}
  (\bibinfo {year} {1998})},\ \Eprint {http://arxiv.org/abs/hep-ph/9802440}
  {hep-ph/9802440} \BibitemShut {NoStop}%
\bibitem [{\citenamefont {Kovchegov}\ and\ \citenamefont
  {Tuchin}(2006)}]{Kovchegov:2006qn}%
  \BibitemOpen
  \bibfield  {author} {\bibinfo {author} {\bibfnamefont {Y.~V.}\ \bibnamefont
  {Kovchegov}}\ and\ \bibinfo {author} {\bibfnamefont {K.}~\bibnamefont
  {Tuchin}},\ }\href {\doibase 10.1103/PhysRevD.74.054014} {\bibfield
  {journal} {\bibinfo  {journal} {Phys.Rev.}\ }\textbf {\bibinfo {volume}
  {D74}},\ \bibinfo {pages} {054014} (\bibinfo {year} {2006})},\ \Eprint
  {http://arxiv.org/abs/hep-ph/0603055} {arXiv:hep-ph/0603055 [hep-ph]}
  \BibitemShut {NoStop}%
\bibitem [{\citenamefont {Kovchegov}\ and\ \citenamefont
  {McLerran}(1999)}]{Kovchegov:1999kx}%
  \BibitemOpen
  \bibfield  {author} {\bibinfo {author} {\bibfnamefont {Y.~V.}\ \bibnamefont
  {Kovchegov}}\ and\ \bibinfo {author} {\bibfnamefont {L.~D.}\ \bibnamefont
  {McLerran}},\ }\href@noop {} {\bibfield  {journal} {\bibinfo  {journal}
  {Phys. Rev.}\ }\textbf {\bibinfo {volume} {D60}},\ \bibinfo {pages} {054025}
  (\bibinfo {year} {1999})},\ \Eprint {http://arxiv.org/abs/hep-ph/9903246}
  {hep-ph/9903246} \BibitemShut {NoStop}%
\bibitem [{\citenamefont {Kovchegov}\ \emph {et~al.}(2009)\citenamefont
  {Kovchegov}, \citenamefont {Kuokkanen}, \citenamefont {Rummukainen},\ and\
  \citenamefont {Weigert}}]{Kovchegov:2008mk}%
  \BibitemOpen
  \bibfield  {author} {\bibinfo {author} {\bibfnamefont {Y.~V.}\ \bibnamefont
  {Kovchegov}}, \bibinfo {author} {\bibfnamefont {J.}~\bibnamefont
  {Kuokkanen}}, \bibinfo {author} {\bibfnamefont {K.}~\bibnamefont
  {Rummukainen}}, \ and\ \bibinfo {author} {\bibfnamefont {H.}~\bibnamefont
  {Weigert}},\ }\href {\doibase 10.1016/j.nuclphysa.2009.03.006} {\bibfield
  {journal} {\bibinfo  {journal} {Nucl. Phys.}\ }\textbf {\bibinfo {volume}
  {A823}},\ \bibinfo {pages} {47} (\bibinfo {year} {2009})},\ \Eprint
  {http://arxiv.org/abs/0812.3238} {arXiv:0812.3238 [hep-ph]} \BibitemShut
  {NoStop}%
\bibitem [{\citenamefont {Albacete}\ \emph {et~al.}(2009)\citenamefont
  {Albacete}, \citenamefont {Armesto}, \citenamefont {Milhano},\ and\
  \citenamefont {Salgado}}]{Albacete:2009fh}%
  \BibitemOpen
  \bibfield  {author} {\bibinfo {author} {\bibfnamefont {J.~L.}\ \bibnamefont
  {Albacete}}, \bibinfo {author} {\bibfnamefont {N.}~\bibnamefont {Armesto}},
  \bibinfo {author} {\bibfnamefont {J.~G.}\ \bibnamefont {Milhano}}, \ and\
  \bibinfo {author} {\bibfnamefont {C.~A.}\ \bibnamefont {Salgado}},\ }\href
  {\doibase 10.1103/PhysRevD.80.034031} {\bibfield  {journal} {\bibinfo
  {journal} {Phys. Rev.}\ }\textbf {\bibinfo {volume} {D80}},\ \bibinfo {pages}
  {034031} (\bibinfo {year} {2009})},\ \Eprint {http://arxiv.org/abs/0902.1112}
  {arXiv:0902.1112 [hep-ph]} \BibitemShut {NoStop}%
\bibitem [{\citenamefont {Albacete}\ \emph {et~al.}(2011)\citenamefont
  {Albacete}, \citenamefont {Armesto}, \citenamefont {Milhano}, \citenamefont
  {Quiroga-Arias},\ and\ \citenamefont {Salgado}}]{Albacete:2010sy}%
  \BibitemOpen
  \bibfield  {author} {\bibinfo {author} {\bibfnamefont {J.~L.}\ \bibnamefont
  {Albacete}}, \bibinfo {author} {\bibfnamefont {N.}~\bibnamefont {Armesto}},
  \bibinfo {author} {\bibfnamefont {J.~G.}\ \bibnamefont {Milhano}}, \bibinfo
  {author} {\bibfnamefont {P.}~\bibnamefont {Quiroga-Arias}}, \ and\ \bibinfo
  {author} {\bibfnamefont {C.~A.}\ \bibnamefont {Salgado}},\ }\href {\doibase
  10.1140/epjc/s10052-011-1705-3} {\bibfield  {journal} {\bibinfo  {journal}
  {Eur. Phys. J.}\ }\textbf {\bibinfo {volume} {C71}},\ \bibinfo {pages} {1705}
  (\bibinfo {year} {2011})},\ \Eprint {http://arxiv.org/abs/1012.4408}
  {arXiv:1012.4408 [hep-ph]} \BibitemShut {NoStop}%
\bibitem [{\citenamefont {Jalilian-Marian}\ \emph {et~al.}(1997)\citenamefont
  {Jalilian-Marian}, \citenamefont {Kovner}, \citenamefont {McLerran},\ and\
  \citenamefont {Weigert}}]{JalilianMarian:1996xn}%
  \BibitemOpen
  \bibfield  {author} {\bibinfo {author} {\bibfnamefont {J.}~\bibnamefont
  {Jalilian-Marian}}, \bibinfo {author} {\bibfnamefont {A.}~\bibnamefont
  {Kovner}}, \bibinfo {author} {\bibfnamefont {L.~D.}\ \bibnamefont
  {McLerran}}, \ and\ \bibinfo {author} {\bibfnamefont {H.}~\bibnamefont
  {Weigert}},\ }\href {\doibase 10.1103/PhysRevD.55.5414} {\bibfield  {journal}
  {\bibinfo  {journal} {Phys. Rev.}\ }\textbf {\bibinfo {volume} {D55}},\
  \bibinfo {pages} {5414} (\bibinfo {year} {1997})},\ \Eprint
  {http://arxiv.org/abs/hep-ph/9606337} {arXiv:hep-ph/9606337 [hep-ph]}
  \BibitemShut {NoStop}%
\bibitem [{\citenamefont {Braun}(2000)}]{Braun:2000bh}%
  \BibitemOpen
  \bibfield  {author} {\bibinfo {author} {\bibfnamefont {M.~A.}\ \bibnamefont
  {Braun}},\ }\href {\doibase 10.1016/S0370-2693(00)00570-0} {\bibfield
  {journal} {\bibinfo  {journal} {Phys. Lett.}\ }\textbf {\bibinfo {volume}
  {B483}},\ \bibinfo {pages} {105} (\bibinfo {year} {2000})},\ \Eprint
  {http://arxiv.org/abs/hep-ph/0003003} {arXiv:hep-ph/0003003} \BibitemShut
  {NoStop}%
\bibitem [{\citenamefont {Kovchegov}\ and\ \citenamefont
  {Tuchin}(2002)}]{Kovchegov:2001sc}%
  \BibitemOpen
  \bibfield  {author} {\bibinfo {author} {\bibfnamefont {Y.~V.}\ \bibnamefont
  {Kovchegov}}\ and\ \bibinfo {author} {\bibfnamefont {K.}~\bibnamefont
  {Tuchin}},\ }\href@noop {} {\bibfield  {journal} {\bibinfo  {journal} {Phys.
  Rev.}\ }\textbf {\bibinfo {volume} {D65}},\ \bibinfo {pages} {074026}
  (\bibinfo {year} {2002})},\ \Eprint {http://arxiv.org/abs/hep-ph/0111362}
  {hep-ph/0111362} \BibitemShut {NoStop}%
\bibitem [{\citenamefont {Kharzeev}\ \emph
  {et~al.}(2003{\natexlab{a}})\citenamefont {Kharzeev}, \citenamefont
  {Kovchegov},\ and\ \citenamefont {Tuchin}}]{Kharzeev:2003wz}%
  \BibitemOpen
  \bibfield  {author} {\bibinfo {author} {\bibfnamefont {D.}~\bibnamefont
  {Kharzeev}}, \bibinfo {author} {\bibfnamefont {Y.~V.}\ \bibnamefont
  {Kovchegov}}, \ and\ \bibinfo {author} {\bibfnamefont {K.}~\bibnamefont
  {Tuchin}},\ }\href@noop {} {\bibfield  {journal} {\bibinfo  {journal} {Phys.
  Rev.}\ }\textbf {\bibinfo {volume} {D68}},\ \bibinfo {pages} {094013}
  (\bibinfo {year} {2003}{\natexlab{a}})},\ \Eprint
  {http://arxiv.org/abs/hep-ph/0307037} {hep-ph/0307037} \BibitemShut {NoStop}%
\bibitem [{\citenamefont {Dominguez}\ \emph
  {et~al.}(2011{\natexlab{a}})\citenamefont {Dominguez}, \citenamefont
  {Marquet}, \citenamefont {Xiao},\ and\ \citenamefont
  {Yuan}}]{Dominguez:2011wm}%
  \BibitemOpen
  \bibfield  {author} {\bibinfo {author} {\bibfnamefont {F.}~\bibnamefont
  {Dominguez}}, \bibinfo {author} {\bibfnamefont {C.}~\bibnamefont {Marquet}},
  \bibinfo {author} {\bibfnamefont {B.-W.}\ \bibnamefont {Xiao}}, \ and\
  \bibinfo {author} {\bibfnamefont {F.}~\bibnamefont {Yuan}},\ }\href {\doibase
  10.1103/PhysRevD.83.105005} {\bibfield  {journal} {\bibinfo  {journal}
  {Phys.Rev.}\ }\textbf {\bibinfo {volume} {D83}},\ \bibinfo {pages} {105005}
  (\bibinfo {year} {2011}{\natexlab{a}})},\ \Eprint
  {http://arxiv.org/abs/1101.0715} {arXiv:1101.0715 [hep-ph]} \BibitemShut
  {NoStop}%
\bibitem [{\citenamefont {Kovchegov}\ and\ \citenamefont
  {Wertepny}(2014)}]{Kovchegov:2013ewa}%
  \BibitemOpen
  \bibfield  {author} {\bibinfo {author} {\bibfnamefont {Y.~V.}\ \bibnamefont
  {Kovchegov}}\ and\ \bibinfo {author} {\bibfnamefont {D.~E.}\ \bibnamefont
  {Wertepny}},\ }\href {\doibase 10.1016/j.nuclphysa.2014.02.021} {\bibfield
  {journal} {\bibinfo  {journal} {Nucl.Phys.}\ }\textbf {\bibinfo {volume}
  {A925}},\ \bibinfo {pages} {254} (\bibinfo {year} {2014})},\ \Eprint
  {http://arxiv.org/abs/1310.6701} {arXiv:1310.6701 [hep-ph]} \BibitemShut
  {NoStop}%
\bibitem [{\citenamefont {Golec-Biernat}\ and\ \citenamefont
  {Wusthoff}(1998)}]{GolecBiernat:1998js}%
  \BibitemOpen
  \bibfield  {author} {\bibinfo {author} {\bibfnamefont {K.~J.}\ \bibnamefont
  {Golec-Biernat}}\ and\ \bibinfo {author} {\bibfnamefont {M.}~\bibnamefont
  {Wusthoff}},\ }\href {\doibase 10.1103/PhysRevD.59.014017} {\bibfield
  {journal} {\bibinfo  {journal} {Phys. Rev.}\ }\textbf {\bibinfo {volume}
  {D59}},\ \bibinfo {pages} {014017} (\bibinfo {year} {1998})},\ \Eprint
  {http://arxiv.org/abs/hep-ph/9807513} {arXiv:hep-ph/9807513 [hep-ph]}
  \BibitemShut {NoStop}%
\bibitem [{\citenamefont {Golec-Biernat}\ and\ \citenamefont
  {Wusthoff}(1999)}]{GolecBiernat:1999qd}%
  \BibitemOpen
  \bibfield  {author} {\bibinfo {author} {\bibfnamefont {K.~J.}\ \bibnamefont
  {Golec-Biernat}}\ and\ \bibinfo {author} {\bibfnamefont {M.}~\bibnamefont
  {Wusthoff}},\ }\href {\doibase 10.1103/PhysRevD.60.114023} {\bibfield
  {journal} {\bibinfo  {journal} {Phys. Rev.}\ }\textbf {\bibinfo {volume}
  {D60}},\ \bibinfo {pages} {114023} (\bibinfo {year} {1999})},\ \Eprint
  {http://arxiv.org/abs/hep-ph/9903358} {arXiv:hep-ph/9903358 [hep-ph]}
  \BibitemShut {NoStop}%
\bibitem [{\citenamefont {Itakura}\ \emph {et~al.}(2004)\citenamefont
  {Itakura}, \citenamefont {Kovchegov}, \citenamefont {McLerran},\ and\
  \citenamefont {Teaney}}]{Itakura:2003jp}%
  \BibitemOpen
  \bibfield  {author} {\bibinfo {author} {\bibfnamefont {K.}~\bibnamefont
  {Itakura}}, \bibinfo {author} {\bibfnamefont {Y.~V.}\ \bibnamefont
  {Kovchegov}}, \bibinfo {author} {\bibfnamefont {L.}~\bibnamefont {McLerran}},
  \ and\ \bibinfo {author} {\bibfnamefont {D.}~\bibnamefont {Teaney}},\ }\href
  {\doibase 10.1016/j.nuclphysa.2003.10.016} {\bibfield  {journal} {\bibinfo
  {journal} {Nucl. Phys.}\ }\textbf {\bibinfo {volume} {A730}},\ \bibinfo
  {pages} {160} (\bibinfo {year} {2004})},\ \Eprint
  {http://arxiv.org/abs/hep-ph/0305332} {arXiv:hep-ph/0305332} \BibitemShut
  {NoStop}%
\bibitem [{\citenamefont {Albacete}\ and\ \citenamefont
  {Kovchegov}(2007)}]{Albacete:2006vv}%
  \BibitemOpen
  \bibfield  {author} {\bibinfo {author} {\bibfnamefont {J.~L.}\ \bibnamefont
  {Albacete}}\ and\ \bibinfo {author} {\bibfnamefont {Y.~V.}\ \bibnamefont
  {Kovchegov}},\ }\href {\doibase 10.1016/j.nuclphysa.2006.09.013} {\bibfield
  {journal} {\bibinfo  {journal} {Nucl. Phys.}\ }\textbf {\bibinfo {volume}
  {A781}},\ \bibinfo {pages} {122} (\bibinfo {year} {2007})},\ \Eprint
  {http://arxiv.org/abs/hep-ph/0605053} {arXiv:hep-ph/0605053} \BibitemShut
  {NoStop}%
\bibitem [{\citenamefont {Aidala}\ \emph {et~al.}(2018)\citenamefont {Aidala}
  \emph {et~al.}}]{Aidala:2017cnz}%
  \BibitemOpen
  \bibfield  {author} {\bibinfo {author} {\bibfnamefont {C.}~\bibnamefont
  {Aidala}} \emph {et~al.} (\bibinfo {collaboration} {PHENIX}),\ }\href
  {\doibase 10.1103/PhysRevLett.120.022001} {\bibfield  {journal} {\bibinfo
  {journal} {Phys. Rev. Lett.}\ }\textbf {\bibinfo {volume} {120}},\ \bibinfo
  {pages} {022001} (\bibinfo {year} {2018})},\ \Eprint
  {http://arxiv.org/abs/1703.10941} {arXiv:1703.10941 [hep-ex]} \BibitemShut
  {NoStop}%
\bibitem [{\citenamefont {Jalilian-Marian}\ and\ \citenamefont
  {Kovchegov}(2004)}]{JalilianMarian:2004da}%
  \BibitemOpen
  \bibfield  {author} {\bibinfo {author} {\bibfnamefont {J.}~\bibnamefont
  {Jalilian-Marian}}\ and\ \bibinfo {author} {\bibfnamefont {Y.~V.}\
  \bibnamefont {Kovchegov}},\ }\href@noop {} {\bibfield  {journal} {\bibinfo
  {journal} {Phys. Rev.}\ }\textbf {\bibinfo {volume} {D70}},\ \bibinfo {pages}
  {114017} (\bibinfo {year} {2004})},\ \Eprint
  {http://arxiv.org/abs/hep-ph/0405266} {hep-ph/0405266} \BibitemShut {NoStop}%
\bibitem [{\citenamefont {Beni\'{c}}\ and\ \citenamefont
  {Hatta}(2019)}]{Hatta2019a}%
  \BibitemOpen
  \bibfield  {author} {\bibinfo {author} {\bibfnamefont {S.}~\bibnamefont
  {Beni\'{c}}}\ and\ \bibinfo {author} {\bibfnamefont {Y.}~\bibnamefont
  {Hatta}},\ }\href {\doibase 10.1103/PhysRevD.99.094012} {\bibfield  {journal}
  {\bibinfo  {journal} {Phys. Rev.}\ }\textbf {\bibinfo {volume} {D99}},\
  \bibinfo {pages} {094012} (\bibinfo {year} {2019})},\ \Eprint
  {http://arxiv.org/abs/1811.10589} {arXiv:1811.10589 [hep-ph]} \BibitemShut
  {NoStop}%
\bibitem [{\citenamefont {Kharzeev}\ \emph
  {et~al.}(2003{\natexlab{b}})\citenamefont {Kharzeev}, \citenamefont {Levin},\
  and\ \citenamefont {McLerran}}]{Kharzeev:2002pc}%
  \BibitemOpen
  \bibfield  {author} {\bibinfo {author} {\bibfnamefont {D.}~\bibnamefont
  {Kharzeev}}, \bibinfo {author} {\bibfnamefont {E.}~\bibnamefont {Levin}}, \
  and\ \bibinfo {author} {\bibfnamefont {L.}~\bibnamefont {McLerran}},\
  }\href@noop {} {\bibfield  {journal} {\bibinfo  {journal} {Phys. Lett.}\
  }\textbf {\bibinfo {volume} {B561}},\ \bibinfo {pages} {93} (\bibinfo {year}
  {2003}{\natexlab{b}})},\ \Eprint {http://arxiv.org/abs/hep-ph/0210332}
  {hep-ph/0210332} \BibitemShut {NoStop}%
\bibitem [{\citenamefont {Albacete}\ \emph {et~al.}(2004)\citenamefont
  {Albacete}, \citenamefont {Armesto}, \citenamefont {Kovner}, \citenamefont
  {Salgado},\ and\ \citenamefont {Wiedemann}}]{Albacete:2003iq}%
  \BibitemOpen
  \bibfield  {author} {\bibinfo {author} {\bibfnamefont {J.~L.}\ \bibnamefont
  {Albacete}}, \bibinfo {author} {\bibfnamefont {N.}~\bibnamefont {Armesto}},
  \bibinfo {author} {\bibfnamefont {A.}~\bibnamefont {Kovner}}, \bibinfo
  {author} {\bibfnamefont {C.~A.}\ \bibnamefont {Salgado}}, \ and\ \bibinfo
  {author} {\bibfnamefont {U.~A.}\ \bibnamefont {Wiedemann}},\ }\href@noop {}
  {\bibfield  {journal} {\bibinfo  {journal} {Phys. Rev. Lett.}\ }\textbf
  {\bibinfo {volume} {92}},\ \bibinfo {pages} {082001} (\bibinfo {year}
  {2004})},\ \Eprint {http://arxiv.org/abs/hep-ph/0307179} {hep-ph/0307179}
  \BibitemShut {NoStop}%
\bibitem [{\citenamefont {Dominguez}\ \emph
  {et~al.}(2011{\natexlab{b}})\citenamefont {Dominguez}, \citenamefont
  {Mueller}, \citenamefont {Munier},\ and\ \citenamefont
  {Xiao}}]{Dominguez:2011gc}%
  \BibitemOpen
  \bibfield  {author} {\bibinfo {author} {\bibfnamefont {F.}~\bibnamefont
  {Dominguez}}, \bibinfo {author} {\bibfnamefont {A.}~\bibnamefont {Mueller}},
  \bibinfo {author} {\bibfnamefont {S.}~\bibnamefont {Munier}}, \ and\ \bibinfo
  {author} {\bibfnamefont {B.-W.}\ \bibnamefont {Xiao}},\ }\href {\doibase
  10.1016/j.physletb.2011.09.104} {\bibfield  {journal} {\bibinfo  {journal}
  {Phys.Lett.}\ }\textbf {\bibinfo {volume} {B705}},\ \bibinfo {pages} {106}
  (\bibinfo {year} {2011}{\natexlab{b}})},\ \Eprint
  {http://arxiv.org/abs/1108.1752} {arXiv:1108.1752 [hep-ph]} \BibitemShut
  {NoStop}%
\bibitem [{\citenamefont {Kuraev}\ \emph {et~al.}(1977)\citenamefont {Kuraev},
  \citenamefont {Lipatov},\ and\ \citenamefont {Fadin}}]{Kuraev:1977fs}%
  \BibitemOpen
  \bibfield  {author} {\bibinfo {author} {\bibfnamefont {E.~A.}\ \bibnamefont
  {Kuraev}}, \bibinfo {author} {\bibfnamefont {L.~N.}\ \bibnamefont {Lipatov}},
  \ and\ \bibinfo {author} {\bibfnamefont {V.~S.}\ \bibnamefont {Fadin}},\
  }\href@noop {} {\bibfield  {journal} {\bibinfo  {journal} {Sov. Phys. JETP}\
  }\textbf {\bibinfo {volume} {45}},\ \bibinfo {pages} {199} (\bibinfo {year}
  {1977})}\BibitemShut {NoStop}%
\bibitem [{\citenamefont {Balitsky}\ and\ \citenamefont
  {Lipatov}(1978)}]{Balitsky:1978ic}%
  \BibitemOpen
  \bibfield  {author} {\bibinfo {author} {\bibfnamefont {I.}~\bibnamefont
  {Balitsky}}\ and\ \bibinfo {author} {\bibfnamefont {L.}~\bibnamefont
  {Lipatov}},\ }\href@noop {} {\bibfield  {journal} {\bibinfo  {journal}
  {Sov.J.Nucl.Phys.}\ }\textbf {\bibinfo {volume} {28}},\ \bibinfo {pages}
  {822} (\bibinfo {year} {1978})}\BibitemShut {NoStop}%
\bibitem [{\citenamefont {Szymanowski}\ and\ \citenamefont
  {Zhou}(2016)}]{Szymanowski:2016mbq}%
  \BibitemOpen
  \bibfield  {author} {\bibinfo {author} {\bibfnamefont {L.}~\bibnamefont
  {Szymanowski}}\ and\ \bibinfo {author} {\bibfnamefont {J.}~\bibnamefont
  {Zhou}},\ }\href {\doibase 10.1016/j.physletb.2016.06.055} {\bibfield
  {journal} {\bibinfo  {journal} {Phys. Lett.}\ }\textbf {\bibinfo {volume}
  {B760}},\ \bibinfo {pages} {249} (\bibinfo {year} {2016})},\ \Eprint
  {http://arxiv.org/abs/1604.03207} {arXiv:1604.03207 [hep-ph]} \BibitemShut
  {NoStop}%
\bibitem [{\citenamefont {Meissner}\ \emph {et~al.}(2007)\citenamefont
  {Meissner}, \citenamefont {Metz},\ and\ \citenamefont
  {Goeke}}]{Meissner:2007rx}%
  \BibitemOpen
  \bibfield  {author} {\bibinfo {author} {\bibfnamefont {S.}~\bibnamefont
  {Meissner}}, \bibinfo {author} {\bibfnamefont {A.}~\bibnamefont {Metz}}, \
  and\ \bibinfo {author} {\bibfnamefont {K.}~\bibnamefont {Goeke}},\ }\href
  {\doibase 10.1103/PhysRevD.76.034002} {\bibfield  {journal} {\bibinfo
  {journal} {Phys. Rev.}\ }\textbf {\bibinfo {volume} {D76}},\ \bibinfo {pages}
  {034002} (\bibinfo {year} {2007})},\ \Eprint
  {http://arxiv.org/abs/hep-ph/0703176} {arXiv:hep-ph/0703176 [HEP-PH]}
  \BibitemShut {NoStop}%
\end{thebibliography}

%

\end{document}